\documentclass[aps,prb,reprint,floatfix,showpacs,superscriptaddress,groupedaddress]{revtex4-1}
\usepackage{hyperref}
\usepackage{graphicx}
\usepackage{amsmath}
\usepackage{amssymb}
\usepackage[usenames,dvipsnames]{xcolor}  
\usepackage{float}
\usepackage{epsfig}
\usepackage{natbib}
\usepackage{bm}
\usepackage[caption=false]{subfig}
\usepackage{amsfonts}
\usepackage{color} 
\begin{document}
\title{ Exploring Novel Quantum Criticality in Strained Graphene }
\author{S. Arya$^{1}$}\email{aryas@imsc.res.in}
\author{M. S. Laad$^{1}$}\email{mslaad@imsc.res.in}
\author{S. R. Hassan$^{1}$}\email{shassan@imsc.res.in}
\affiliation{$^{1}$Institute of Mathematical Sciences (Homi Bhabha National Institute), Taramani, Chennai 600113, India}

\begin{abstract}
  Strain tuning is increasingly being recognized as a clean tuning parameter to 
induce novel behavior in quantum matter.  Motivated by the possibility of straining graphene up to $20$ percent, we investigate novel quantum criticality due to 
interplay between strain-induced anisotropic band structure and critical 
antiferromagnetic spin fluctuations (AFSF) in this setting.  We detail how this interplay drives $(i)$ a 
quantum phase transition (QPT) between the Dirac-semimetal-incoherent 
pseudogapped metal-correlated insulator as a function of strain ($\epsilon$), 
and $(ii)$ critical AFSF-driven divergent nematic susceptibility near critical 
strain ($\epsilon_{c}$) manifesting as critical singularities in 
magneto-thermal expansion and Gr\"uneisen co-efficients.  The correlated band 
insulator at large strain affords realization of a 
two-dimensional dimerized spin-singlet state due to this interplay, and we 
argue how doping such an insulator can lead to a spin-charge separated metal, 
leading to anomalous metallicity and possible unconventional superconductivity.
On a wider front, our work serves to illustrate the range of novel states 
realizable by strain-tuning quantum materials.    
  
\end{abstract}
\maketitle
\vspace{1.4cm}

   The exciting discovery of single-layer graphene has spawned a tremendous burst of activity~\cite{RMP-graphene} on fundamental and applied grounds.  It is the strongest electronic material, capable of sustaining reversible elastic deformations in excess of $20$ percent.~\cite{liu}.  Besides, graphene possesses rich electronic properties, originating from gapless, linearly dispersing Dirac-like
excitations~\cite{geim}.  Attractive features like high mobility, absence of backscattering and strong field effects hold out the promise for future devices, 
but such possibilities are hindered by the gapless spectrum.  Inducement of 
a gap by quantum confinement is possible, but this also gives rise to edge 
roughness, with deleterious effect on electronic properties.  

   Recently, strain engineering has been studied~\cite{castro} as a way to 
profitably unite seemingly independent mechanical and electronic properties 
of graphene.  Experimental studies indicate that reversible and controlled 
strain up to $20$ percent can be produced in graphene~\cite{zxshen}, opening 
a unique opportunity to realize this aspect.  While detailed investigation of 
this interplay within one-electron band structure has been carried 
out~\cite{castro}, possibility of novel physics arising due to interplay 
between strain-modified electronic structure and electron-electron interactions
 has not received much attention.  Observed breakdown of the Wiedemann-Franz 
law~\cite{sachdev} near the charge neutrality point betrays the 
effect of sizable $e-e$ interactions in graphene.  In this light, one may 
anticipate that a strain-modified band structure could give rise to novel 
quantum phase transition(s) and novel physical responses as a result of 
interplay between an anisotropic electronic structure and local Coulomb 
interactions.  In this letter, we explore
 the possibility of inducing novel quantum phase transitions in ($20$ percent)
 strained graphene as a consequence of interplay between strong 
antiferromagnetic spin fluctuations (AFSF) and anisotropic band structure under
strain.  We stress that our analysis has potentially wider consequences for 
materials with honeycomb lattice structures, but with correlations not so 
strong as to induce a Mott transition (which requires complementary DMFT-like 
analyses~\cite{craco}).

   Our starting point is the electronic structure of graphene under 
strain~\cite{castro}.  We have repeated 
their earlier analysis.  As shown in detail in Supplementary Information (SI) \cite{SI} and in Fig.~\ref{fig1}, zig-zag strain induces sizable reconstruction of 
the electronic structure: it moves the Dirac point away from $K$ toward $M$ 
for finite strain, and $(ii)$ beyond a critical zig-zag strain 
($\epsilon_{c}\approx 0.23$), a band-insulating gap develops in graphene and the 
``Dirac point'' gets shifted to the $M$ point.  In contrast, armchair strain 
has very little effect on the electronic structure, so we will restrict 
ourselves to zig-zag strain hereafter.
\begin{figure}[ht]
\centering
\includegraphics[scale = 0.4]{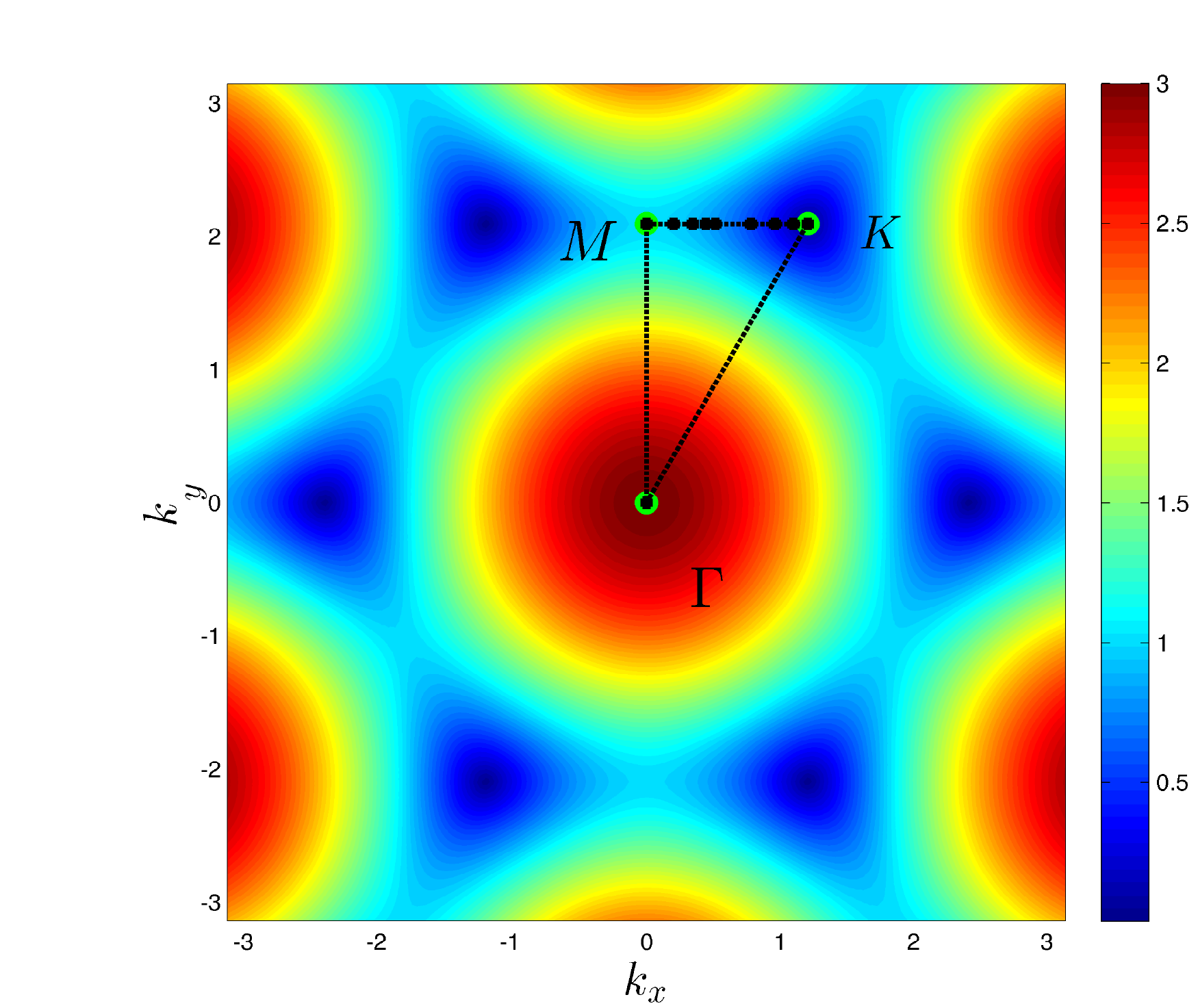}
\caption{Contour plot of the valence band for the unstrained case. As the strain is increased, the Dirac point shifts from the $K$ point to the $M$ point.}
\label{fig1}
\end{figure}
  Though scarcely studied~\cite{wehling,sheehy}, $e-e$ interactions are not 
negligible in graphene.  Realistic estimates put the intra-orbital Hubbard 
$U=U_{pp}\simeq 8-10$~eV within first-principles studies.  Our strategy is to 
vary $U/t$ from small to realistic values, where $t$ is the tight-binding hopping integral between nearest-neighbor $p_{z}$ orbitals on the honeycomb lattice.
Under zig-zag strain, the zig-zag hoppings $t_{x,x}=t_{y,y}$ progressively 
reduce, while the $z-z$ hopping $t_{z,z}$ increases.  Thus, interplay with 
Hubbard correlations is, very generally, expected to $(i)$ enhance 
strain-induced anisotropy, $(ii)$ produce a tendency to Mott localization 
beyond a critical $U_{c}/t$, and $(iii)$ perhaps most importantly, to induce 
exotic semi-metallic/insulating states around critical strain 
$\epsilon=\epsilon_{c}$ when $U$ is varied across a critical $U_{c}$.  In 
the $D=2$ case that applies in graphene, long-range AFSF should be very 
relevant.  While dynamic-mean-field theory (DMFT)~\cite{craco} or its cluster 
extensions can reveal ``Mottness'' effects, they cannot, by construction, 
access the effects of long-ranged critical AFSF and possible AF quantum 
criticality at low $T$.

   Here, we investigate the role of long-range critical AFSF in strained 
graphene.  We employ the Two-Particle Self-Consistent
(TPSC) theory: TPSC exploits the Pauli principle and exact sum 
rules~\cite{tremblay} relating wave-vector dependent susceptibilities to 
{\it local} correlators to formulate a non-trivially renormalized version of 
the random-phase approximation (RPA).  It is similar in spirit to the 
self-consistent renormalization (SCR) theory of Moriya and 
co-workers~\cite{moriya}.  Use of the renormalized Hubbard U (different for 
charge and spin channels), $U_{sp}=U \langle n_{\uparrow}n_{\downarrow}\rangle/\langle n_{\uparrow}\rangle \langle n_{\downarrow}\rangle$ leads to a negative 
feedback in RPA, preventing the divergence of the spin susceptibility at 
{\it any} finite $T$, in accord with stringent Mermin-Wagner restrictions in 
$D=2$.  Further, TPSC also recovers quantum critical (QC)-``renormalized 
classical (RC)'' and QC-quantum disordered (QD) crossovers at low but finite 
$T$, and yields low-$T$ AF correlation lengths in good accord with non-linear 
sigma model (NLSM) predictions~\cite{CHN}.  It is thus an excellent 
approximation for $D=2$ when coupling of carriers to long-range magnetic 
fluctuations is dominant.  Indeed, TPSC has been extensively 
employed~\cite{d-wave,cooling} within the itinerant spin-fluctuation theory 
to study AF quantum criticality in magnetic metals within the framework of 
Hubbard models.  

\begin{figure}[!ht]
\centering
\subfloat{\includegraphics[scale = 0.4]{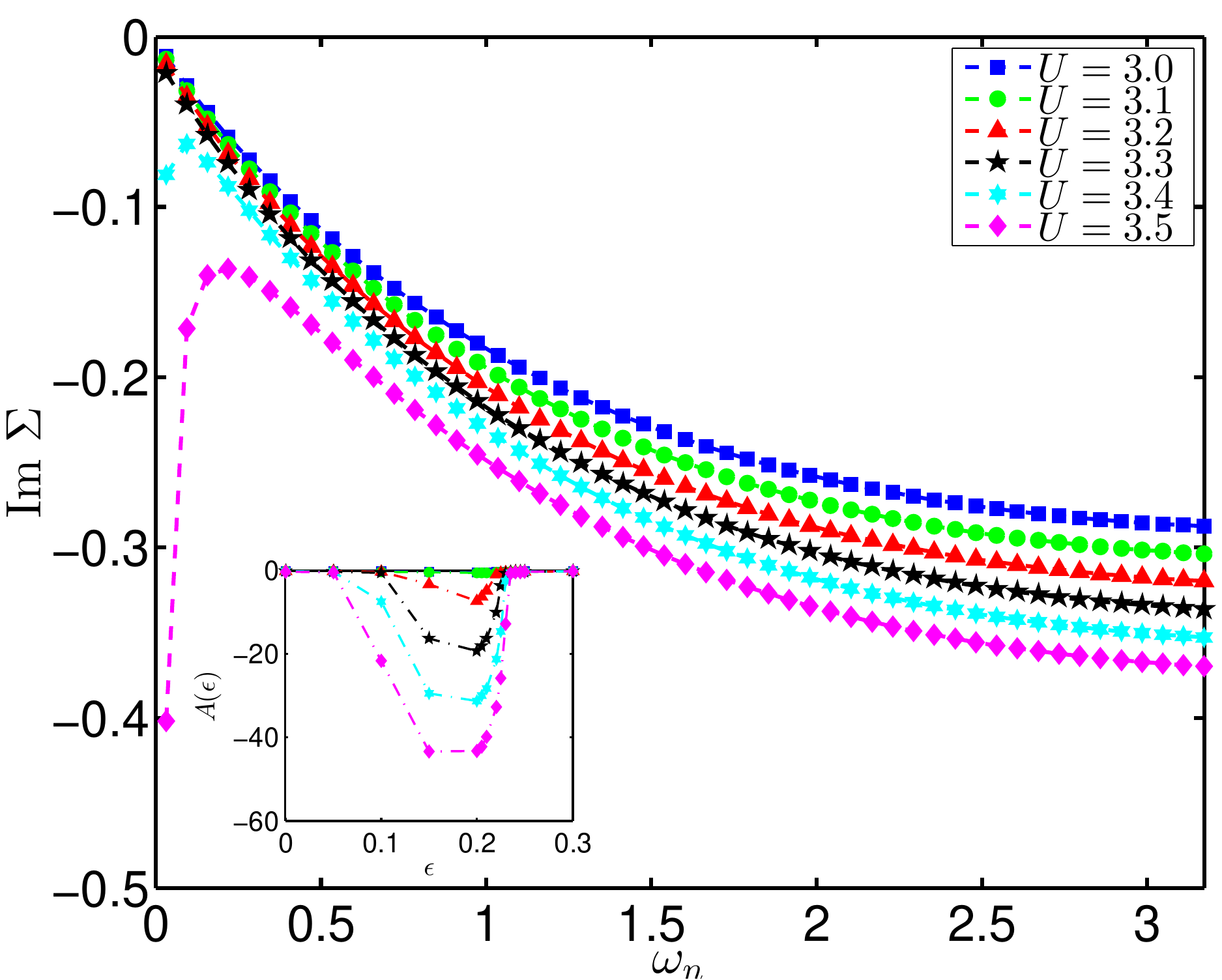}} 
\caption{Imaginary part of the self-energy $\Sigma$ as a function of Matsubara frequency $\omega_{n}$, for $\epsilon = 0.23$. The inset shows how the $A(\epsilon)$ co-efficient of -Im$\Sigma(\omega_{n})$ develops clear anomaly around $U/t=3.4$.}
\label{fig2}
\end{figure}

\begin{figure}[!ht]
\centering
\includegraphics[scale = 0.4]{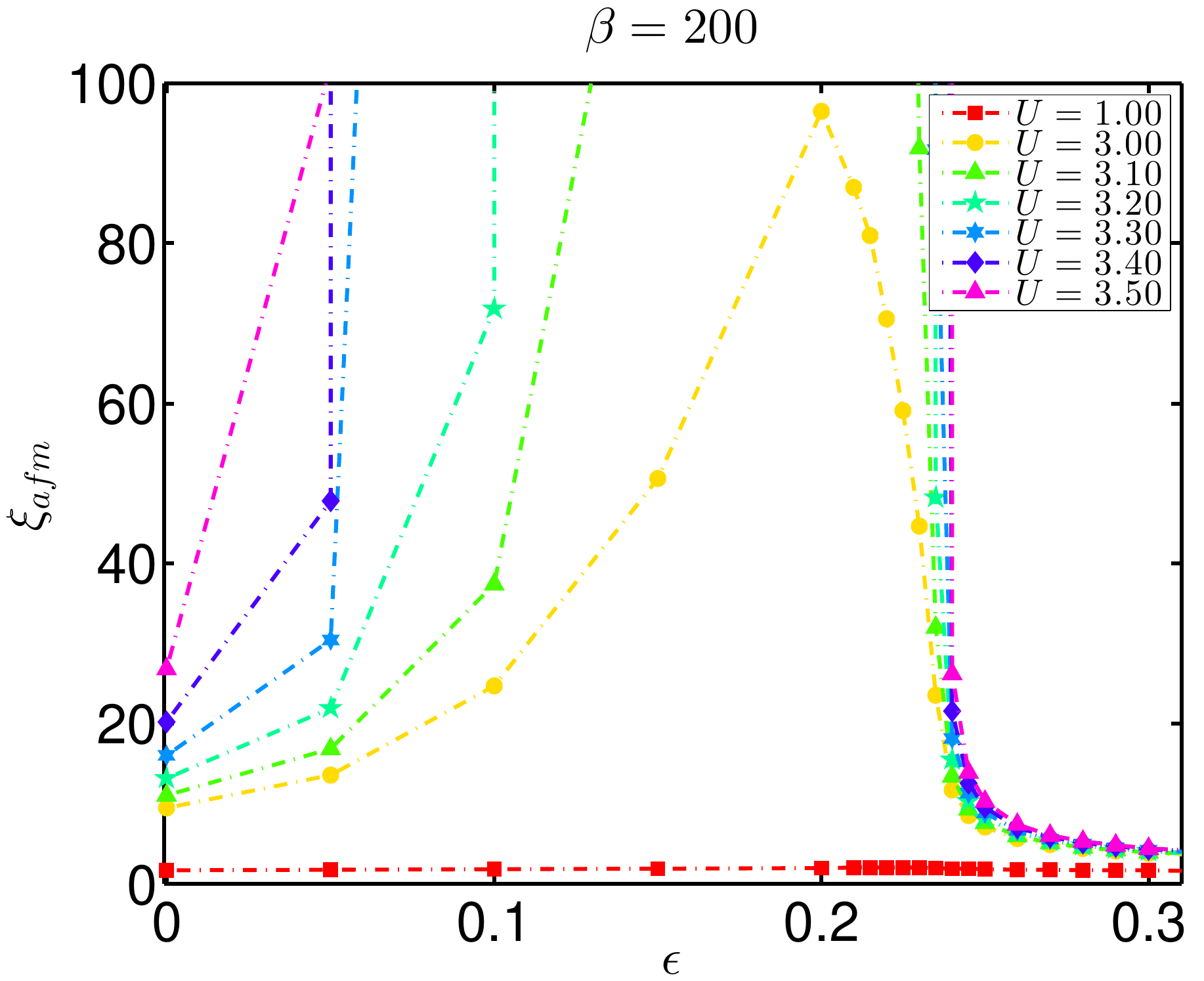}
\caption{Antiferromagnetic correlation length $\xi_{afm}$ plotted as a function of strain $\epsilon$ for various values of interaction, at $\beta = 200$. Antiferromagnetic fluctuations rapidly become critical with increasing $U$ as seen from the figure.}
\label{fig3}
\end{figure}

   We now present our results for strained graphene, relegating technical 
details to SI\cite{SI}.  For each value of strain ($\epsilon$), we vary the Hubbard 
$U$ in the interval $0<U<3.5$~eV choosing an unstrained value of $t=1.0$~eV.  
In Fig.~\ref{fig2}, we show the imaginary part of the self-energy, 
Im$\Sigma(i\omega_{n})$ \cite{SI} at the strain shifted ``Dirac'' point
as a function of Matsubara frequency $i\omega_{n}$.  For zero or small 
($\epsilon=0.05$) strain, the Dirac LFL metal remains stable for the range of 
$U/t$ we consider, as seen by the fact that -Im$\Sigma(i\omega_{n}) =A(i\omega_{n})$ for small $U$ \cite{SI}.  For $0.1 <\epsilon < 0.23$, however,
we find Im$\Sigma(i\omega_{n})=A(\epsilon)i\omega_{n}$, implying correlated 
Landau Fermi liquid (LFL), or Dirac liquid metallicity, for $U<U_{c}(\epsilon)$. \cite{SI}
  However, for larger $\epsilon$-dependent $U>U_{c}(\epsilon)$, 
Im$\Sigma(i\omega_{n})$ shows anomalous behavior: -Im$\Sigma(i\omega_{n})$ 
{\it increases} as $\omega_{n}\rightarrow 0$, reflecting appearance of a 
{\it pseudogap} in the one-particle excitation spectrum.  In the inset, we 
show how $A(\epsilon)$ develops an anomaly around $U_{c}/t=3.5$.  Further, 
since gapless electronic states exist only at the ``Dirac'' point, 
Im$\Sigma(\omega_{n})$ also reflects the $T$-dependence of the $dc$ 
resistivity: thus, given $\rho_{dc}(T)=AT^{2}$ in the LFL regime, 
we expect the $A$-co-efficent of the resistivity to show a similar anomaly,
reflecting a change from LFL metal to an incoherent non-LFL metal as 
$U\rightarrow U_{c}$ around the critical strain.  
  Within TPSC, the origin of this feature is associated with the coupling of 
itinerant $p_{z}$-band carriers to long-ranged, critical AF spin fluctuations
(AFSF).  In Fig.~\ref{fig3}, we show that while the AF correlation length 
($\xi_{afm}$) is finite for $U<U_{c}$, it indeed diverges (for low $T$, 
$\beta=1/T=200$) at $T\rightarrow 0$ for $U\geq U_{c}$, confirming this view.  
It is known that $\xi_{afm}\simeq$ e$^{a(U)/T}$ in the quantum critical region
within TPSC (see Fig.~\ref{fig3}), in accord with predictions of the NLSM 
approach.  These features are thus manifestations of 
the crossover from QD to the QC regime of AFSF as $U$ increases: it is the 
coupling of itinerant $p_{z}$ carriers (encoded in the bare one-electron propagator, $G_{0}(\bm{k},i\omega_{n})$) to nearly critical AFSF, encoded in the spin 
susceptibility $\chi_{s}({\bf q},i\omega_{n})\simeq [i\omega_{n}+ q + \xi^{-1}(T)]^{-1}$ with $\xi_{afm}(T)$ as above that drives the 
QD-QC crossover.   
   As expected, in Fig.~\ref{fig3}, we show that increasing strain (toward 
$\epsilon_{c}$) reduces $U_{c}$ for the QD-QC crossover.  This is due to the 
fact that reduction in $t_{x,x}$ along the zig-zag direction under strain 
enhances the effective $U/t$ anisotropically (it becomes larger for the zig-zag
 relative to $z-z$ bonds), resulting in enhanced AFSF for smaller $U/t_{x,x}$.  
This also implies that the renormalized AFSF under strain will be enhanced
by nearly-divergent $\chi_{s}({\bf q},i\omega_{n})$, and that the feedback 
effect of the anisotropic AFSF will in turn enhance the electronic anisotropy.
While this issue requires a self-consistent extension of TPSC in its current 
form, we consider this issue (see below) while investigating strain-enhanced 
nematicity.
\begin{figure}[!ht]
\centering
\includegraphics[scale = 0.4]{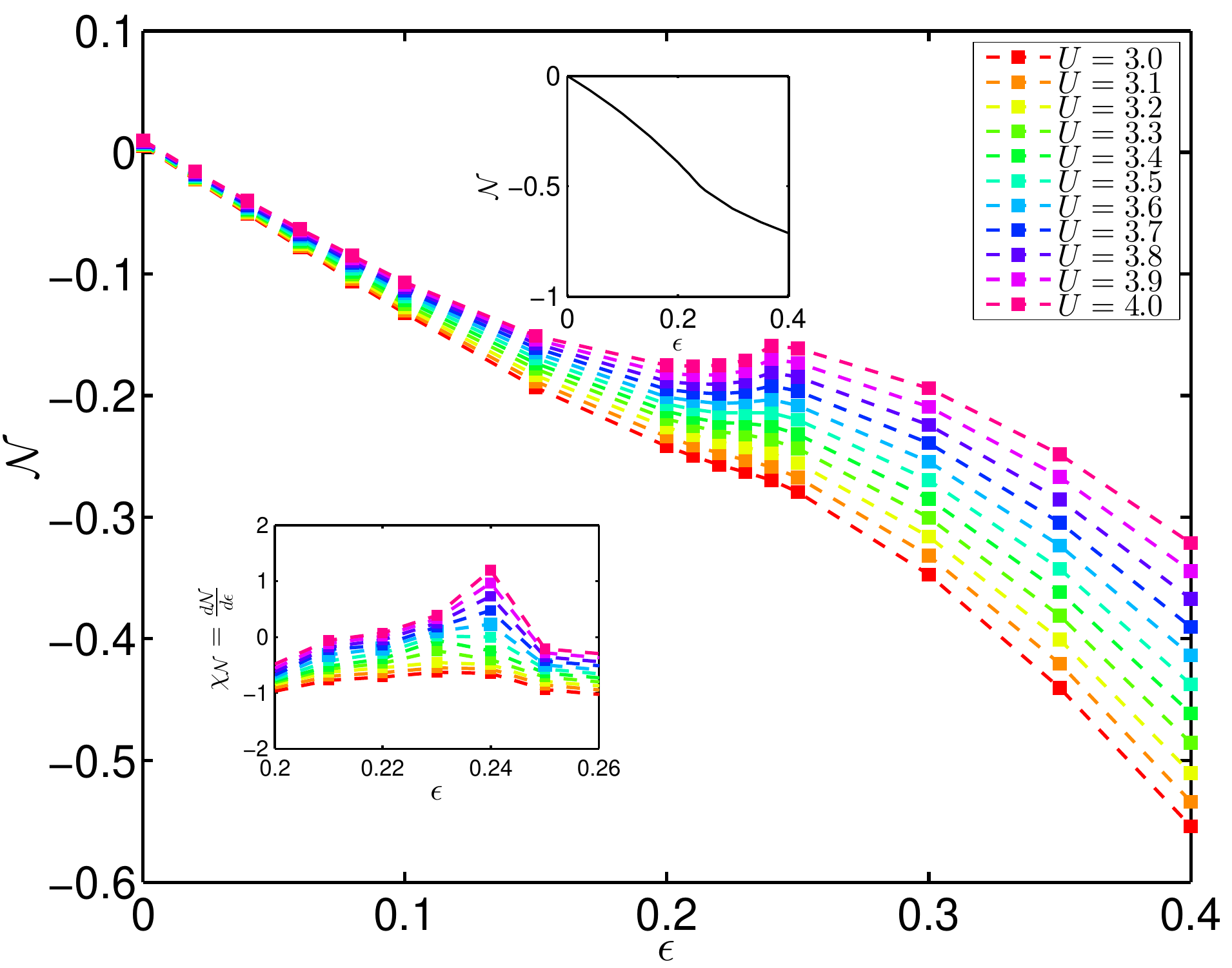}
\caption{$\mathcal{N}=\frac{\langle(T_{xx}-T_{zz})\rangle}{\langle(T_{xx}+T_{zz})\rangle}$, plotted as a function of strain $\epsilon$ for given values of interaction. The upper inset shows $\mathcal{N}$ for $U=0$.  The lower inset shows the nematic susceptibility, 
$\chi_{N} = \frac{d \mathcal{N}}{d \epsilon}$, which shows singular behavior around critical strain due to interplay between anisotropic band strcuture and critical AFSF.}
\label{fig4}
\end{figure}

   Anisotropic electronic structure modification under strain has profound 
consequences.  Since uniaxial strain acts as a field conjugate to the nematic
order parameter, electronic nematic (EN) criticality itself will be washed out 
by strain, ruling out EN itself as a candidate for the quantum phase transition
(QPT) under strain.  However, since strong AFSF directly feed back on the TPSC
self-energy~\cite{tremblay} and hence the renormalized propagator, the 
non-trivial renormalization effects do show up in $G({\bf k},i\omega_{n})=[G_{0}^{-1}(\bf{k},i\omega_{n})-\Sigma(\bf{k},i\omega_{n})]^{-1}$.  Computing the average 
(renormalized) kinetic energies along $xx,zz$ bonds from 
$G({\bf k},i\omega_{n})$ thus allows for a direct estimate of the influence of 
strong AFSF in enhancing strain-induced electronic anisotropy,
and hence on {\it bond} EN state.  In Fig.~\ref{fig4}, we show variation of 
$\mathcal{N} =\frac{\langle(T_{xx}-T_{zz})\rangle}{\langle(T_{xx}+T_{zz})\rangle}$, with $\langle T_{\mu\mu}\rangle=\langle c_{i\sigma}^{\dag}c_{i+\mu,\sigma}\rangle$ with $\mu=x,z$ without and with the Hubbard 
$U$.  It is rather clear that EN ``order'' is significantly enhanced by strong 
AFSF, the origin of which is anisotropic suppression of $\langle T_{xx}\rangle$ 
relative to $\langle T_{zz}\rangle$, the latter itself being a direct 
consequence of AFSF-driven anisotropic renormalization of $t_{x,x}<t_{z,z}$ 
under strain.  Finally, notwithstanding 
the absence of quantum criticality associated with EN order, we find that 
$d\langle N\rangle/d\epsilon$ shows an abrupt change precisely around 
$U_{c}(\epsilon)$.
Thus, there {\it is} evidence for nematic criticality as a function of strain: 
to characterize it, we observe that $(i)$ if bond EN order would exist without 
strain, it would break the discrete ($C_{6v}$ for graphene) rotational symmetry 
of the crystal lattice, implying spontaneous breaking of an Ising-like 
symmetry, but $(ii)$ strain washes out this criticality.  However, since strain
 is a field conjugate to $\langle N\rangle$, nematic critiality can still 
occur~\cite{fisher,nevidomskyy} at finite strain, as in Fe arsenides.  Such a 
criticality should show up as a divergent nematic susceptibility, 
$\chi_{N}(\epsilon)=d \mathcal N/d \epsilon$, near $U_{c}$.  
In the inset to Fig.~\ref{fig4}, we show that this is indeed what 
we find.  This leads us to the central conclusion of this work: {\it a 
nematic quantum criticality of the quantum liquid-gas variety, driven by strong
 AFSF around the critical $U_{c}$, underlies the ``transition'' from Dirac-LFL 
to an incoherent metal near critical strain}. 

  Lack of LFL quasiparticles for $U>U_{c}(\epsilon)$ as above implies that
one should expect an (anisotropic) breakdown of the Wiedemann-Franz (WF) law.  
This has recently been shown to occur for neutral graphene, leading to 
possibility of a novel ``Dirac liquid''~\cite{sachdev} with transport due to 
collective excitations.  Thermal and electrical transport under strain could 
help in determining the fate of the WF law under strain.  
However, use of strain as a tuning parameter suggests that studying
magnetic-fluctuation contributions to magneto-volume, thermal expansion and
Gr\"uneisen co-efficents could more naturally facilitate observation of the 
above QPT.  That it is possible to reversibly tune the strain to about $20$ 
percent could make the above scenario realistically possible.  Motivated 
hereby, we now propose that critical divergences in thermal expansion 
and magnetic Gr\"uneisen co-efficients due to divergent AFSF near $\epsilon_{c}$
 can be fruitfully employed to unearth the novel criticality proposed here. 

   In our picture, coupling of strain to AFSF results in a magneto-elastic 
interaction.  This will have novel consequences for spin fluctuation 
contribution to the magneto-volume, defined as $\omega_{m}(T)=\delta V(T)/V$  
and thermal expansion co-efficient, 
$\alpha_{m}(T)=\partial \omega_{m}/\partial T$.  Such effects have long been 
well-documented in $d$-band transition metals and compounds
and explicated by the well-known Moriya-Usami approach~\cite{moriyausami} 
within SCR theory.  Given that TPSC is an advanced variant of SCR theory, we 
can use the SCR formulation to investigate magneto-elastic effects in strained 
graphene.

\begin{figure}[!ht]
\centering
\includegraphics[scale = 0.4]{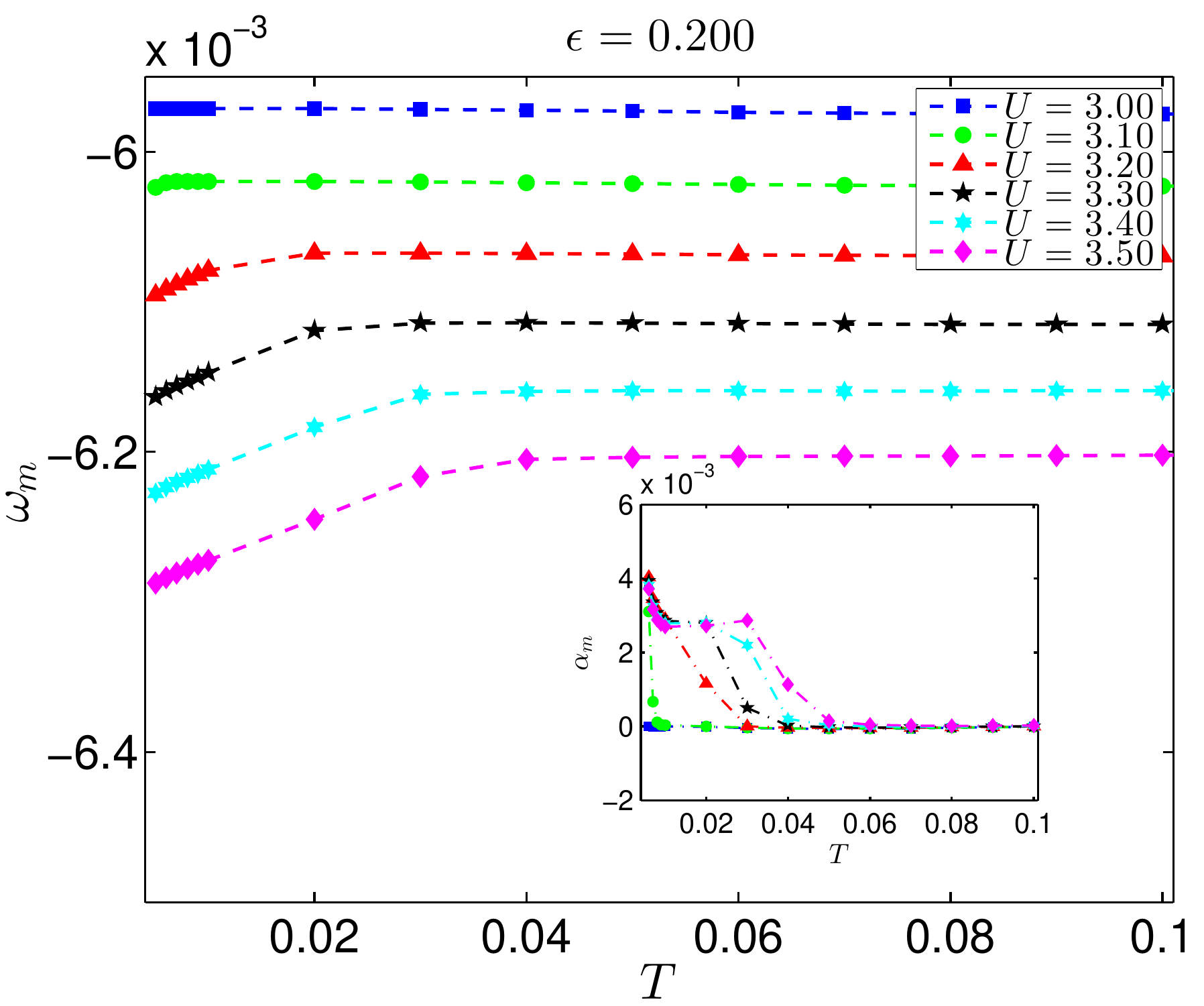}
\caption{Magneto-volume, $\omega_{m}$ as a function of temperature $T$. The inset shows a plot of thermal expansion co-efficient, $\alpha_{m}$ as a function of $T$, showing how it diverges as $T \rightarrow 0$ due to coupling of strin to critical AFSF for $U\geq 3.2$ near critical strain ($\epsilon=0.20$).}
\label{fig5}
\end{figure}

  Moriya-Usami theory asserts that electronic correlation or spin fluctuation 
contribution to the magneto-volume in the quantum paramagnetic state ($T>T_{N}(=0)$ in $D=2$) is related, apart from parameters $D_{0}$ (related to the strain 
dependence of the band-width via $D_{0}=\frac{\partial ln(W)}{\partial ln \epsilon}$) and $B$, the bulk modulus, to the magnetic fluctuations, 
$\langle \delta M^{2}\rangle=(\langle S_{z}^{2}\rangle$, the latter evaluated 
in the quantum paramagnetic phase.  This can be evaluated using 
$D=\langle n_{\uparrow}n_{\downarrow}\rangle =(n-\langle S_{z}^{2}\rangle)/2$,
 with $D$, the double occupancy evaluated from TPSC (from $\chi({\bf q},i\omega_{n})$) using the 
TPSC sum rule, providing a direct way to study magnetoelastic effects near the 
QCP above.  In Fig.\ref{fig5}, we show $\omega_{m}(T)$ and $\alpha_{m}(T)$ for 
a range of $U$ near critical strain to illustrate the main features of 
interest. $D_{0}$ is negative and of the order of 1, while $B \approx 200 Nm^{-1}$.
        
   Interestingly, though the magnetic fluctuation contribution to 
$\omega_{m}(T), \alpha_{m}(T)$ is sizable at small strain, we find that 
{\it both} contributions vanish as $T \rightarrow 0$.  Beyond a 
strain-dependent $U_{c}(\epsilon)$, however, marked changes show up: most 
interestingly, the influence of long-range critical AFSF reveal themselves 
in a critical divergence in $\alpha_{m}(T\rightarrow 0)$.  Further, this 
should have direct bearing on the divergence of the {\it magnetic} 
Gr\"uneisen co-efficient, defined as $\Gamma_{m}(T)=\alpha_{m}(T)/C_{el}(T)$, 
where $C_{el}(T)$ is the electronic contribution to the specific heat.  It is 
numerically involved to compute $C_{el}$ within TPSC: however, we expect 
$C_{el}(T)\simeq T^{2}$ in $D=2$ up to logarithmic corrections (in fact,
within the field-theoretic RG work of Sheehy {\it et al}~\cite{sheehy}, one 
finds $C_{el}(T)\simeq T^{2}/(1+b$log$T)^{2}$ with $b$ a constant.  If this 
were to qualitatively hold right up to $U_{c},\epsilon_{c}$ from the LFL side,    it would imply that $\Gamma_{m}(T\rightarrow 0)$ would also show a critical 
divergence.  Furthermore, these hallmarks of quantum criticality disappear 
beyond critical strain \cite{SI}, pinning them to the influence of critical 
AFSF.

   Anomalies in $\omega_{m}(T), \alpha_{m}(T)$ and $\Gamma_{m}(T)$ have recently
 received intense attention in the context of quantum criticality in $d$- and
$f$-band materials~\cite{gegenwart}.  We can now rationalize the 
critical divergence of $\alpha_{m}, \Gamma_{m}$ near critical strain at $U$ 
close to $U_{c}(\epsilon)$.  These are directly related to singular 
$T$-dependence of $\partial\langle S_{z}^{2}\rangle/\partial T$ due to 
diverging AFSF at sizable $U$ in the vicinity of $\epsilon_{c}$ as shown above.
  In light of inducement of reversible strain~\cite{zxshen} up to $20$ percent,
such a novel QCP with diverging nematic susceptibility at finite strain 
is potentially realizable in practice.  More generally, {\it metallic} 
honeycomb materials involving $Cu$ in $d^{9}$ configuration~\cite{baker} 
(whence modelled by one-band Hubbard model) could exhibit such anomalies as we 
find near a pressure-driven Mott transition between a spin-liquid insulator 
and (incoherent) metal.
\begin{figure}[ht!]
\centering
\subfloat[]{\includegraphics[scale = 0.3]{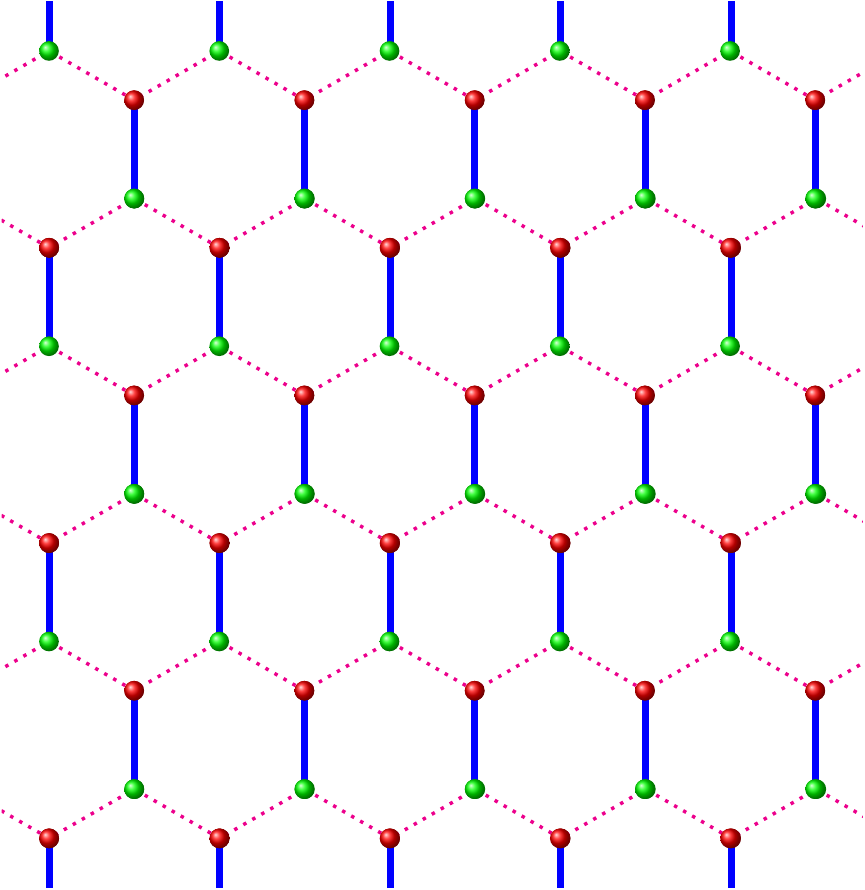}}\qquad
\subfloat[]{\includegraphics[scale = 0.3]{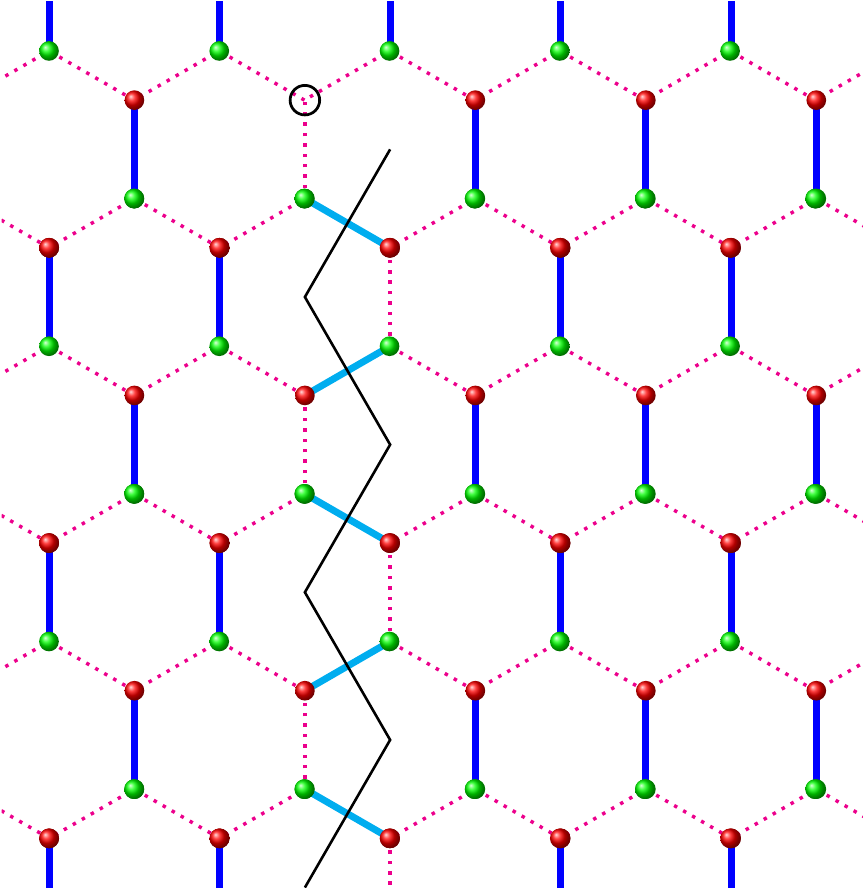}}\\
\subfloat[]{\includegraphics[scale = 0.3]{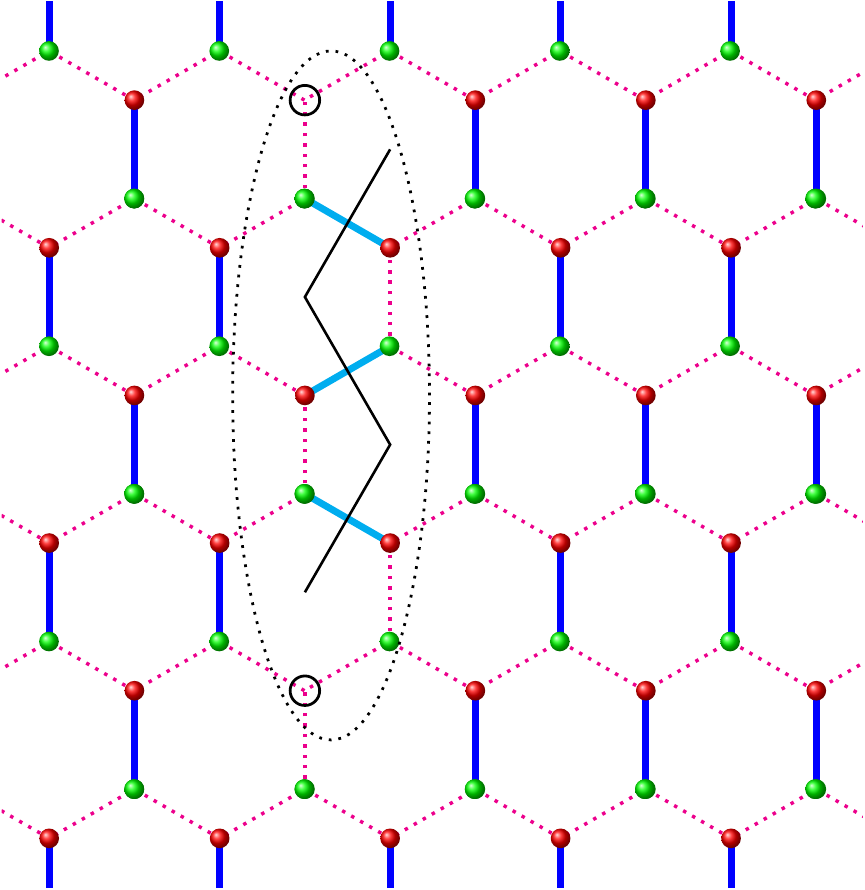}}
\caption{Combination of anisotropic strain and strong AFSF beyond $\epsilon_{c}$ produces a valence bond solid correlated insulator (panel (a)), with preferential dimerization of $z-z$ bonds of the honeycomb lattice.  A single doped hole (panel (b)) generates a ``defect string''
of singlets with local Kekule like structure, inhibiting coherent one-hole propagation.  With two doped holes (panel (c)), the defect string of singlets is healed.  This allows the hole pair, along with the intervening three singlets (shown as the dashed loop in panel (c)), to propagate coherently through the lattice. }
\label{fig7}
\end{figure}
   We conclude by discussing the novel prospect of fractionalization of 
added electrons (holes) in heavily strained graphene.  Since $t_{x,x}<t_{z,z}$ 
is enhanced by AFSF under strain, it is reasonable to expect that at sizable 
$U>U_{c}$ $i.e$, in the 
correlated ``band'' insulator beyond $\epsilon_{c}$, the combined effect
will be to readily promote spin-singlet 
valence-bond pairing on $zz$-bonds.  As shown in Fig.~\ref{fig7}, it then 
follows that this state can be viewed as a $D=2$ generalization of (dimerized) 
polyacetylene.  In our case, it is the interplay between strain and strong 
AFSF (which also promotes the nematic state) which produces the dimerized state
 in the insulator.  In the spinful case of relevance here, a single doped hole 
corresponds to a missing state in the valence band, and unoccupied, singly 
and doubly occupied states correspond~\cite{franz} to charge and spin quantum 
numbers $(Q=-e,S=0)$, $(Q=0,S=1/2)$ and $(Q=e,S=0)$.  In panel (b) of Fig.~\ref{fig7}, we schematically show how a doped hole results in
generation of an extended ``string defect'' of singlets emanating from the bond hosting the hole: clearly the unpaired spin, originally in the vicinity of the hole, separates from it by creating such a string.  Thus, a doped hole will 
fractionalize into a spinless {\it holon} and a spin-$1/2$ spinon!  In the continuum limit, this could be
 seen by solving the Bogoliubov-de Gennes like equation 
for a doped hole, which results in only {\it one} renormalizable solution for a vortex or antivortex.  This bares an alternative route
to create fractionalized excitations by strain tuning in honeycomb lattice
models.  These solitonic excitations are defects in the dimerized ground state,
and have exciting consequences.  Specifically, beginning with such an insulator, doping should lead to a fractionalized state, wherein electrical current is 
carried by holons but the Hall current by spinons, as is implicit in the two-relaxation rate scenario~\cite{anderson}.  Further, if two doped holes sit on different dimers, the hole pair, dressed by the dynamically fluctuating short-range singlet correlations (see the dashed region in Fig.~\ref{fig7}, panel (c)) could hop as a composite pair without scrambling the background 
spin configuration, leading to unconventional superconductivity in a 
spin-gapped background (this would be a $D=2$ version of the Luther-Emery scenario in $D=1$).  
Our study points toward such exotic possibilities, but we leave more detailed exploration of these themes for the future.

   To conclude, we have presented a specific scenario to realize a range of 
unconventional states of matter by use of ``clean'' strain-tuning in graphene.
Our work motivates study of other interesting materials like phosphorene, 
silicene, among other honeycomb systems, whose unstrained structures bear
some resemblance to graphene.  It should also be an attractive tool to 
investigate novel strain-induced QPTs in topological and Weyl systems, where 
the role of sizable Coulomb interactions remains largely unexplored, and 
should be a theme of potential interest.   
     
%
\pagebreak
\onecolumngrid
\begin{center}
\textbf{\large Supplementary Information for `Exploring Novel Quantum Criticality in Strained Graphene'}
\end{center}
\setcounter{equation}{0}
\setcounter{figure}{0}
\setcounter{table}{0}
\setcounter{page}{1}
\makeatletter
\renewcommand{\theequation}{S\arabic{equation}}
\renewcommand{\thefigure}{S\arabic{figure}}
\renewcommand{\bibnumfmt}[1]{[S#1]}
\renewcommand{\citenumfont}[1]{S#1}

\section {Tight-binding analysis of graphene under strain}
The application of uniaxial strain on graphene has been studied in the tight-binding approximation by Pereira and co-workers \cite{Pereira2009}. The major results are i) the application of strain opens up a gap in the bandstructure ii) the strain has to cross a threshold value for the gap to open up and iii) the application of strain along every direction does not result in a bandgap, for instance a strain along the armchair direction never opens a gap. Along the zigzag direction, a threshold strain of $\epsilon_{c} > 0.23$ is needed for the gap to open up. 

\begin{figure}[!ht]
\centering
\subfloat{\includegraphics[scale = 0.33]{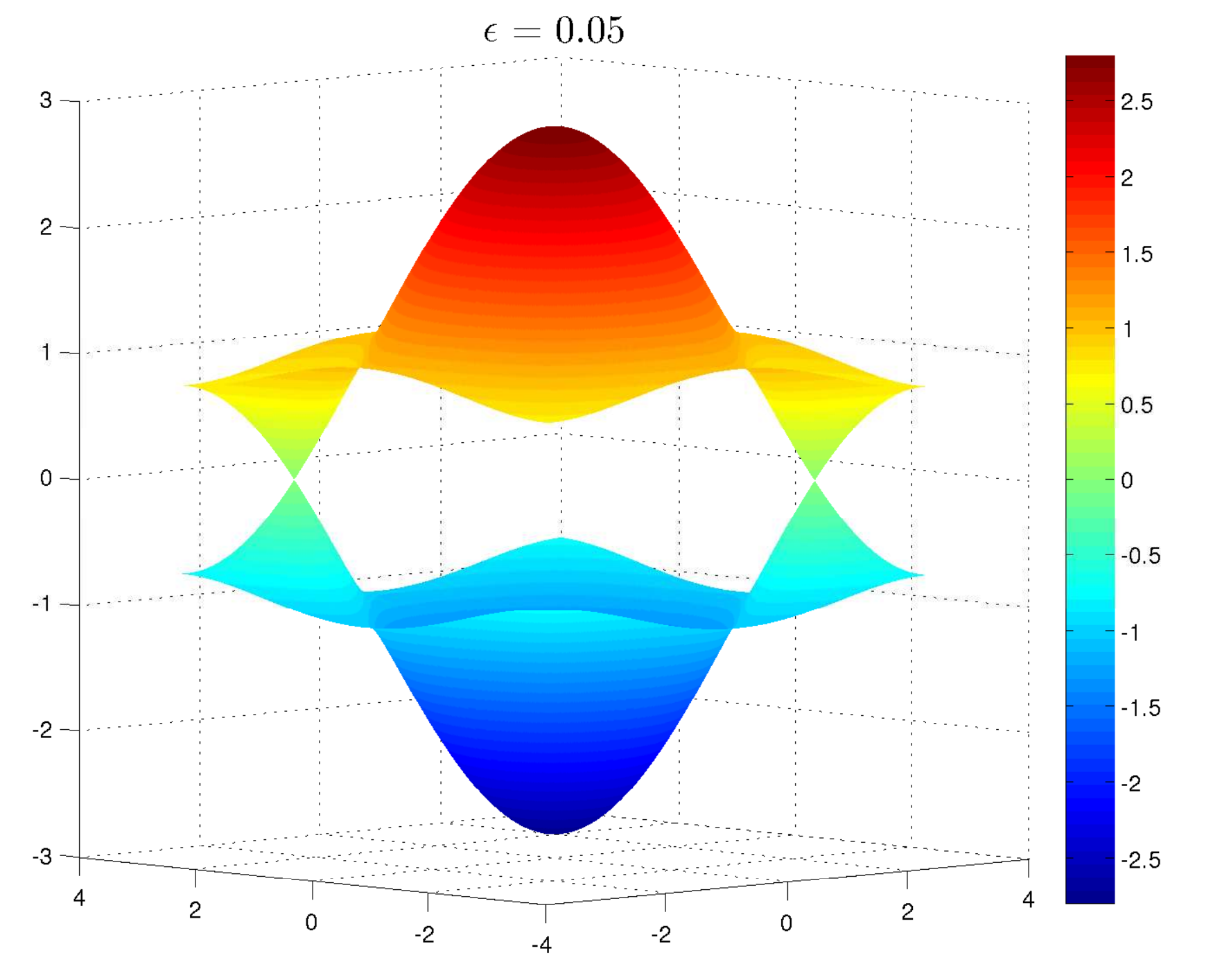}}
\subfloat{\includegraphics[scale = 0.33]{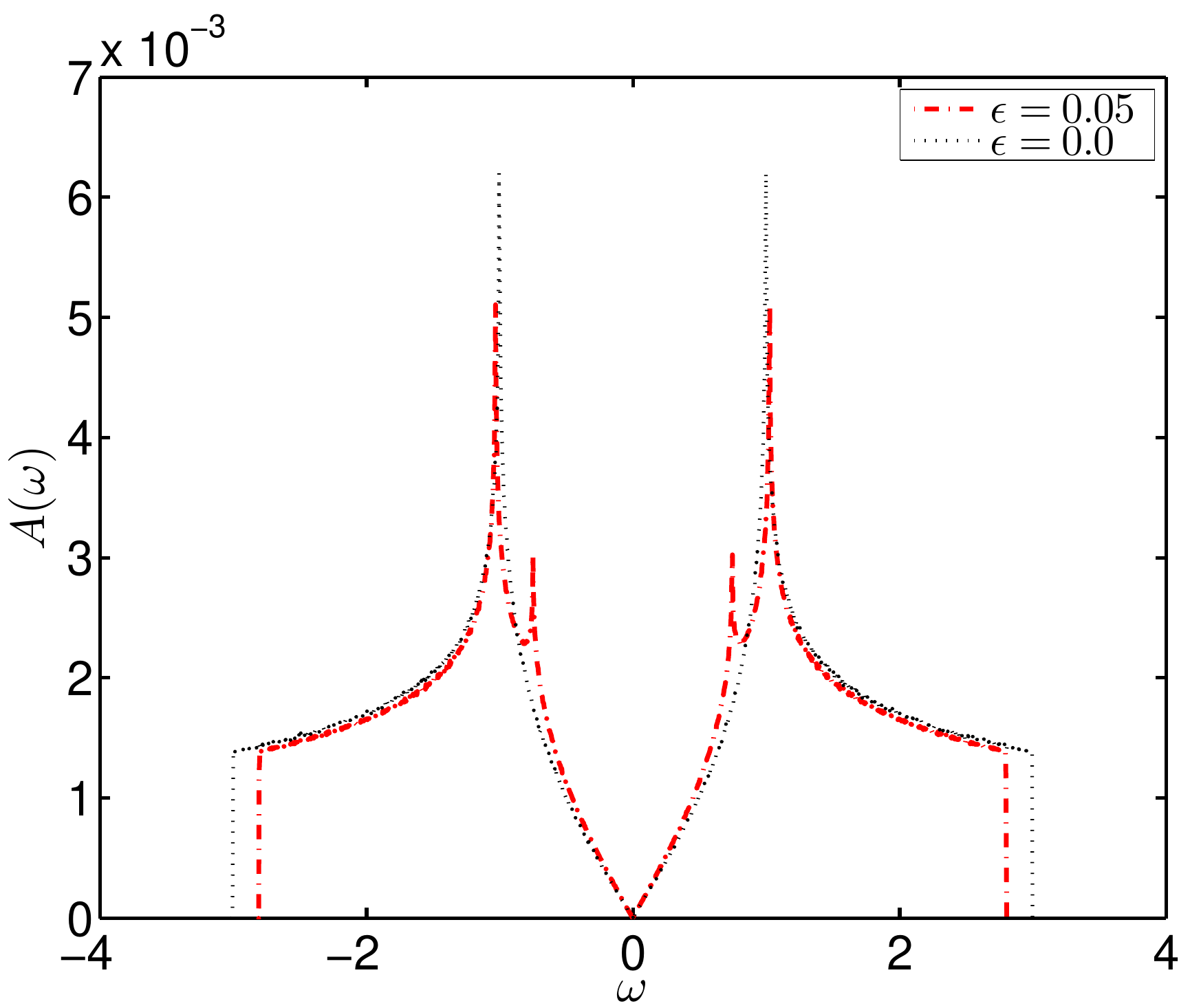}}\\
\subfloat{\includegraphics[scale = 0.33]{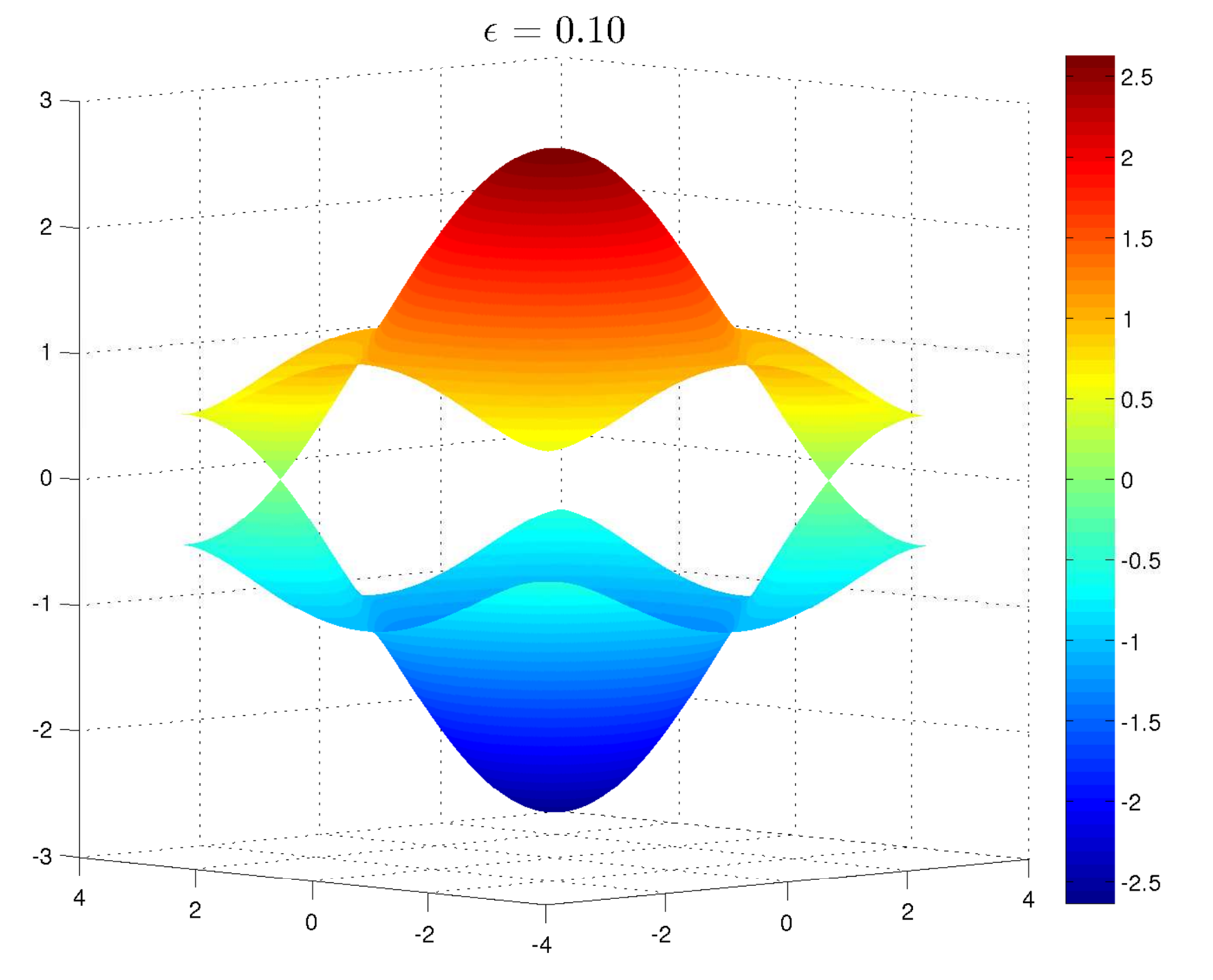}}
\subfloat{\includegraphics[scale = 0.33]{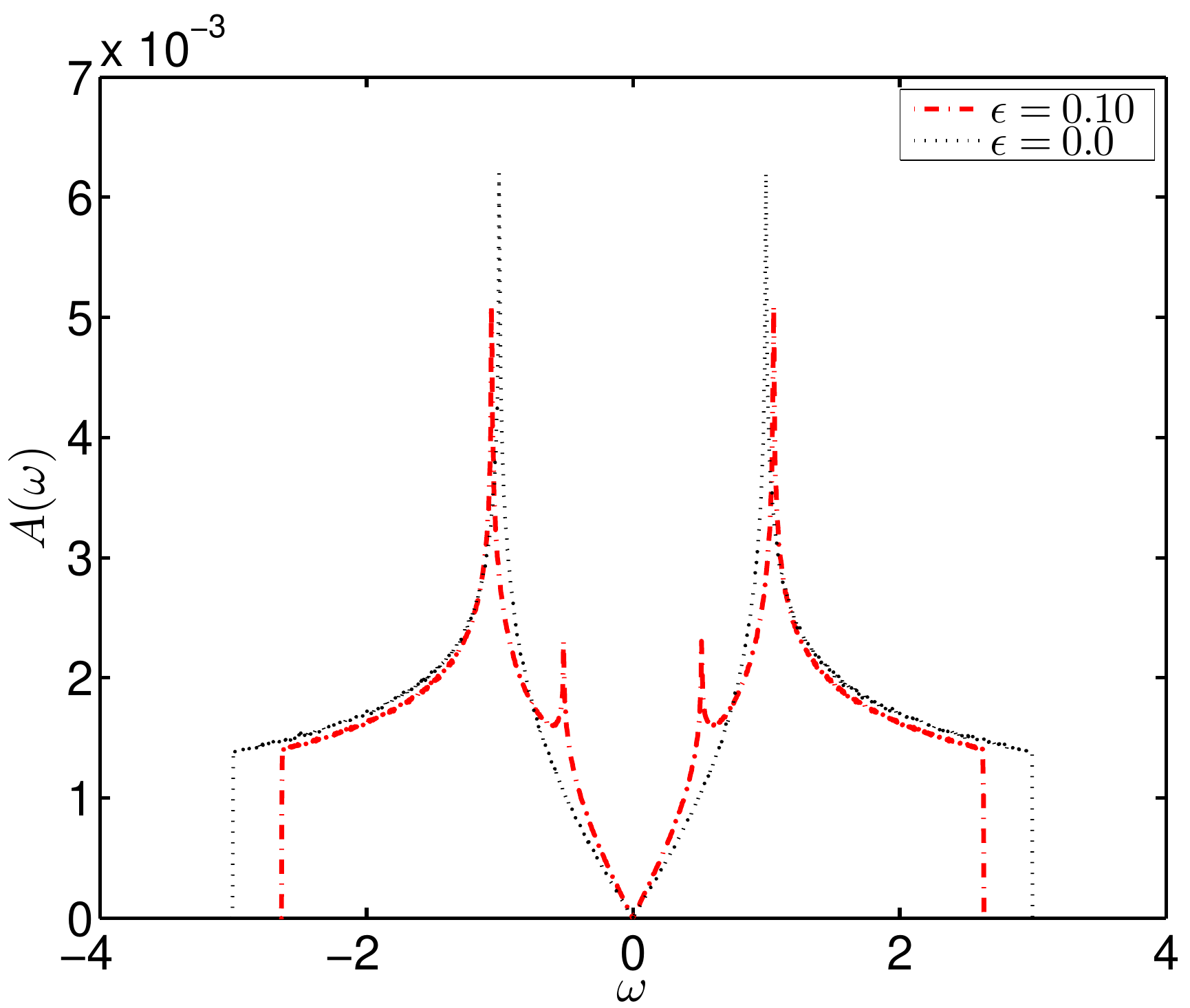}}\\
\subfloat{\includegraphics[scale = 0.33]{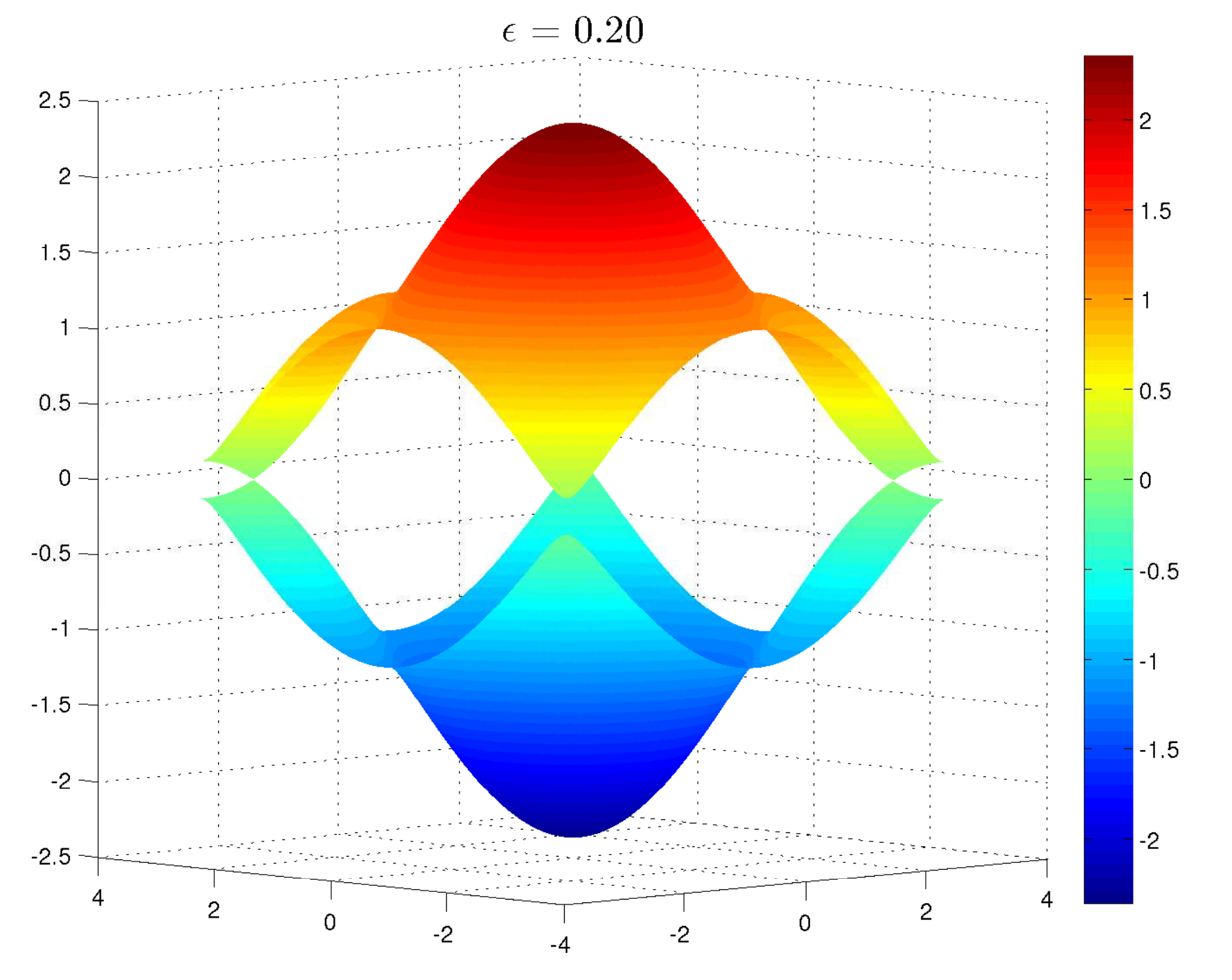}}
\subfloat{\includegraphics[scale = 0.33]{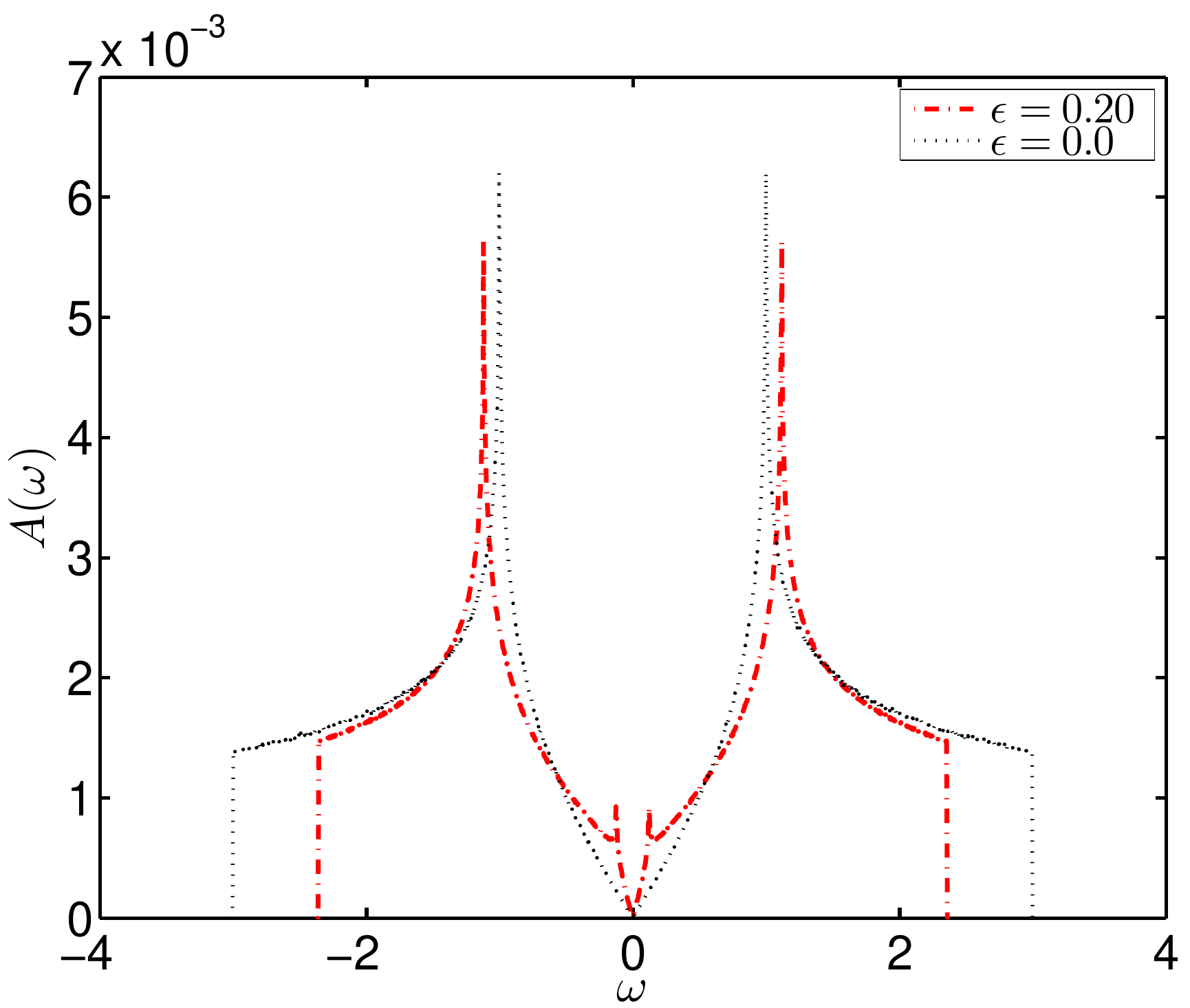}}\\
\subfloat{\includegraphics[scale = 0.33]{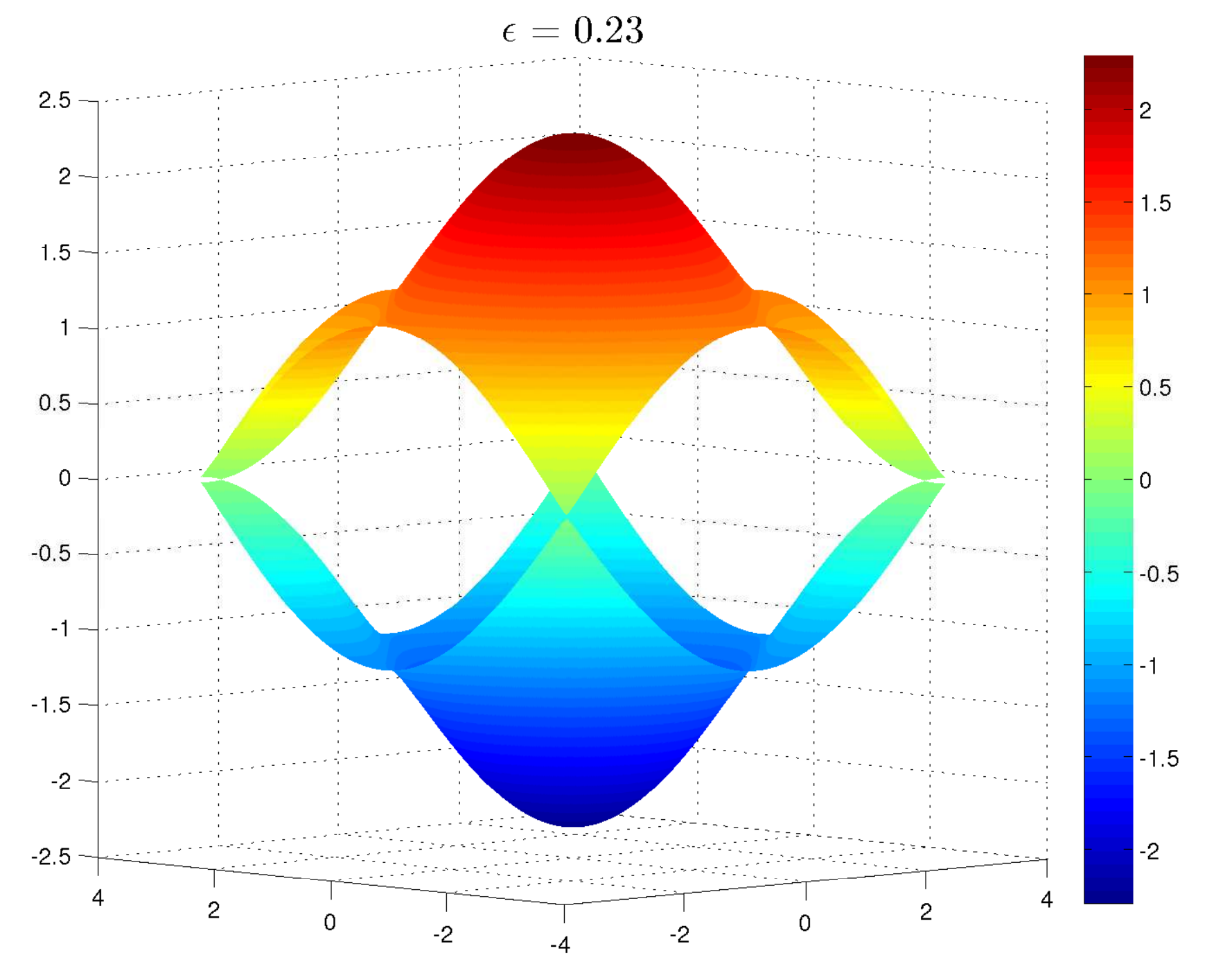}}
\subfloat{\includegraphics[scale = 0.33]{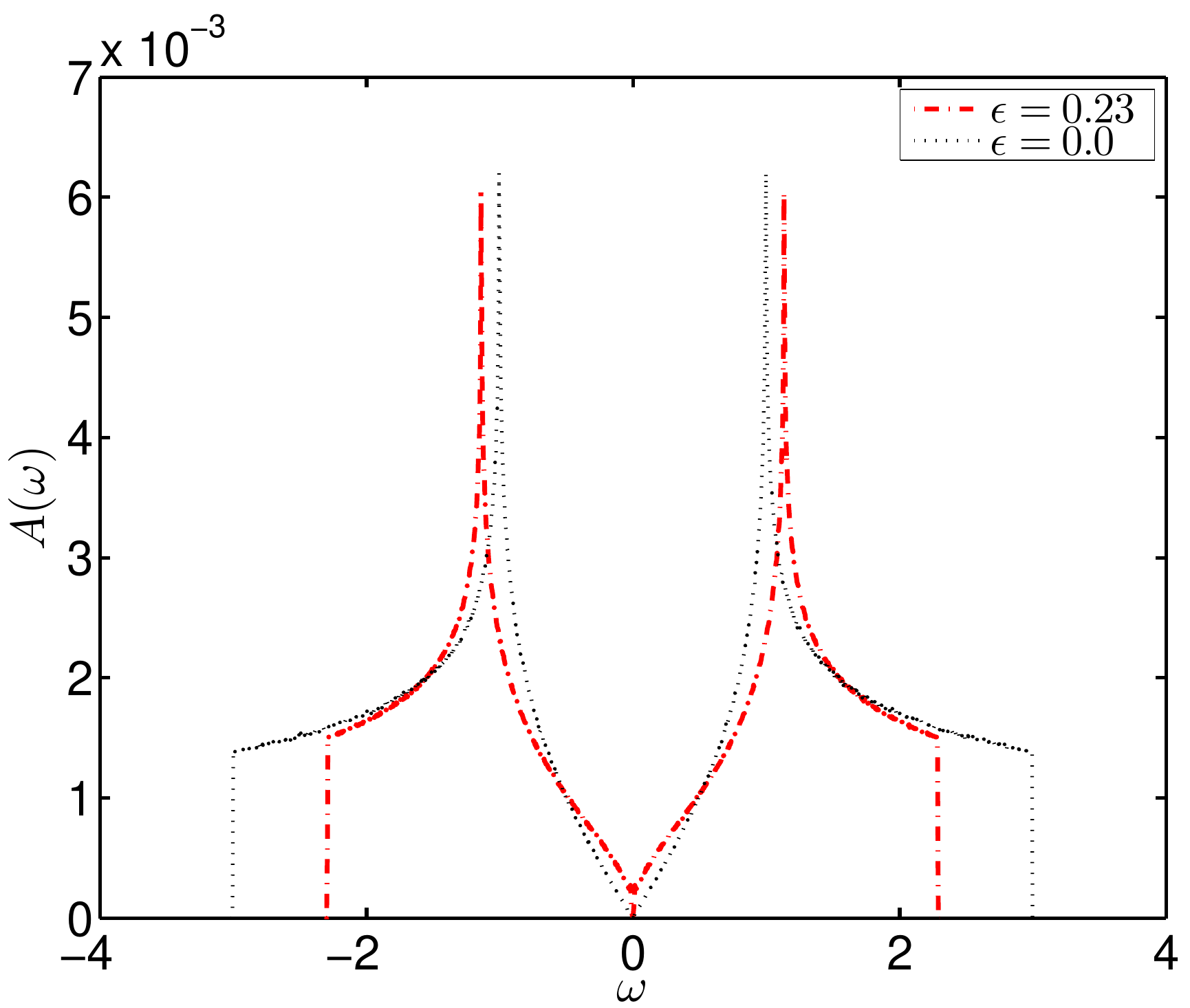}}
\caption{Plots of bandstructure and density of states (DOS) $A(\omega)$ for graphene under strain $\epsilon$ along the zigzag direction. The values of strain shown here are less than the critical strain needed to induce a bandgap. The DOS for the unstrained case is also plotted in black dotted lines for comparison.}
\label{fig_bs_dos_till_eps_crit}
\end{figure}

\begin{figure}[!ht]
\centering
\subfloat{\includegraphics[scale = 0.25]{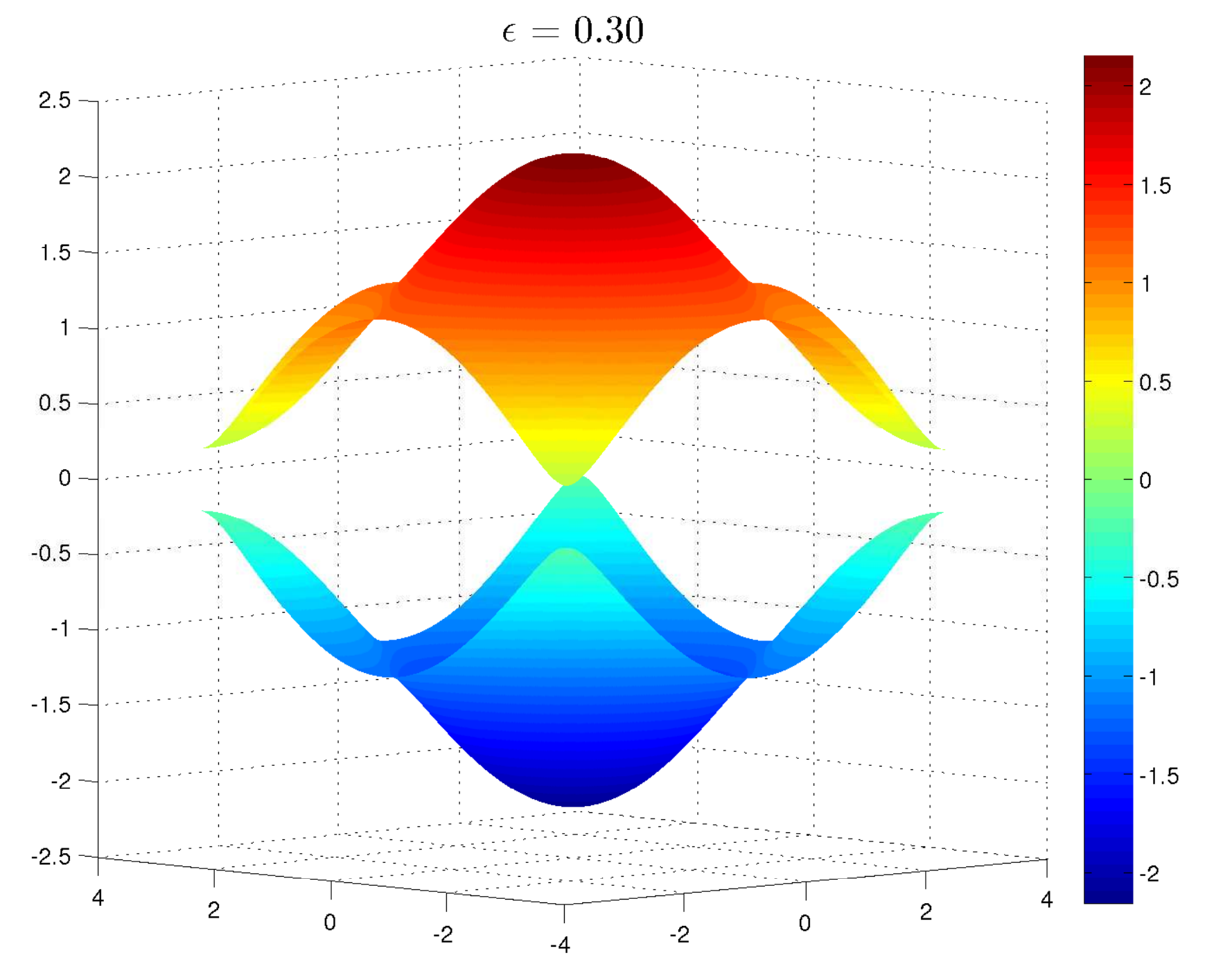}}
\subfloat{\includegraphics[scale = 0.25]{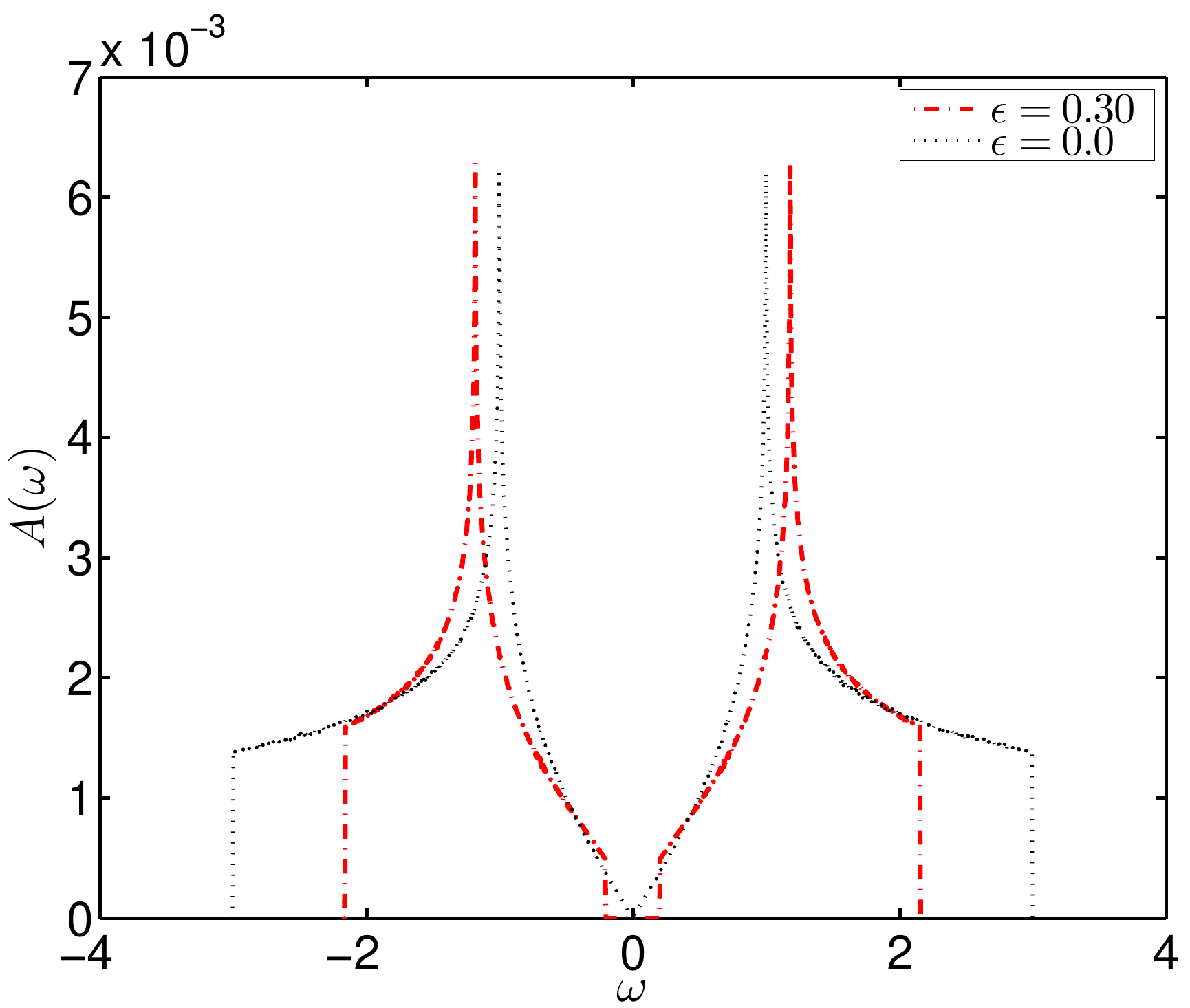}}
\subfloat{\includegraphics[scale = 0.25]{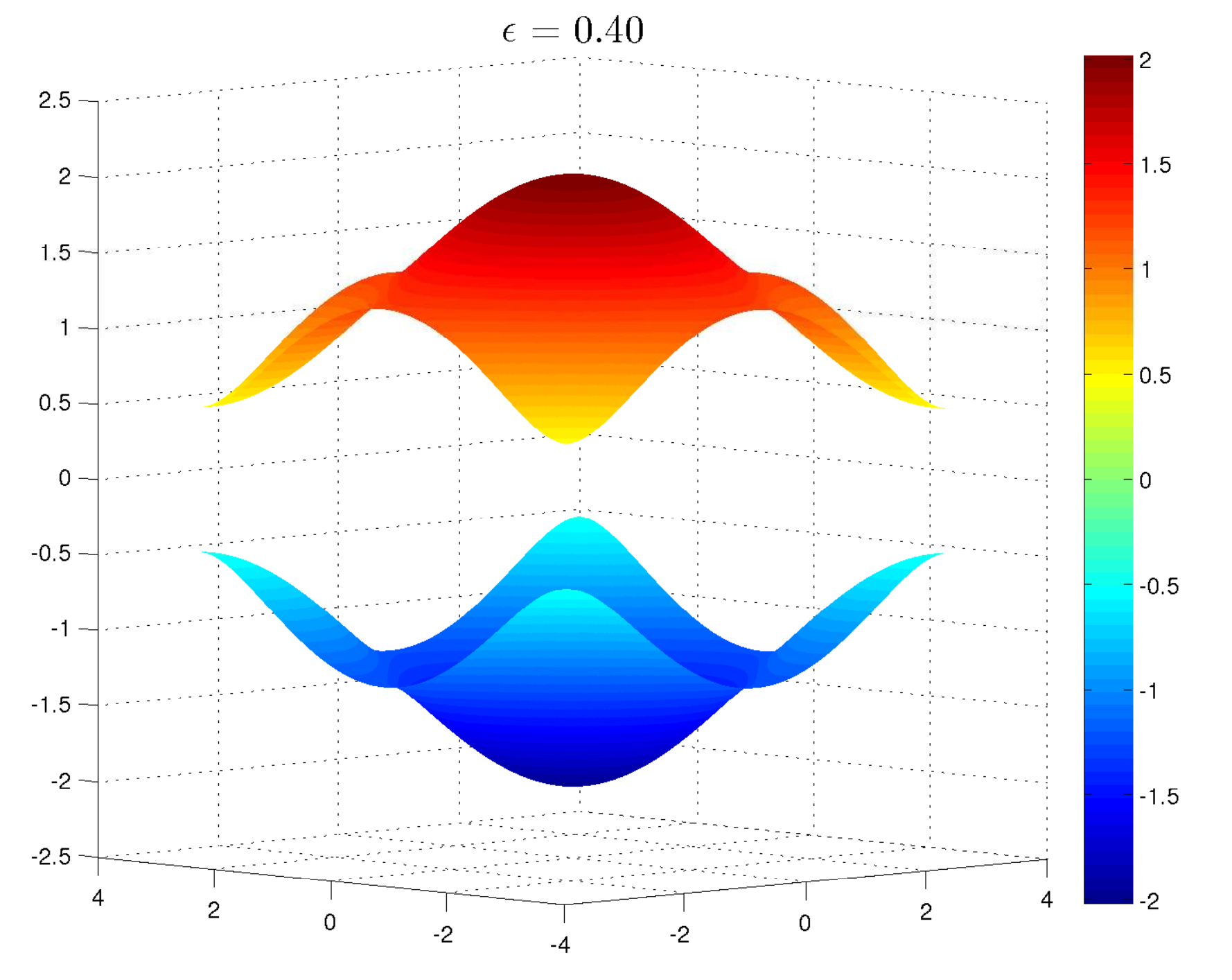}}
\subfloat{\includegraphics[scale = 0.25]{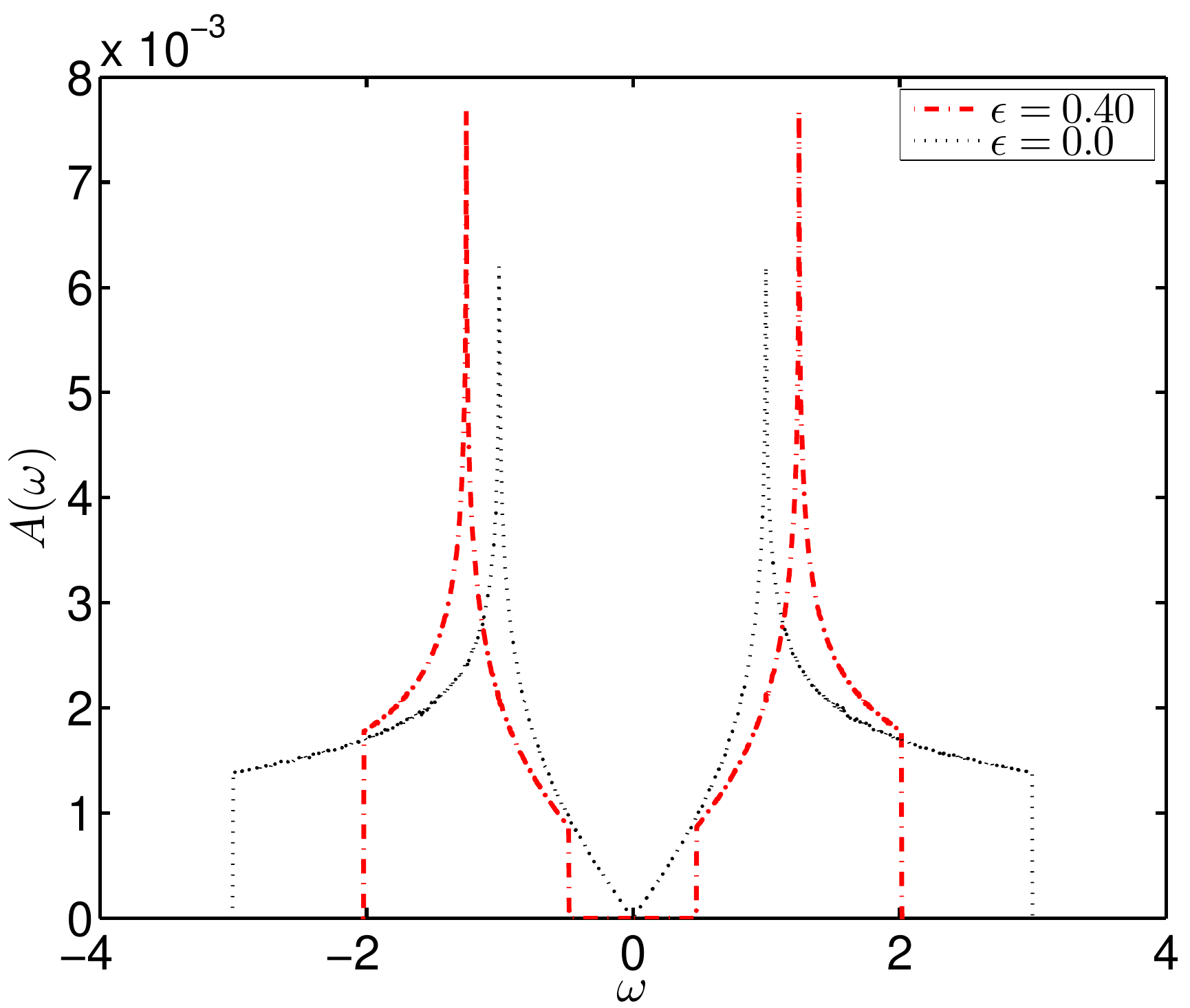}}
\caption{Plots of bandstructure and density of states (DOS) $A(\omega)$ for graphene under strain $\epsilon$ along the zigzag direction. The values of strain shown here are greater than the critical strain. The gap is clearly visible in both the bandstructure and DOS plots.}
\label{fig_bs_dos_beyond_eps_crit}
\end{figure}

Figs. \ref{fig_bs_dos_till_eps_crit} and \ref{fig_bs_dos_beyond_eps_crit} show plots of the bandstructure and the density of states (DOS) corresponding to various values of strain along the zigzag direction. The plots in Fig. \ref{fig_bs_dos_till_eps_crit} are for strain values less than the critical strain required to open a bandgap, whereas in Fig. \ref{fig_bs_dos_beyond_eps_crit} the strain is larger than the critical strain.
\section{ Two-Particle Self-Consistent method for the Hubbard model on the honeycomb lattice under uniaxial strain}

As mentioned in the main text, in order to study the Hubbard model on the honeycomb lattice under uniaxial strain, the method we choose is the two-particle self-consistent approach (TPSC). This nonperturbative, semi-analytical technique is valid from weak to intermediate values of interaction. Initially developed as an approximation scheme to study the single-band Hubbard model on the square lattice \cite{Vilk1997,Tremblay2012}, TPSC was later extended to study the multi-band case of the Hubbard model \cite{Miyahara2013,Ogura2015,Arya2015}. In particular, the semi-metal to antiferromagnet transition of the half-filled Hubbard model on the honeycomb lattice was studied using this method \cite{Arya2015}.

TPSC satisfies many important physical constraints like conservation laws, Pauli principle and  local sum rules for spin and charge susceptibilities. Most importantly, this approach obeys the Mermin-Wagner theorem, which prevents a finite temperature phase transition in two dimensions, while aptly capturing the physics of long wavelength antiferromagnetic fluctuations in the system. Once the antiferromagnetic correlation length becomes larger than the thermal de Broglie wavelength, there is a crossover to the renormalized classical regime. Since the quasiparticles get destroyed by these large antiferromagnetic fluctuations, a pseudogap opens up \cite{Vilk1997,Tremblay2012}.

\subsection{Method and Formalism}
In TPSC, we start with the noninteracting susceptibility and define Bethe-Salpeter equations for the interacting spin and charge susceptibilities. Although TPSC looks similar to RPA, the irreducible vertices for the spin and charge channels are different, and are different from the bare interaction\cite{Vilk1997,Tremblay2012}. In the case of the honeycomb lattice, for our purposes we have a non-interacting susceptibility which is a $2 \times 2$ matrix,
\begin{align}
 \mathbf{\bm{\chi}}_{0}(q) = \begin{bmatrix}
\chi_{0}^{aaaa}(q)  &\chi_{0}^{aabb}(q) \\
\chi_{0}^{bbaa}(q) &\chi_{0}^{bbbb}(q) \\
\end{bmatrix}
\end{align}
where $q \rightarrow (\bm{q},i\nu_{n})$, $\bm{q}$ is the momentum and $i\nu_{n}$ is the bosonic Matsubara frequency.
An element of the noninteracting susceptibility in the momentum-imaginary frequency representation is given by,
\begin{align}
 \chi^{\rho \rho \lambda \lambda}_{0}(q) = -\frac{T}{N^{2}} \sum_{k \sigma} \; G^{(0)\lambda \rho}_{\sigma}(k) \;  G^{(0) \rho \lambda}_{\sigma}(k + q) \label{eq_nonint_susc}
\end{align}
where $\rho, \lambda = a,b$ are the sublattice indices on the honeycomb lattice and the lattice / momentum grid is of size $N \times N$. Here $k \rightarrow (\bm{k},i\omega_{n})$, $\bm{k}$ is the momentum and $i\omega_{n}$ is the fermionic Matsubara frequency.

The spin and charge susceptibilities in the position-imaginary time representation are,
\begin{align}
\chi^{\rho \rho \lambda \lambda}_{sp}(1,2) &= \langle T_{\tau} S^{z}_{\rho}(1) S^{z}_{\lambda}(2) \rangle \label {eq_chi_sp} \\ 
\chi^{\rho \rho \lambda \lambda}_{ch}(1,2) &= \langle T_{\tau} n_{\rho}(1) n_{\lambda}(2) \rangle - \langle n_{\rho}\rangle \langle n_{\lambda}\rangle \label {eq_chi_ch}
\end{align}
Here $1$ is a shorthand for $(\bm{r}_{1},\tau_{1})$ the position and the imaginary time. 
From Eq.s \eqref{eq_chi_sp} and \eqref{eq_chi_ch}, the local spin and charge sum rules are obtained when we set $1 \rightarrow 2, \lambda = \rho$, 
\begin{align}
\frac{T}{N^{2}}\sum_{q} \, \chi_{sp}^{\rho \rho \rho \rho}(q) &= \langle n_{\rho \uparrow} \rangle + \langle n_{\rho \downarrow} \rangle  - 2 \langle n_{\rho \uparrow} n_{\rho \downarrow} \rangle  \label{eq_sumrule_sp_graphene} \\
\frac{T}{N^{2}}\sum_{q} \, \chi_{ch}^{\rho \rho \rho \rho}(q) &= \langle n_{\rho \uparrow} \rangle + \langle n_{\rho \downarrow} \rangle  + 2 \langle n_{\rho \uparrow} n_{\rho \downarrow} \rangle - n_{\rho}^{2}  \label{eq_sumrule_ch_graphene} 
\end{align}

The interacting spin and charge susceptibilities are given by
\begin{align}
\bm{\chi}_{sp}(q) = \left[ \bm{1} - \frac{1}{2} \bm{\chi}_{0}(q)\mathbf{U}_{sp} \right]^{-1} \bm{\chi}_{0}(q)   \label{eq_spin_susc_graphene} \\
\bm{\chi}_{ch}(q) = \left[ \bm{1} + \frac{1}{2} \bm{\chi}_{0}(q)\mathbf{U}_{ch} \right]^{-1} \bm{\chi}_{0}(q)   \label{eq_charge_susc_graphene}
\end{align}
 in terms of the noninteracting susceptibility. This is the matrix version of the scalar equation appearing in the single-band case \cite{Vilk1997,Tremblay2012,Arya2015}. The elements of the matrix spin ($\mathbf{U}_{sp}$) and charge ($\mathbf{U}_{ch}$) vertices appearing in the above expression can be obtained as functional derivatives of the elements of the self-energy matrix with respect to the elements of the Green function matrix. 

Using the TPSC ansatz, we can write a first approximation to the self-energy which is momentum and frequency independent \cite{Vilk1997,Tremblay2012,Arya2015} and from this form of the self-energy we get a local spin vertex where the diagonal $aaaa$ and $bbbb$ elements are non-zero. This local spin vertex is given by
\begin{align}
 U_{sp} = U \frac{ \langle n_{\uparrow} n_{\downarrow} \rangle}{\langle n_{\uparrow} \rangle \langle n_{\downarrow} \rangle} \label{eq_tpsc_ansatz}
\end{align}
Because of sublattice symmetry $ \langle n_{a \uparrow} n_{a \downarrow} \rangle =  \langle n_{b \uparrow} n_{b \downarrow} \rangle$ and $\langle n_{a \sigma} \rangle = \langle n_{b \sigma} \rangle$. Although the charge vertex is not local, and involves higher order correlation functions which are hard to compute, we assume that the charge vertex is also local ($U_{ch}$).

We define ferromagnetic and the antiferromagnetic spin susceptibilities for both noninteracting and interacting cases as,
\begin{align}
\chi^{fm}_{0,sp} &= \chi^{aaaa}_{0,\,sp} - \sqrt{ \chi^{aabb}_{0,\,sp} \; \chi^{aabb}_{0,\,sp}} \label{eq_chi_fm}\\
\chi^{afm}_{0,sp} &= \chi^{aaaa}_{0,\,sp} + \sqrt{ \chi^{aabb}_{0,\,sp} \; \chi^{aabb}_{0,\,sp}} \label{eq_chi_afm}
\end{align}

The interacting susceptibilities are related to the noninteracting ones by the relation
\begin{align}
\chi_{afm,fm}^{s} &= \frac{ \chi^{afm,fm}_{0} } {1 - \frac{U_{sp}}{2} \chi^{afm,fm}_{0}}
\end{align}
that resembles the scalar Bethe-Salpeter equations in the single-band case.

Focusing only on antiferromagnetic correlations, the correlation length $\xi_{afm}$ can be quantified in terms of the ratio of the maximum of the interacting antiferromagnetic susceptibility to the maximum of the noninteracting antiferromagnetic susceptibility at $(\bm{q} = 0, i\nu_{n} = 0)$.
\begin{align}
\xi_{afm} &= \frac{\chi^{s}_{afm}(\bm{q} = \bm{0}, i\nu_{n} = 0)}{\chi^{0}_{afm}(\bm{q} = \bm{0}, i\nu_{n} = 0)} \label{eq_corr_len_afm} \\
\end{align}

The second approximation to self-energy including the effects of spin and charge fluctuations can be made as 
\begin{align}
\Sigma _{\sigma}(k) &= U n_{\rho, -\sigma} +  \frac{U}{4} \frac{T}{N^{2}} \sum  G^{aa }_{\sigma}(k + q) \left[ U_{sp} \chi^{aaaa}_{sp}(q) + U_{ch} \chi^{aaaa}_{ch}(q) \right] \label{eq_se_graphene}
\end{align}

We work at half-filling, and start by finding the noninteracting susceptibility $\bm{\chi}_{0}(q)$ using a combination of FFT and cubic splines for a momentum grid $N \times N$ for $N_{\nu}$ number of Matsubara frequencies. For a particular value of interaction $U$, a guess value of double occupancy $\langle n_{\uparrow} n_{\downarrow}\rangle$ is chosen. Using the TPSC ansatz \eqref{eq_tpsc_ansatz} and the spin sum rule \eqref{eq_sumrule_sp_graphene}, the double occupancy, and thereby the spin vertex ($U_{sp}$) can be determined self-consistently. Once the double occupancy is obtained, using the charge sum rule \eqref{eq_sumrule_ch_graphene}, we can solve for the charge vertex ($U_{ch}$). Again by using a combination of FFTs and cubic splines we can compute the self-energy using \eqref{eq_se_graphene}.

\subsection{Graphene under strain}
The application of strain to the honeycomb lattice changes bond lengths. This results in a change in the reciprocal lattice vectors as well as a change in the hopping parameters. In our calculations, we assume the reciprocal lattice remains the same as the one corresponding to the undistorted lattice. We take into account only the change in the hopping parameters. In the case of unstrained graphene, the hopping parameters along all the three bonds are equal. Upon the application of strain, the hopping parameters are no longer isotropic. When a strain is applied along the zigzag direction, then $t_{x,x} = t_{y,y} <  t_{z,z}$. This in turn changes the dispersion relation. When we evaluate the noninteracting Green functions and the noninteracting susceptibilities, this information is taken into account. The rest of the procedure remains intact as mentioned in the previous section.

\subsection{Results}
\begin{figure}[!ht]
\centering
\subfloat{\includegraphics[scale = 0.3]{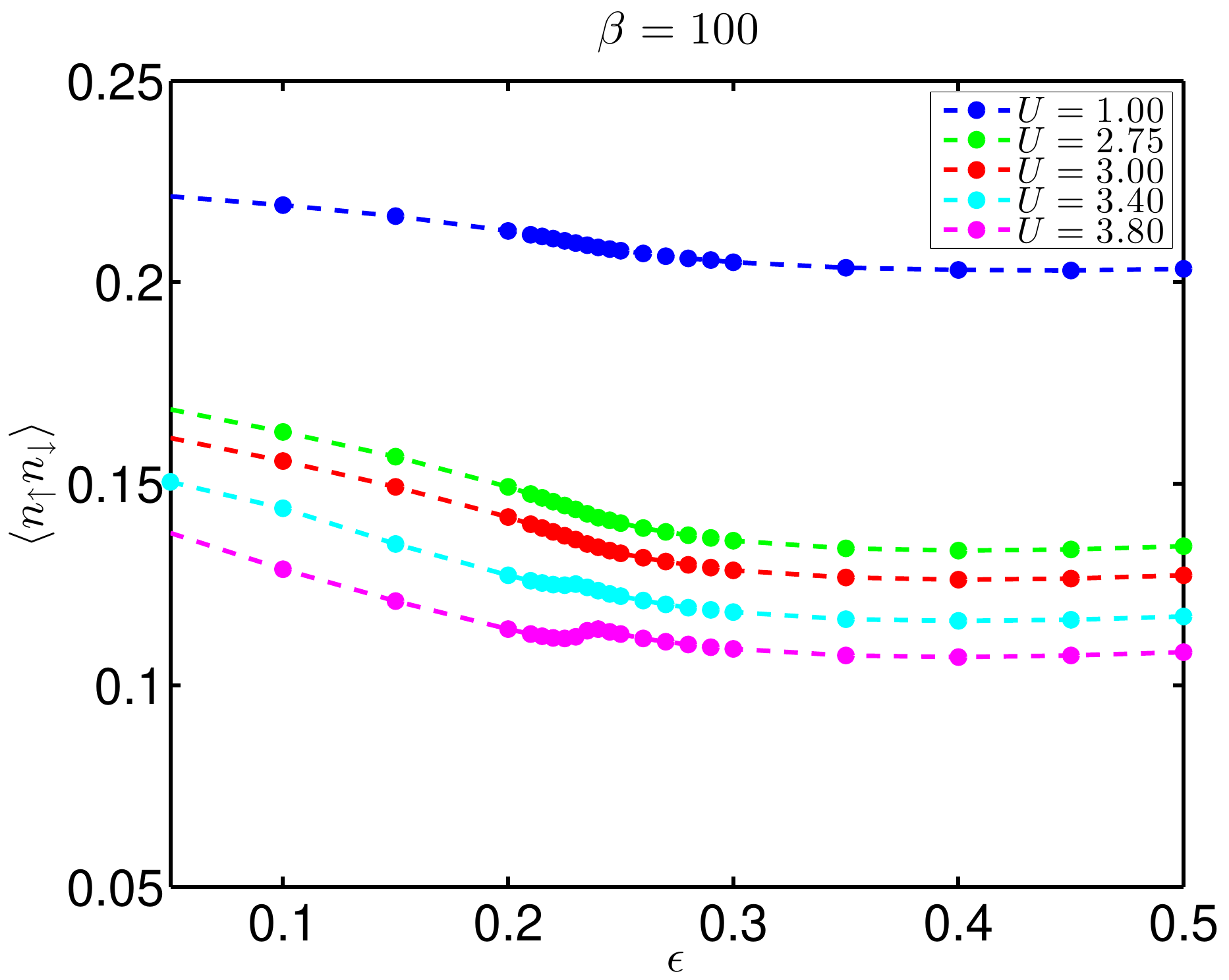}}
\subfloat{\includegraphics[scale = 0.3]{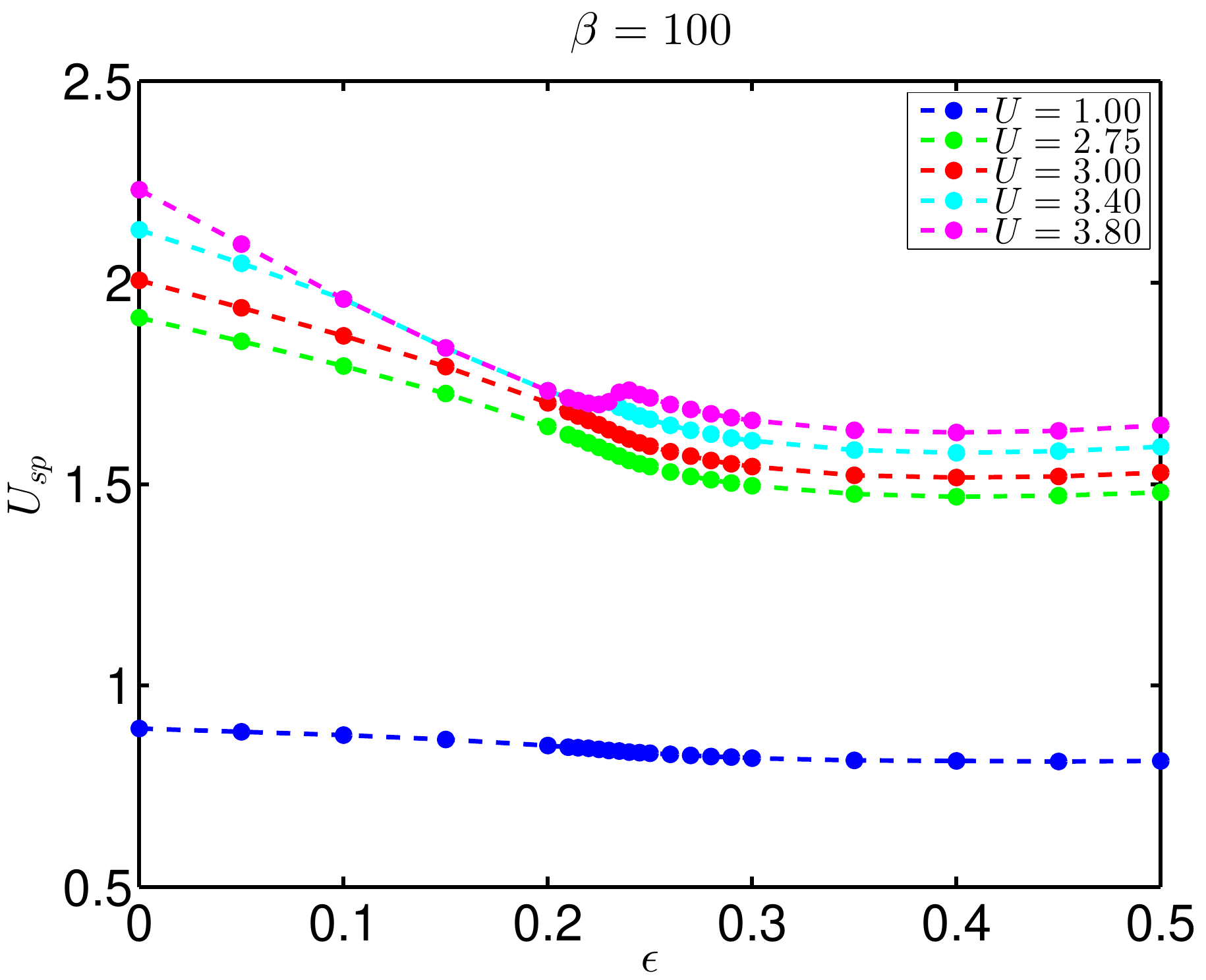}}\\
\subfloat{\includegraphics[scale = 0.3]{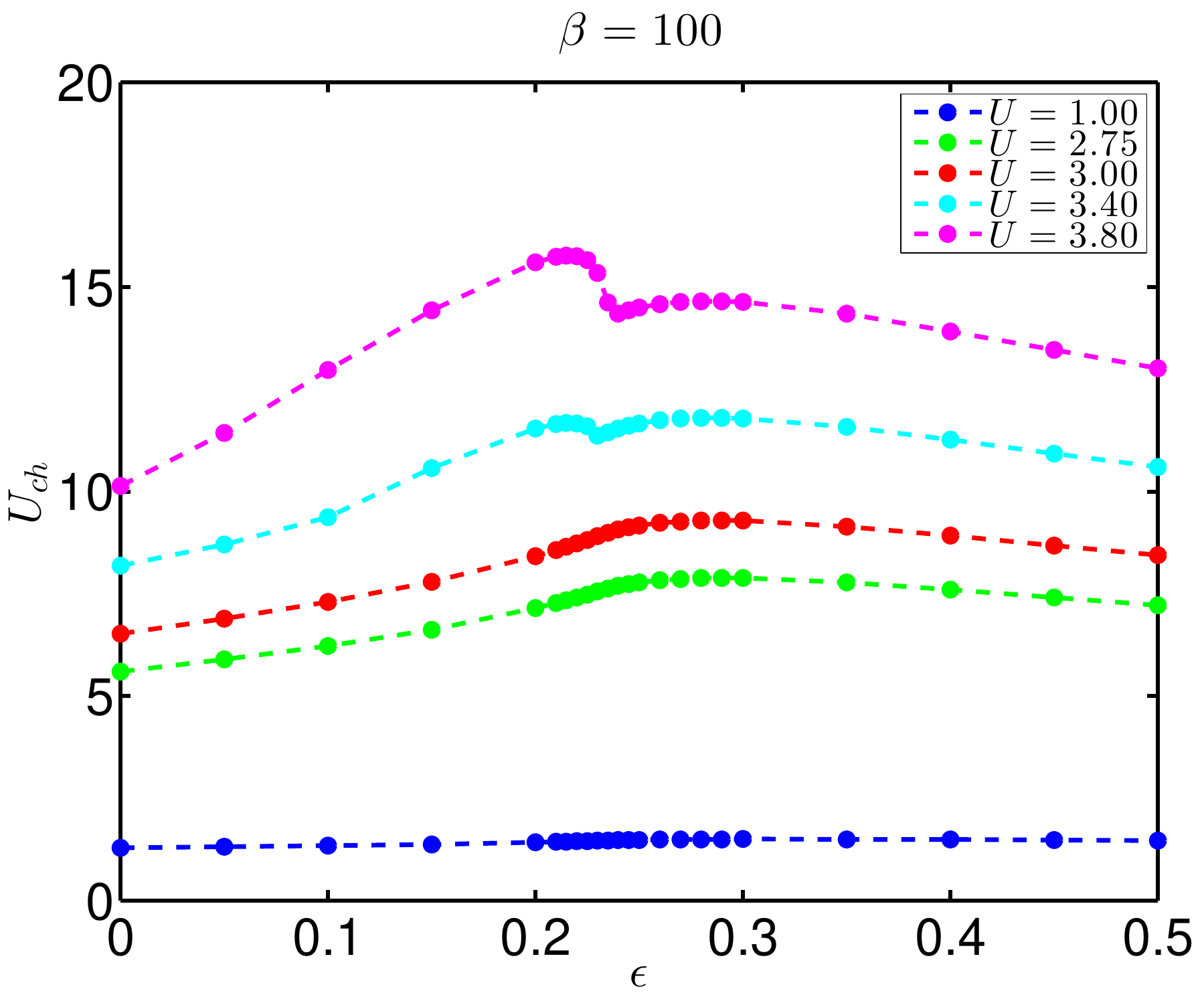}}
\subfloat{\includegraphics[scale = 0.3]{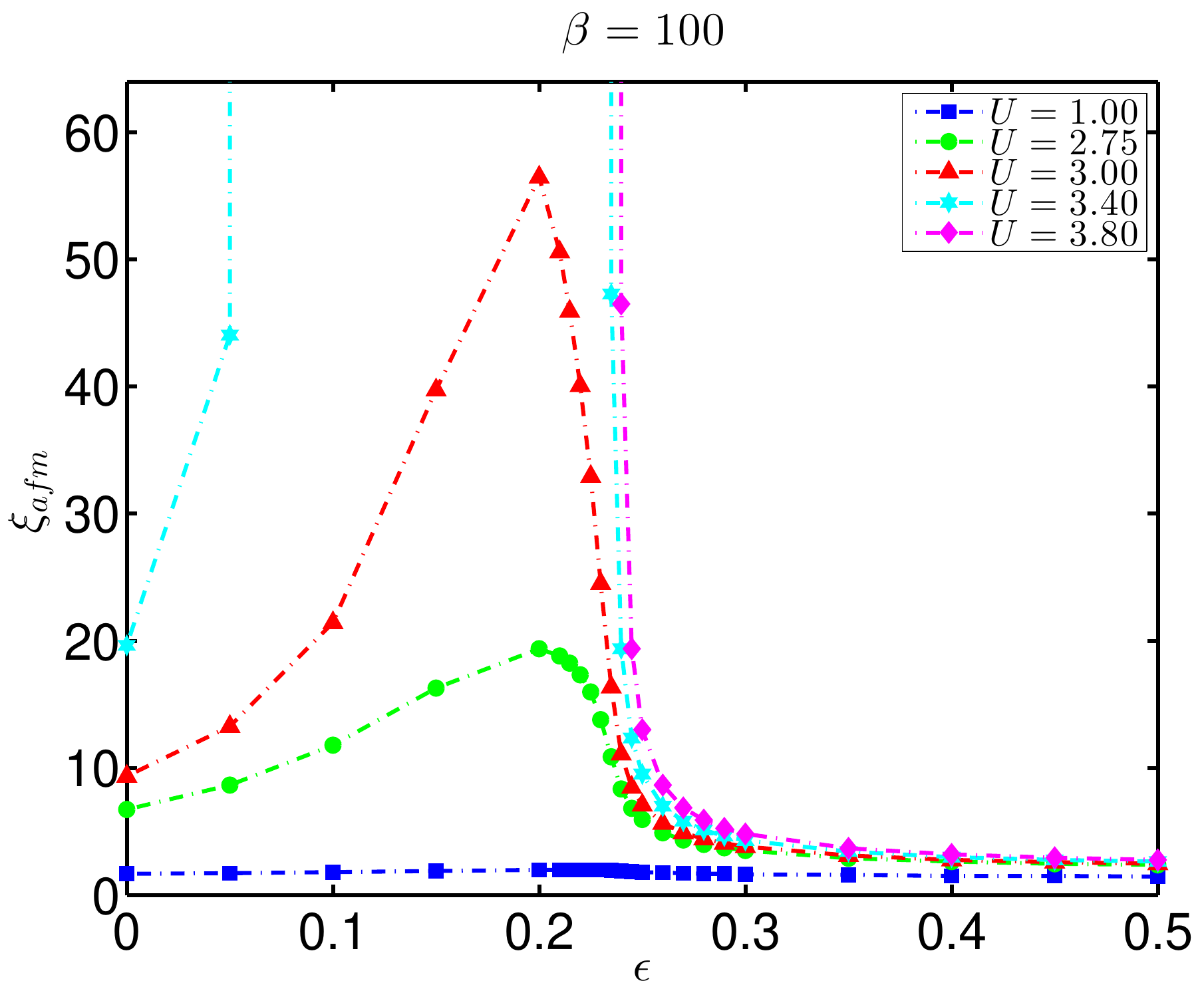}}
\caption{Plots of double occupancy ($\langle n_{\uparrow} n_{\downarrow} \rangle$) , irreducible vertices ($U_{sp}$, $U_{ch}$) and antiferromagnetic correlation length $\xi_{afm}$ as functions of strain $\epsilon$ at $\beta = 100$. Interaction values ($U$) are indicated in the legend.}
\label{fig_prelim_results}
\end{figure}

The results for double occupancy $\langle n_{\uparrow} n_{\downarrow}\rangle$, irreducible spin and charge vertices $U_{sp}$ and $U_{ch}$ and antiferromagnetic correlation length ($\xi_{afm}$) for $T = 0.01$ are shown in Fig. \ref{fig_prelim_results}. These quantities have been plotted as functions of strain $\epsilon$ for various values of $U$. As we can see, the antiferromagnetic fluctuations grow in magnitude with increasing interaction, at low values of temperature. 

\begin{figure}[!ht]
\centering
\subfloat{\includegraphics[scale = 0.3]{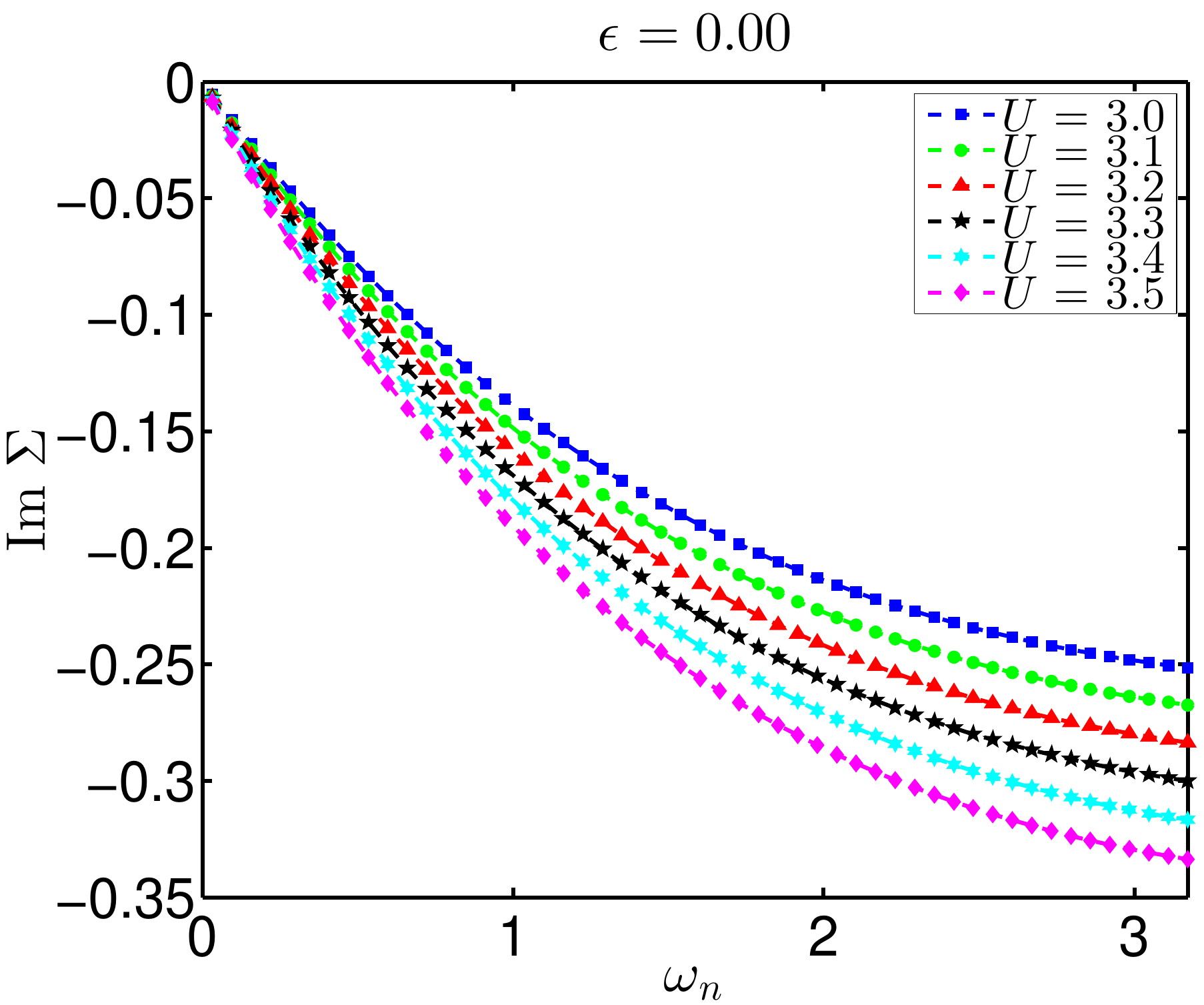}}
\subfloat{\includegraphics[scale = 0.3]{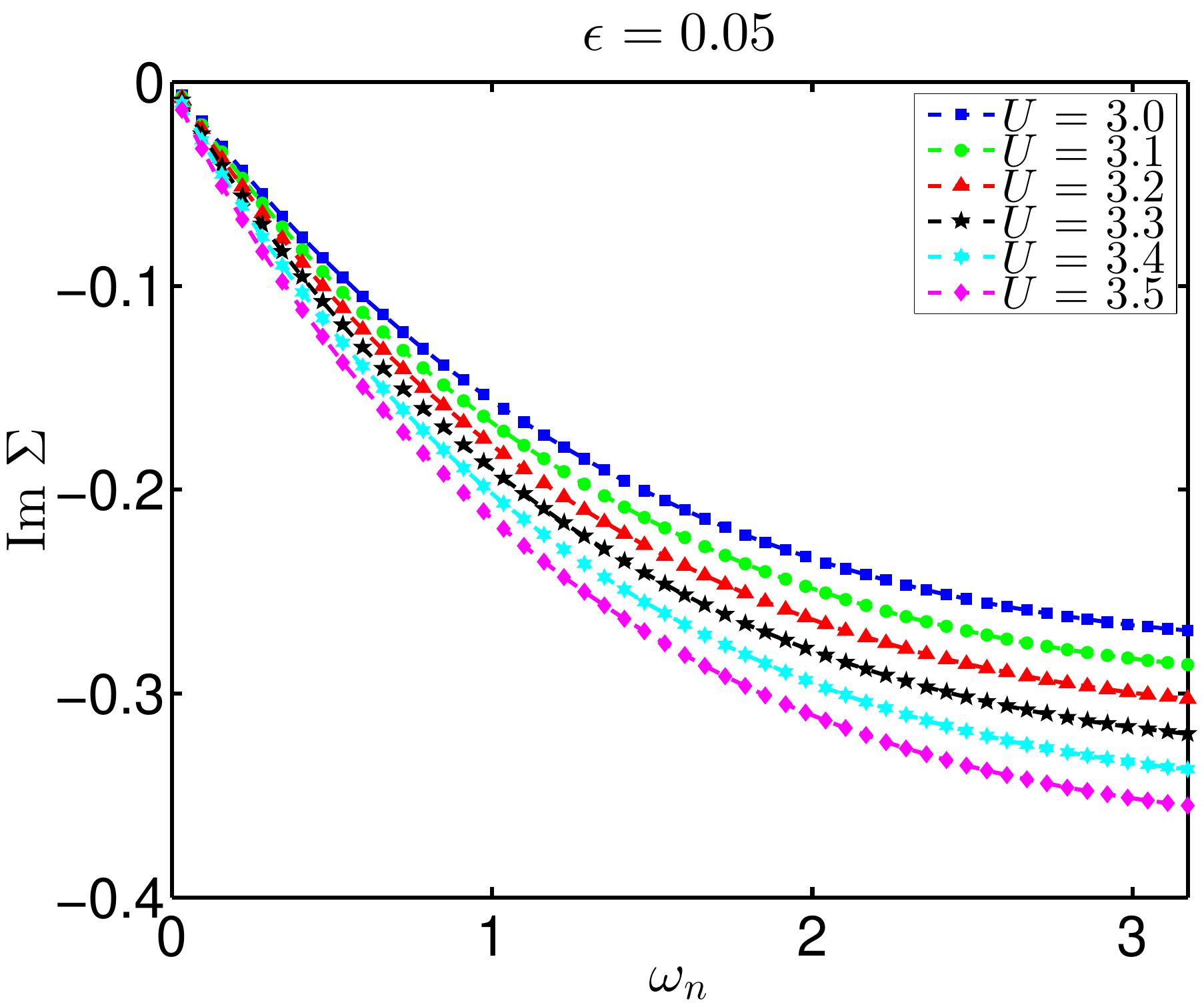}}
\subfloat{\includegraphics[scale = 0.3]{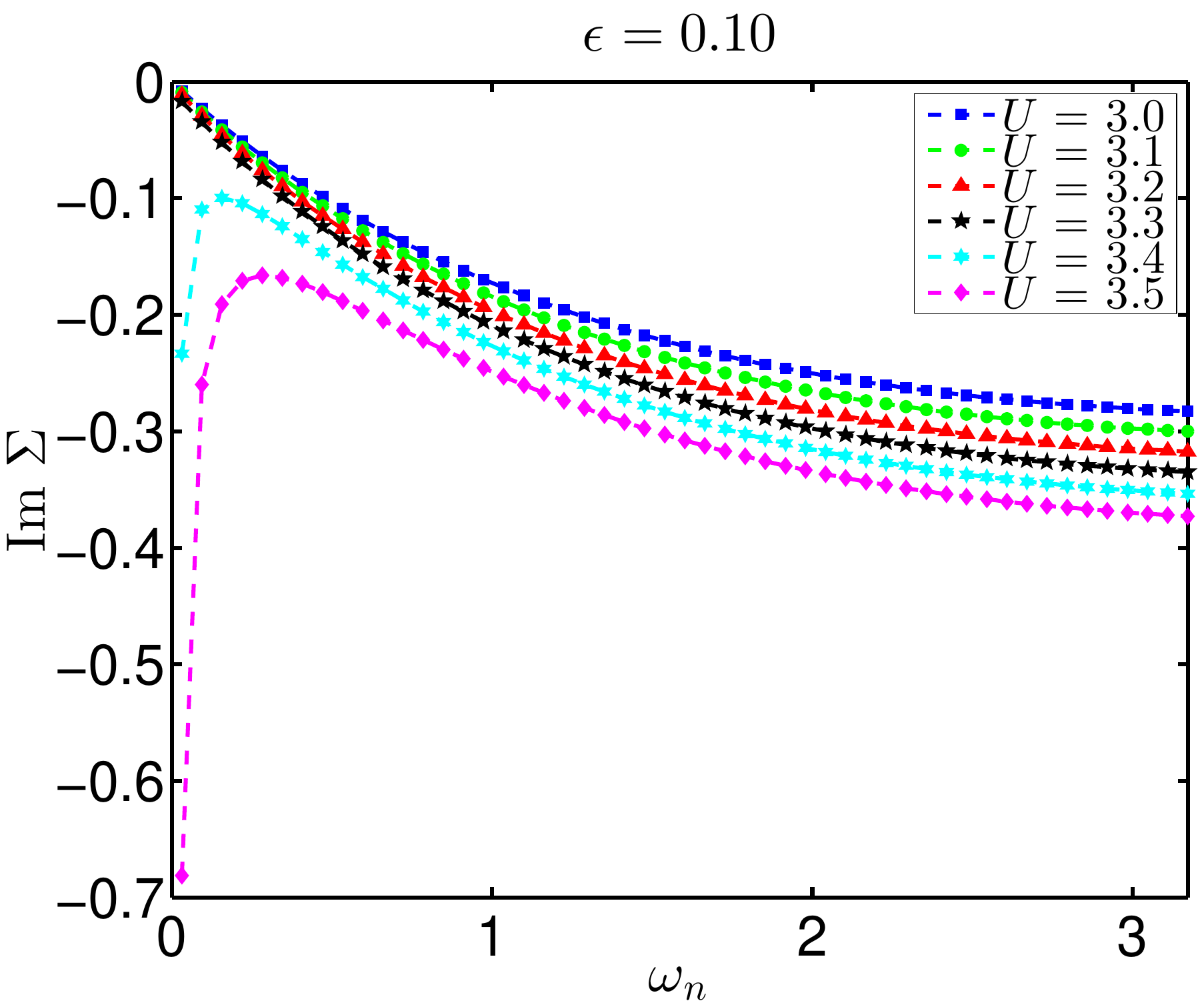}}\\
\subfloat{\includegraphics[scale = 0.3]{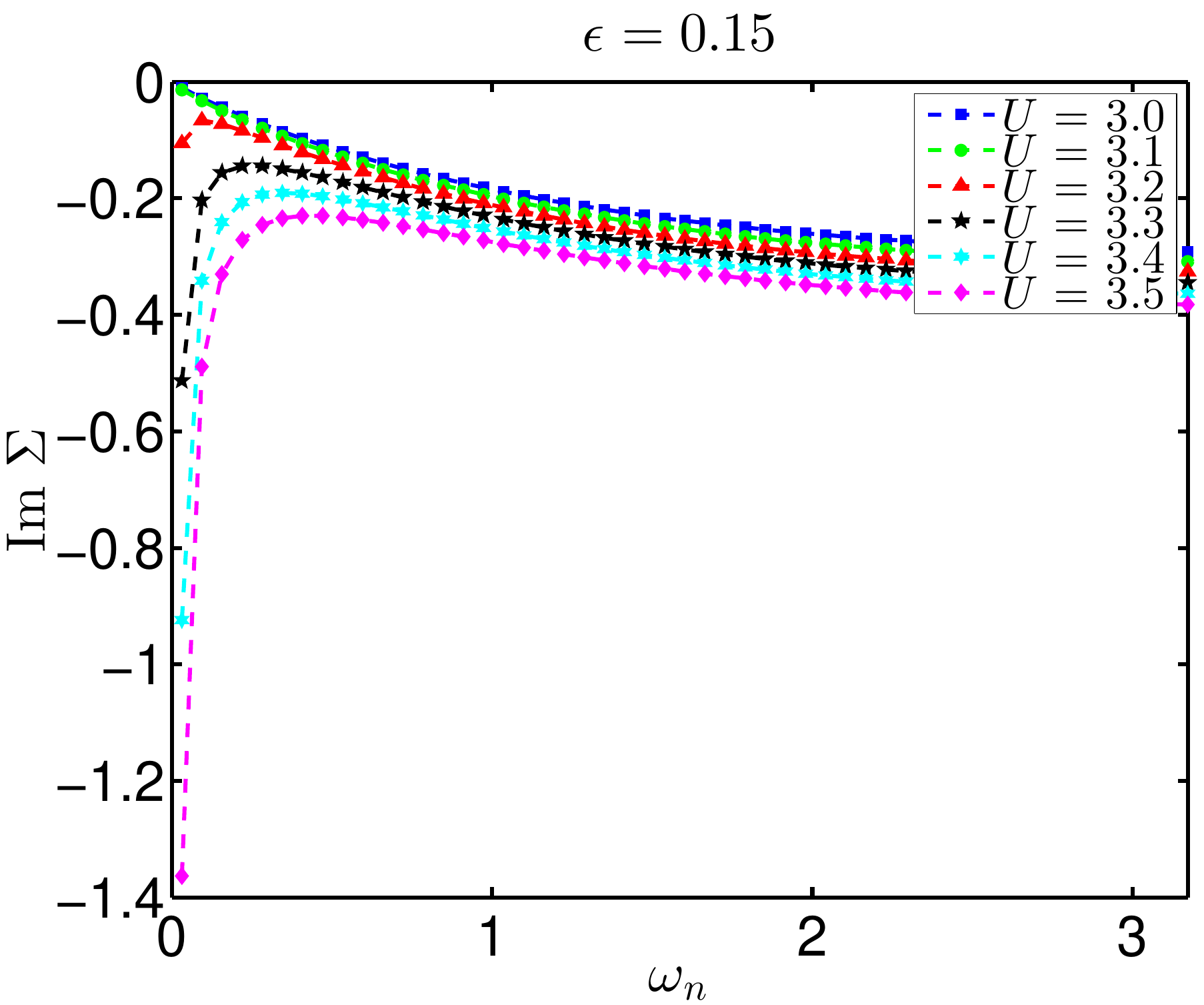}}
\subfloat{\includegraphics[scale = 0.3]{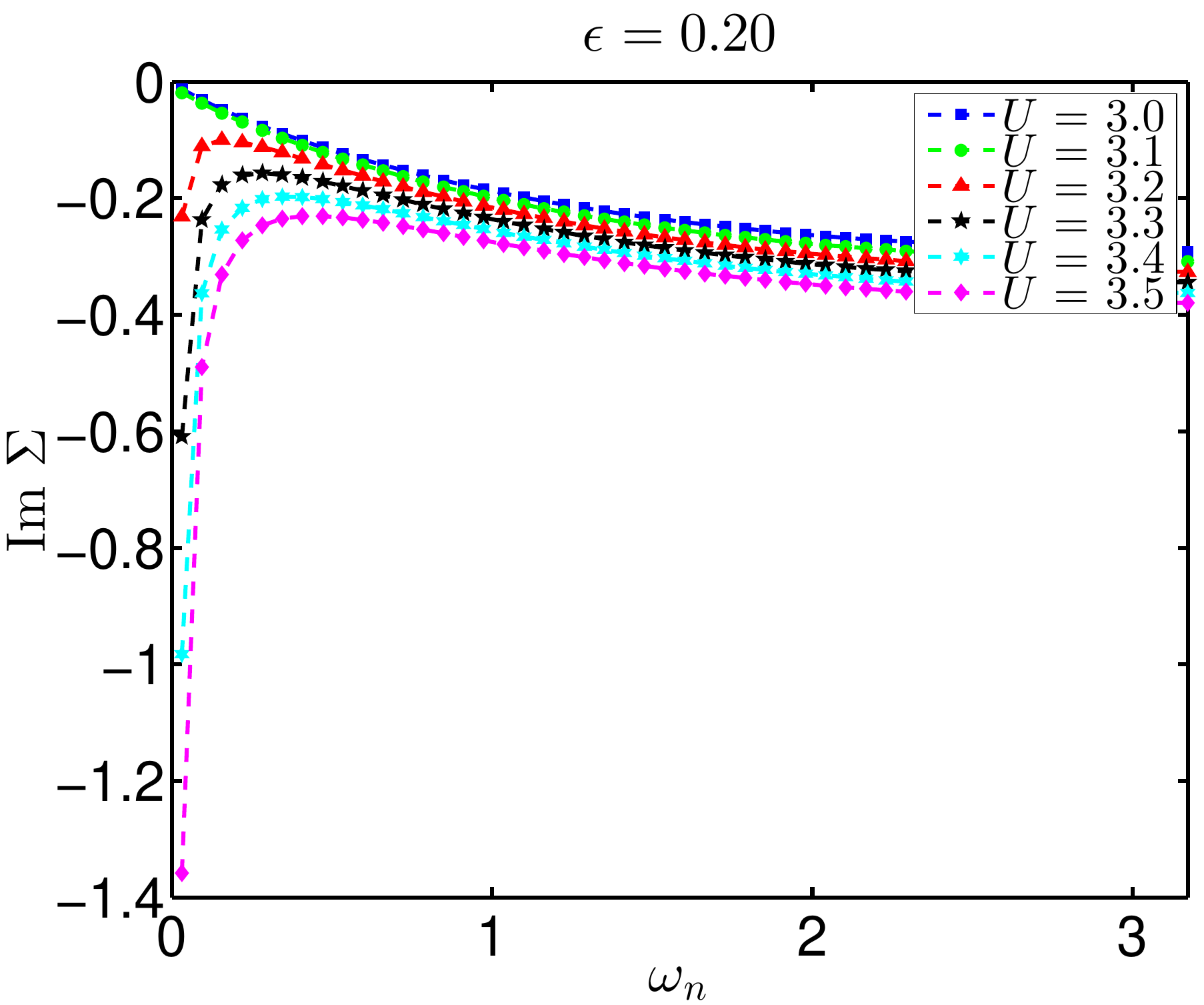}}
\subfloat{\includegraphics[scale = 0.3]{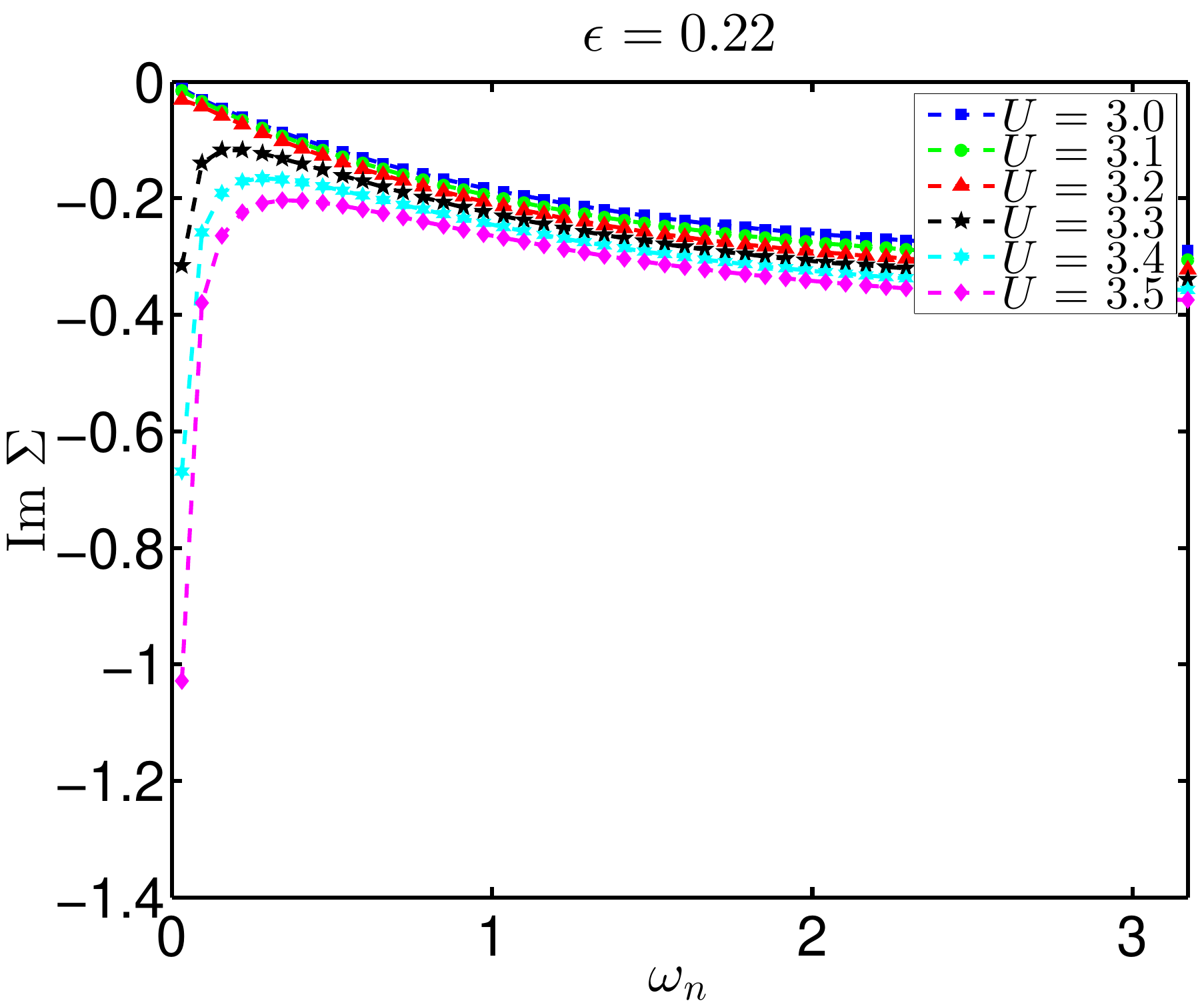}}
\caption{Imaginary part of the self energy $\Sigma$ plotted as a function of Matsubara frequency $\omega_{n}$, for various values of strain $0.0 \leq \epsilon < 0.23$ for $\beta = 100$. The legend indicates the values of interaction $U$.}
\label{fig_se_till_eps_crit}
\end{figure}

The imaginary part of self-energy as a function of the fermionic Matsubara frequency $\omega_{n}$ has been plotted for various values of strain less than the critical strain. As mentioned in the main text, for no strain or very low strain ($\epsilon = 0.05$), the Dirac Landau Fermi liquid remains stable till $U = 3.5$, whereas for $0.1 \leq \epsilon \leq 0.23$, for $U > U_{c}(\epsilon)$, the imaginary part of self-energy shows anomalous behavior. All these details can be seen from the plots in Fig. \ref{fig_se_till_eps_crit}.

\begin{figure}[!ht]
\centering
\subfloat{\includegraphics[scale = 0.3]{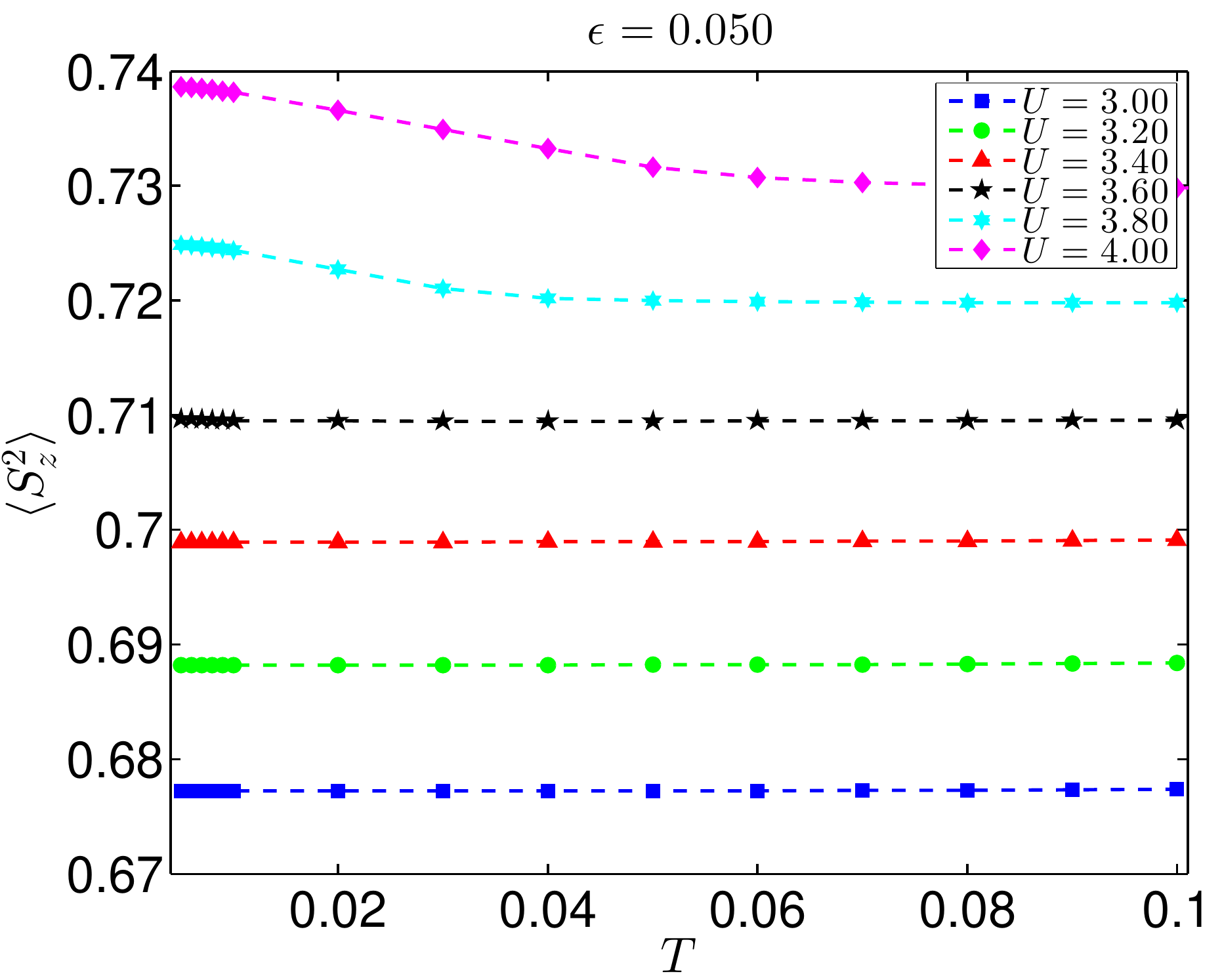}}
\subfloat{\includegraphics[scale = 0.3]{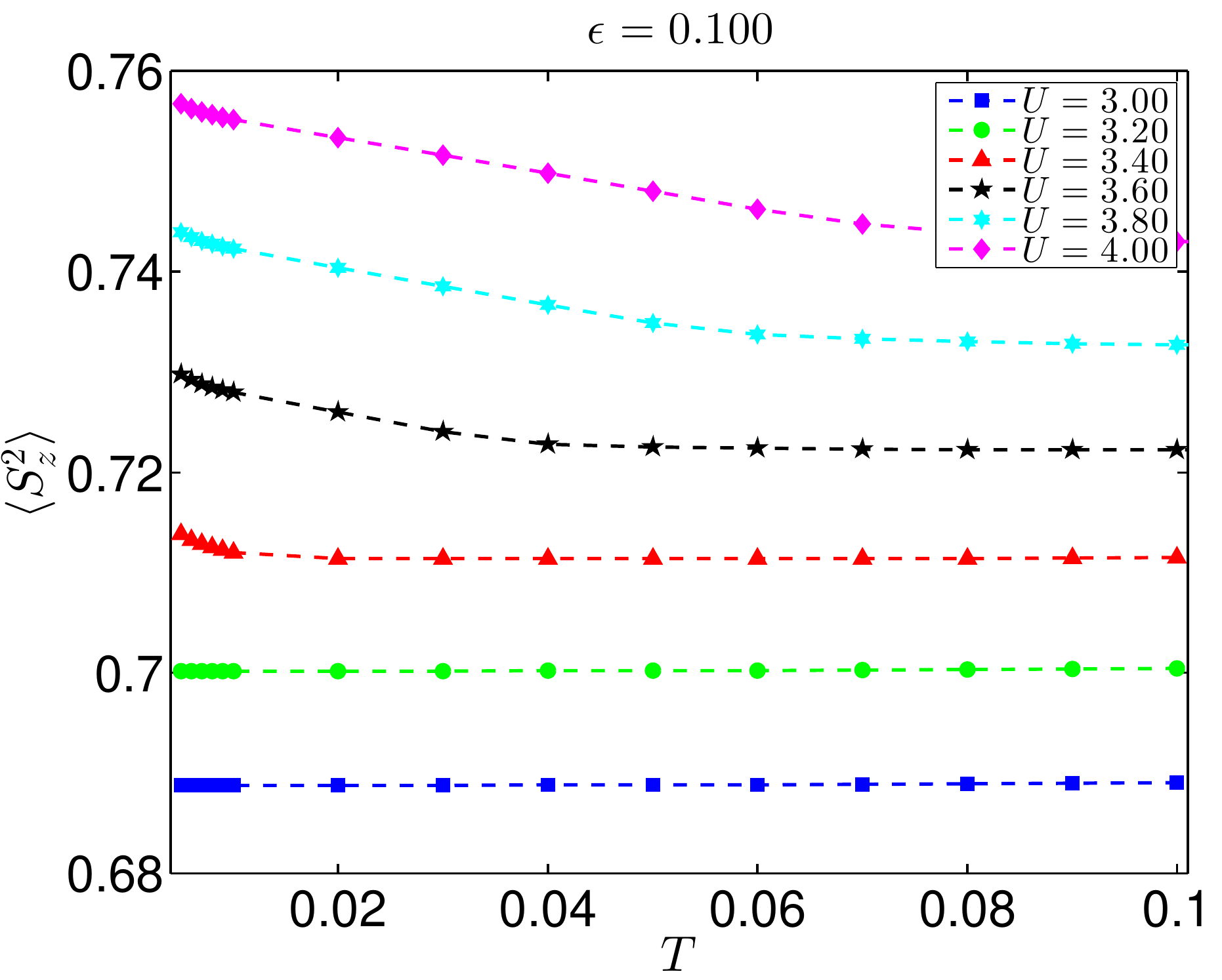}}
\subfloat{\includegraphics[scale = 0.3]{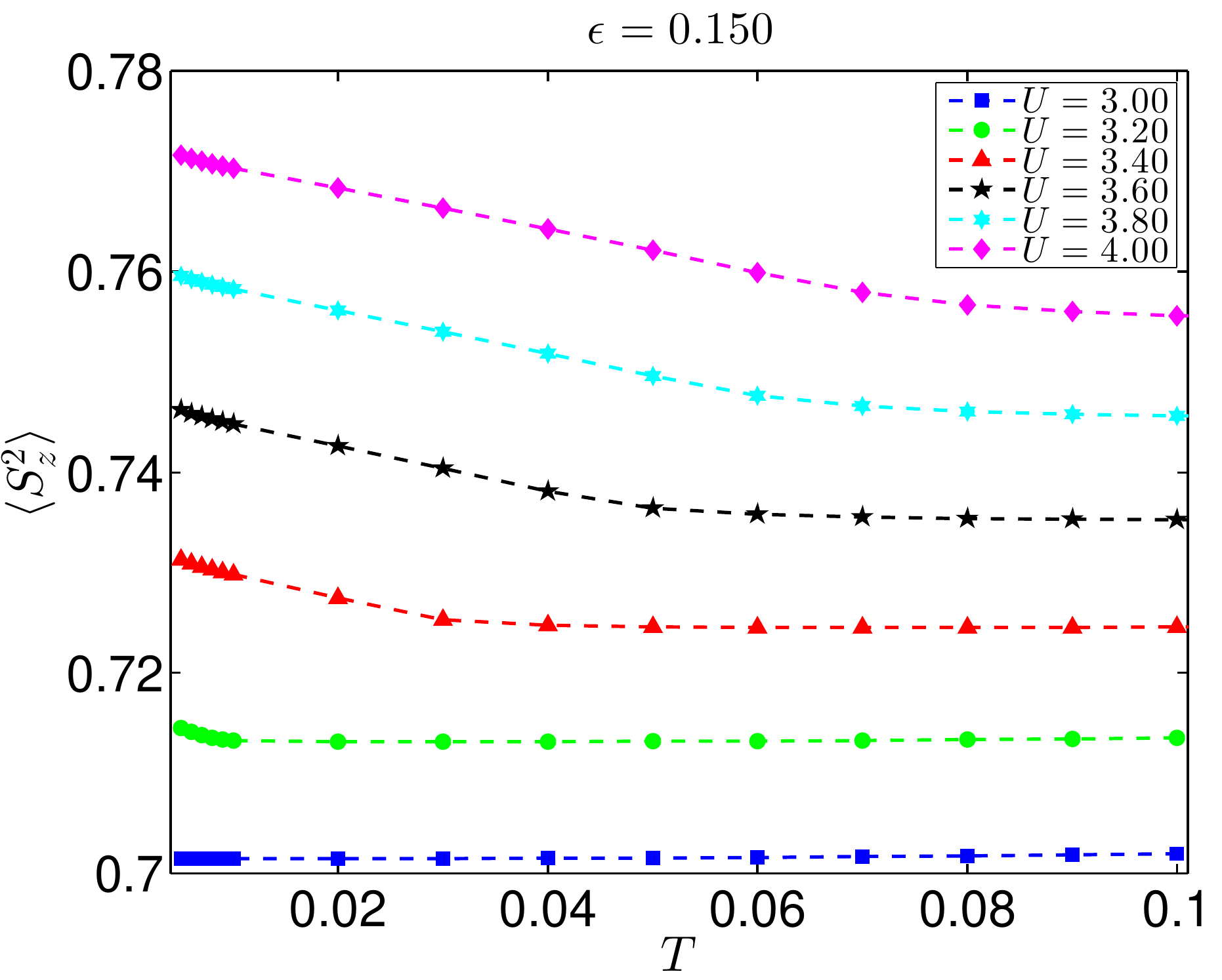}}\\
\subfloat{\includegraphics[scale = 0.3]{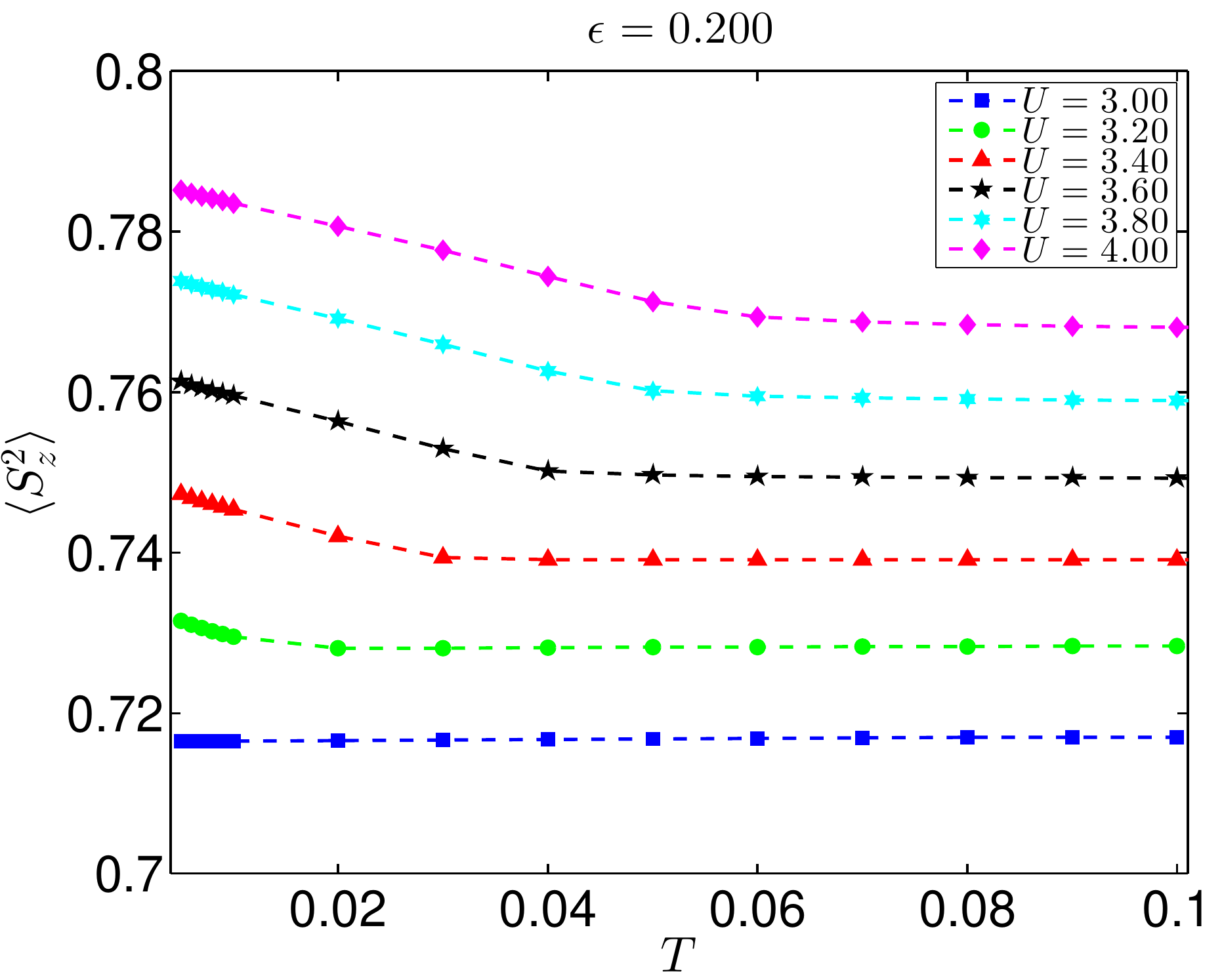}}
\subfloat{\includegraphics[scale = 0.3]{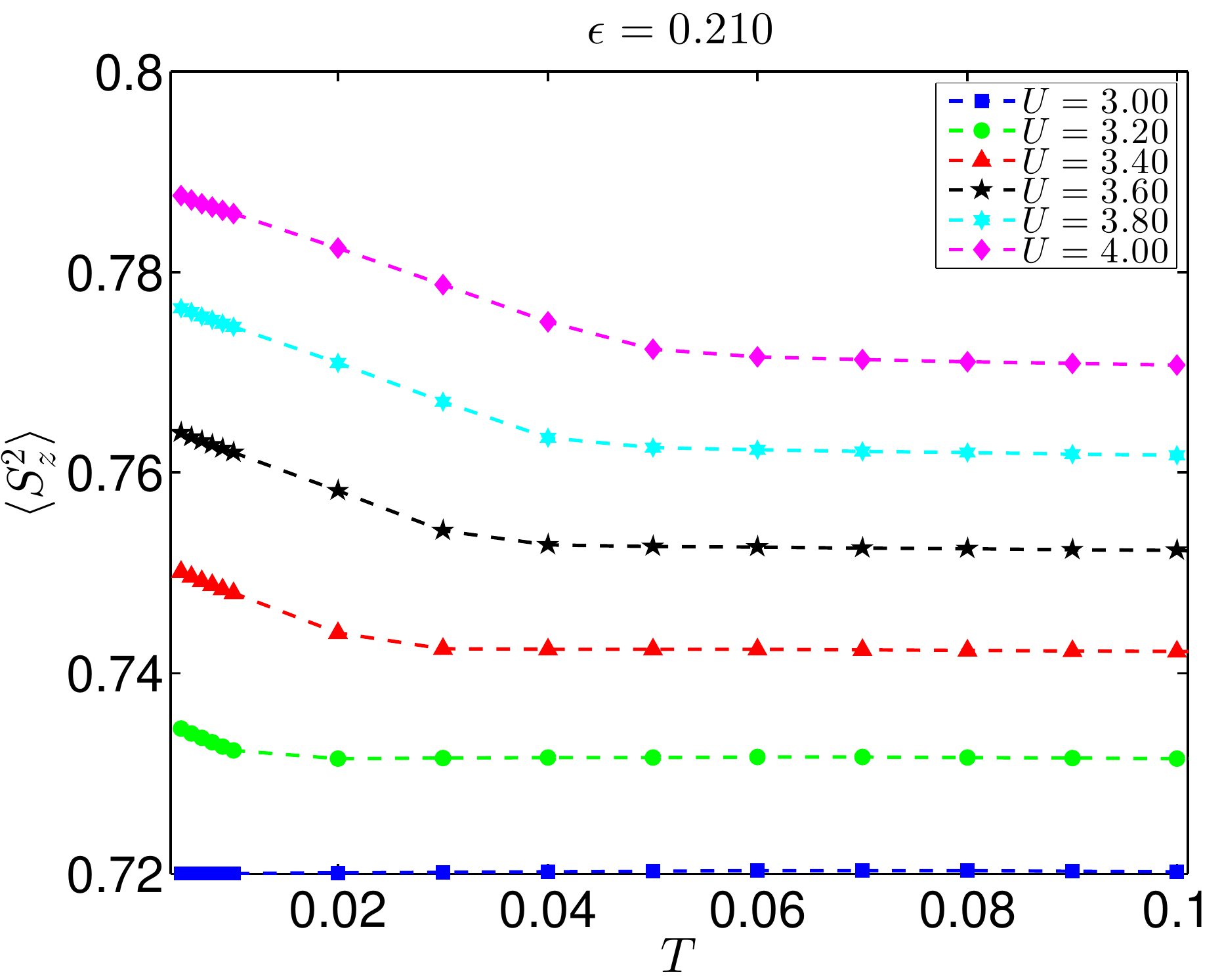}}
\subfloat{\includegraphics[scale = 0.3]{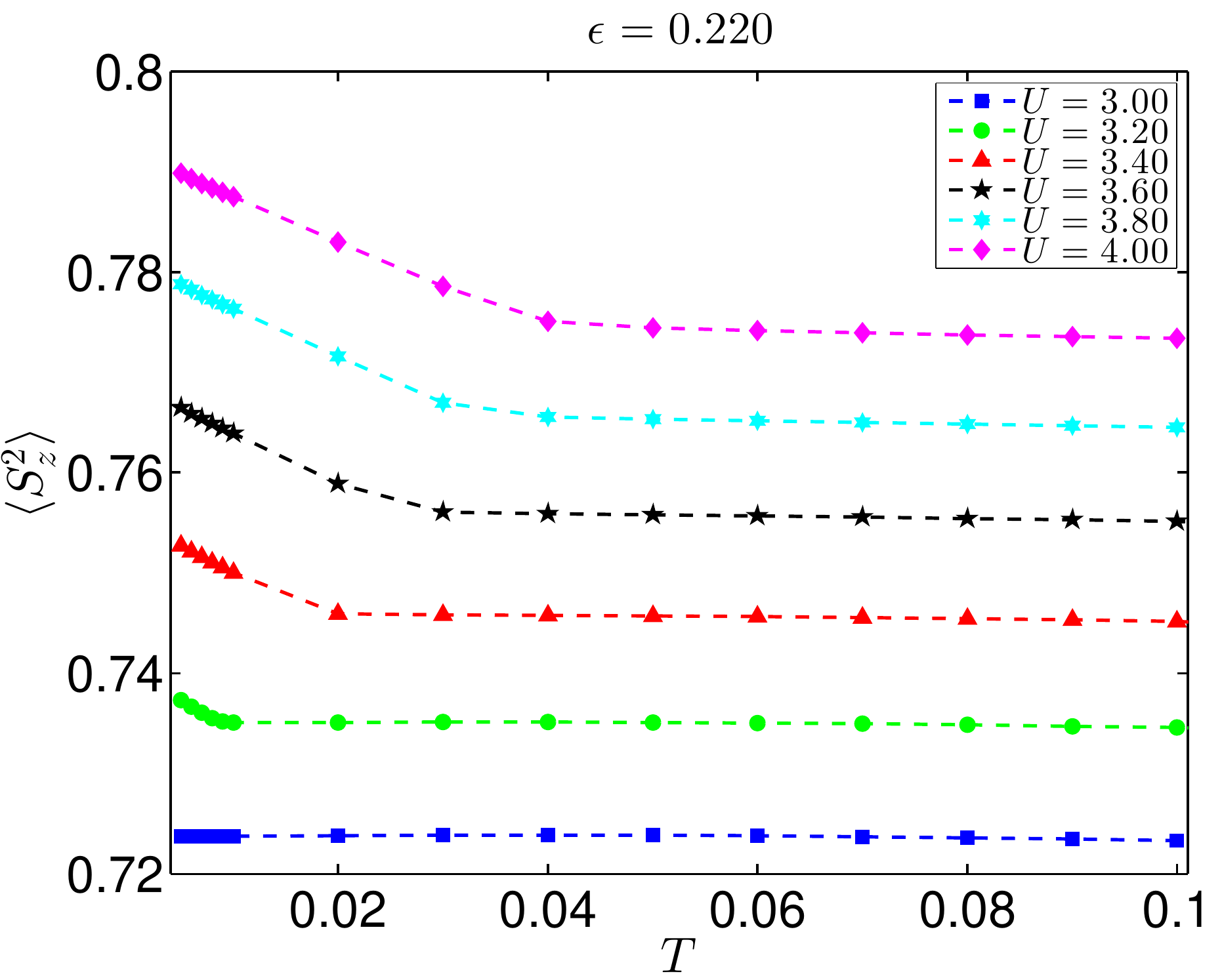}}
\caption{Plots of $\langle S_{z}^{2} \rangle = 1 - 2 \langle n_{\uparrow} n_{\downarrow} \rangle$ as a function of temperature $T$, for various values of strain $\epsilon$ below the critical strain. The legends indicate the values of interaction $U$.}
\label{fig_Z_till_eps_crit}
\end{figure}

The spin fluctuation contribution to the magneto-volume was discussed in the main text. From the self-consistent value of double occupancy obtained, we can calculate $\langle S_{z}^{2} \rangle = 1 - 2 \langle n_{\uparrow} n_{\downarrow} \rangle$. The plots of this quantity as a function of temperature for various values of strain are given in \ref{fig_Z_till_eps_crit}. As temperature decreases, the antiferromagnetic fluctuations in the system become large and $\langle S_{z}^{2}\rangle$ also increases as a consequence.

\begin{figure}[!ht]
\centering
\subfloat{\includegraphics[scale = 0.3]{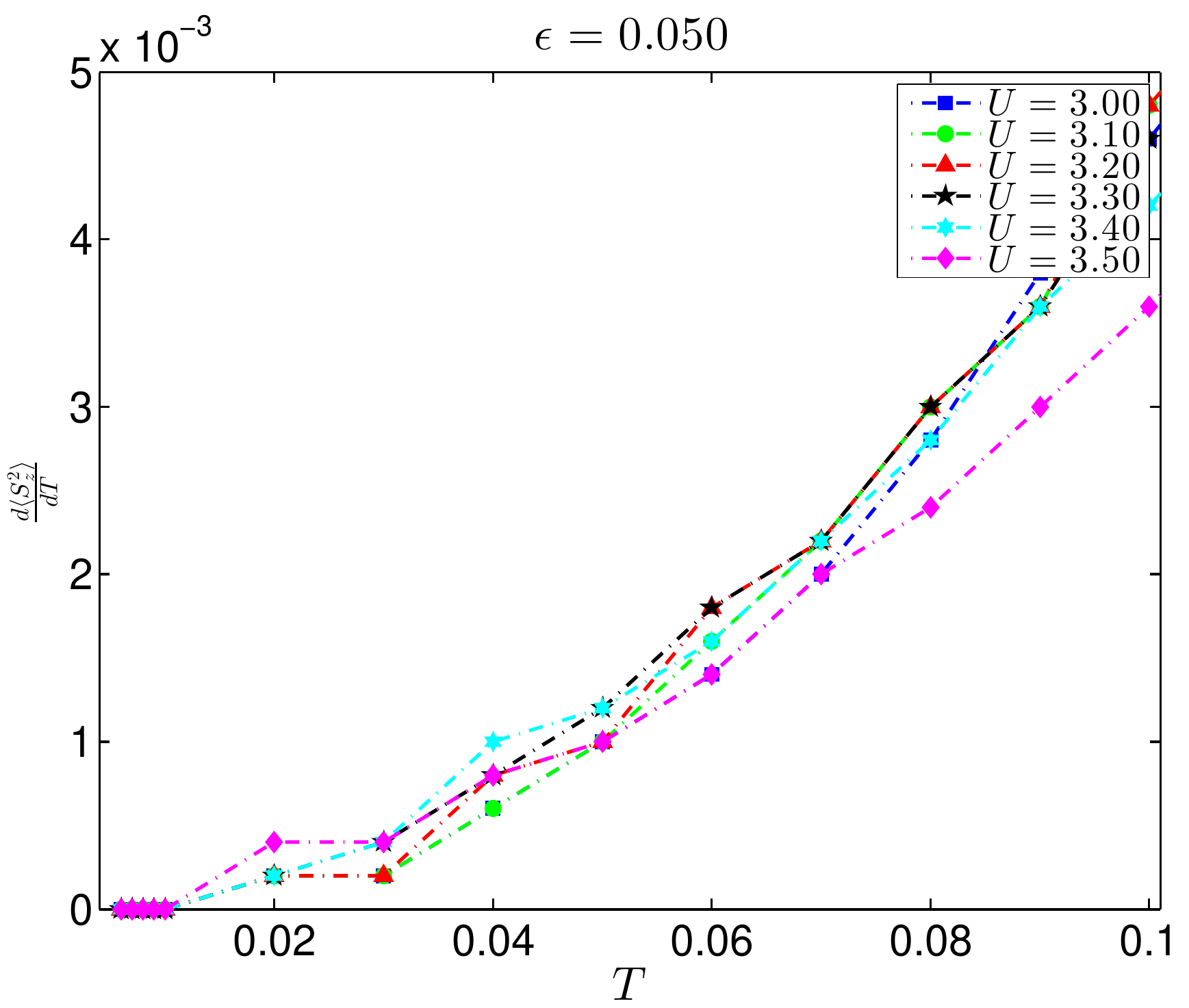}}
\subfloat{\includegraphics[scale = 0.3]{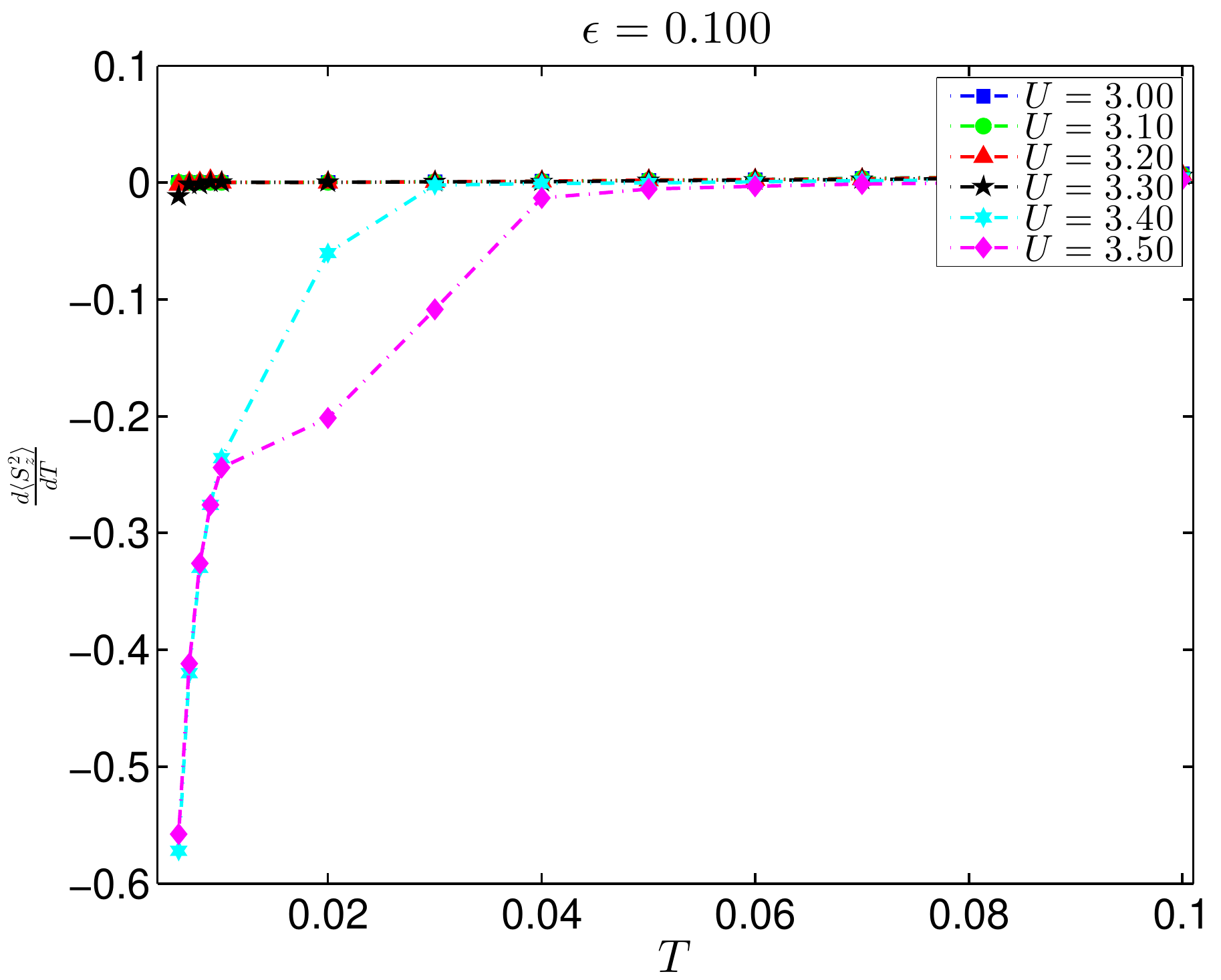}}
\subfloat{\includegraphics[scale = 0.3]{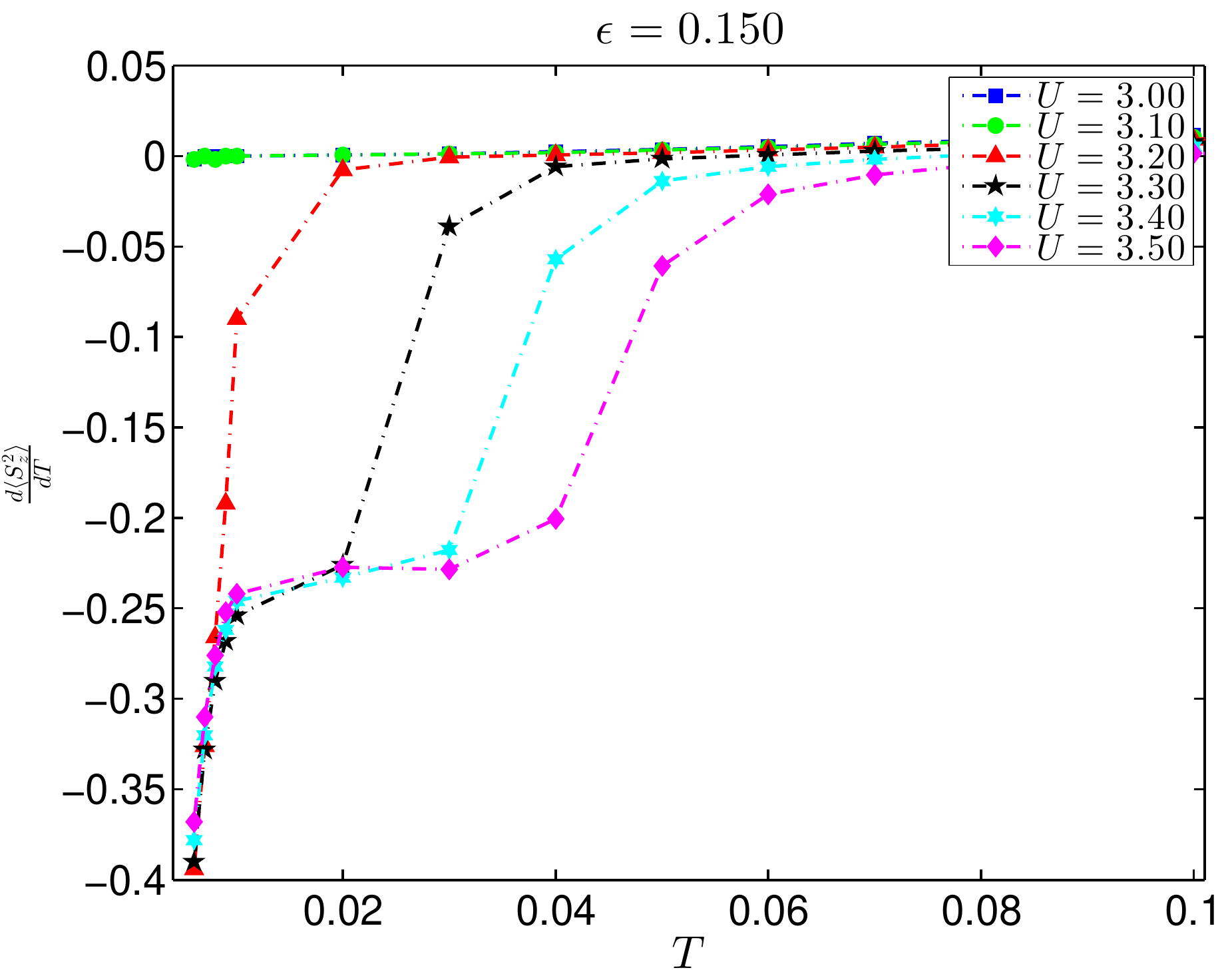}}\\
\subfloat{\includegraphics[scale = 0.3]{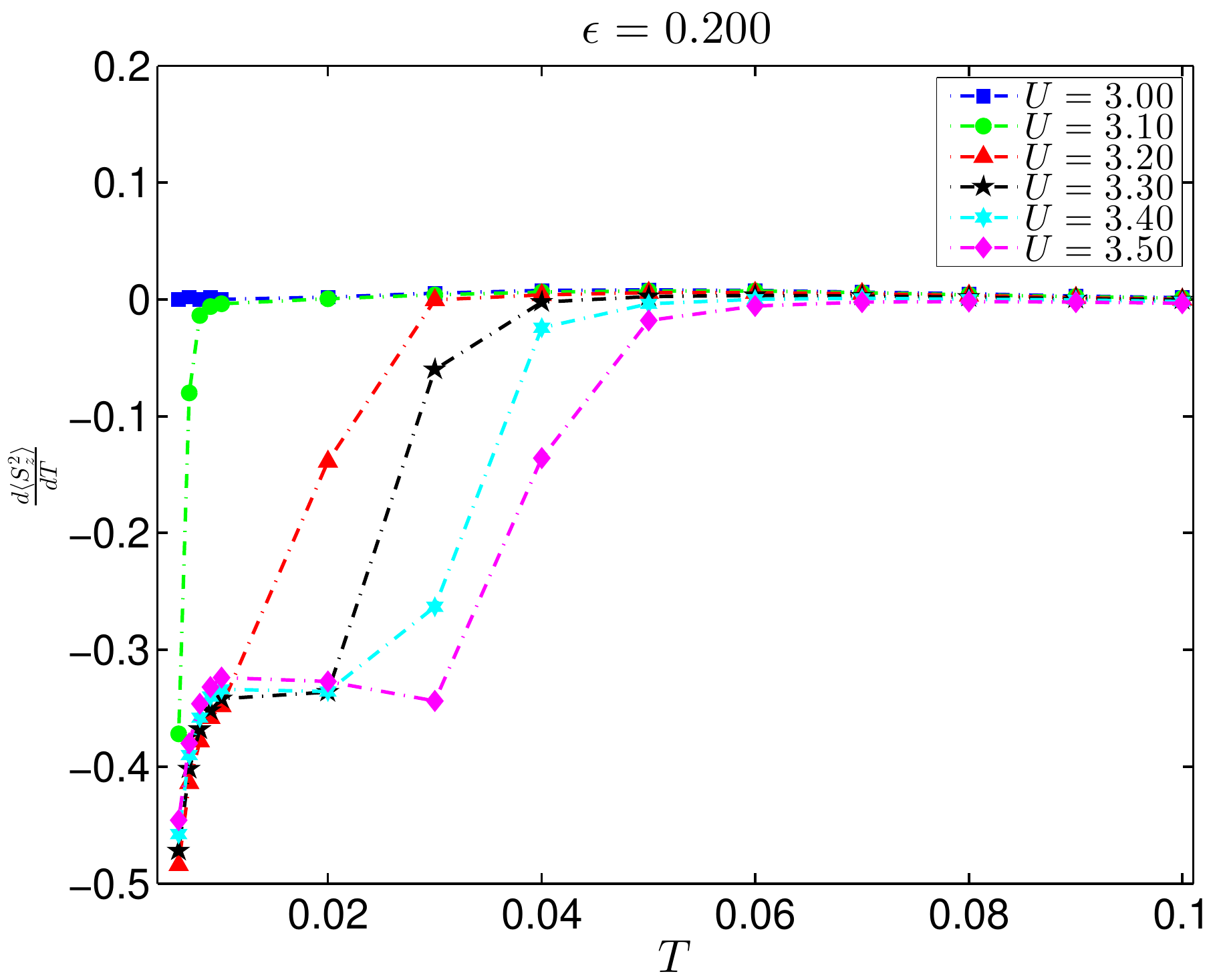}}
\subfloat{\includegraphics[scale = 0.3]{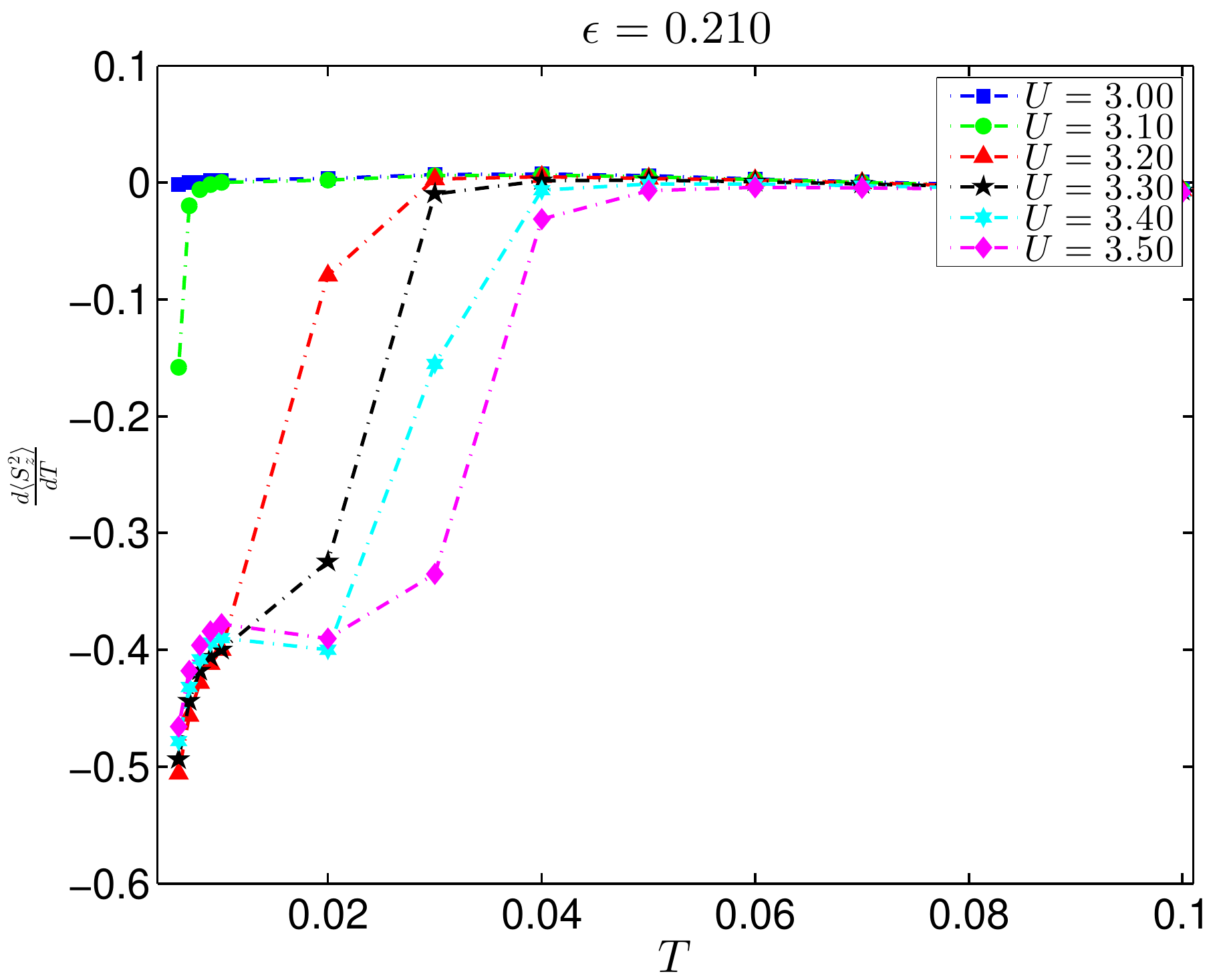}}
\subfloat{\includegraphics[scale = 0.3]{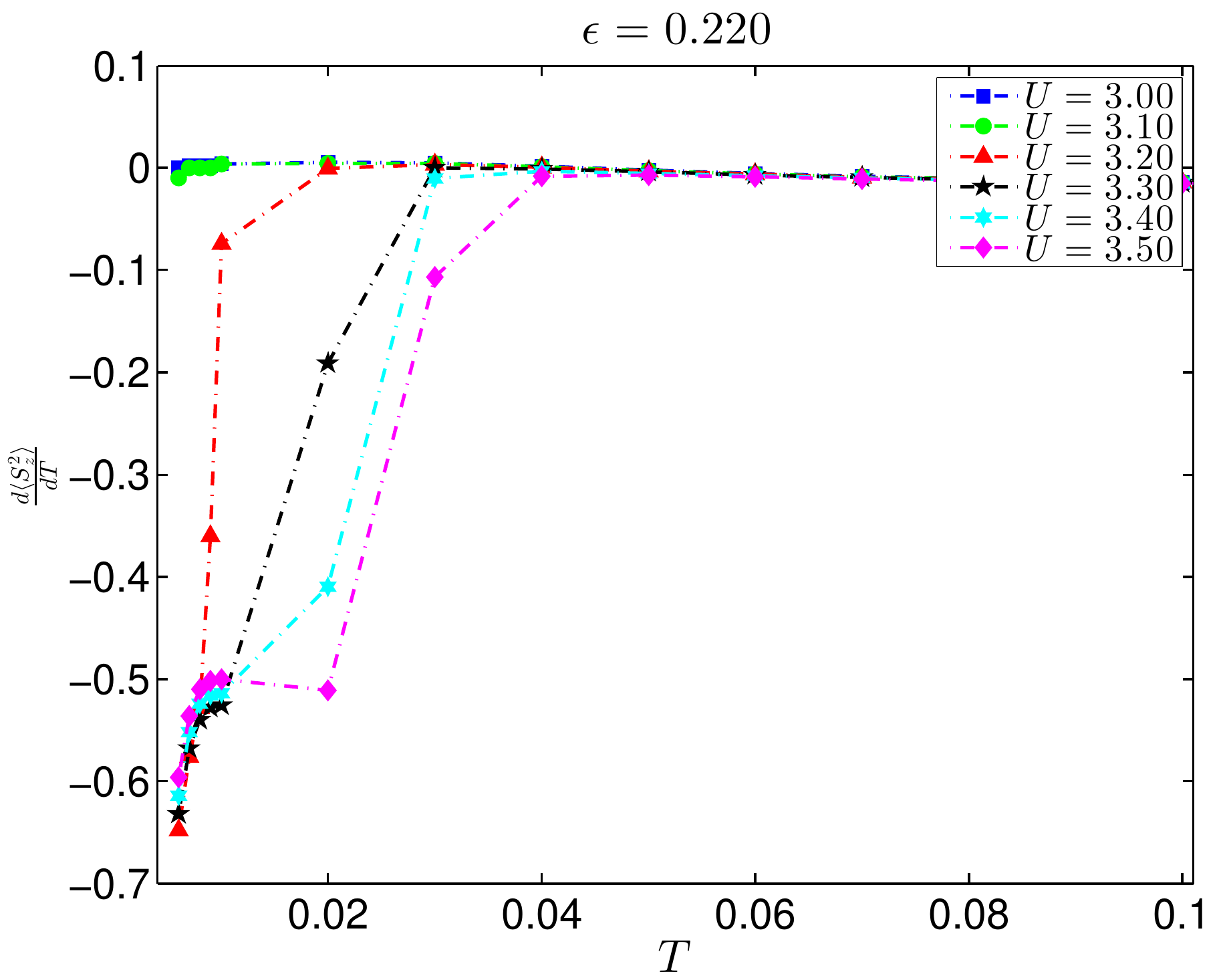}}
\caption{Plots of $\frac {d \langle S_{z}^{2} \rangle}{dT}$ as a function of temperature $T$, for various values of strain $\epsilon$ below the critical strain. The legend indicates the values of interaction $U$.}
\label{fig_der_Z_till_eps_crit}
\end{figure}

Similarly, the derivative of $\langle S_{z}^{2}\rangle$ with respect to temperature $T$ ($\frac {d \langle S_{z}^{2} \rangle}{dT}$)is plotted in Fig. \ref{fig_der_Z_till_eps_crit} for various values of strain. For $T \rightarrow 0$, for $0.1 < \epsilon \leq 0.23$, the derivative tends to diverge beyond $U > U_{c} (\epsilon)$. This, as previously stated in the main text refer to quantum criticality associated with large antiferromagnetic fluctuations.  

\begin{figure}[!ht]
\centering
\subfloat{\includegraphics[scale = 0.45]{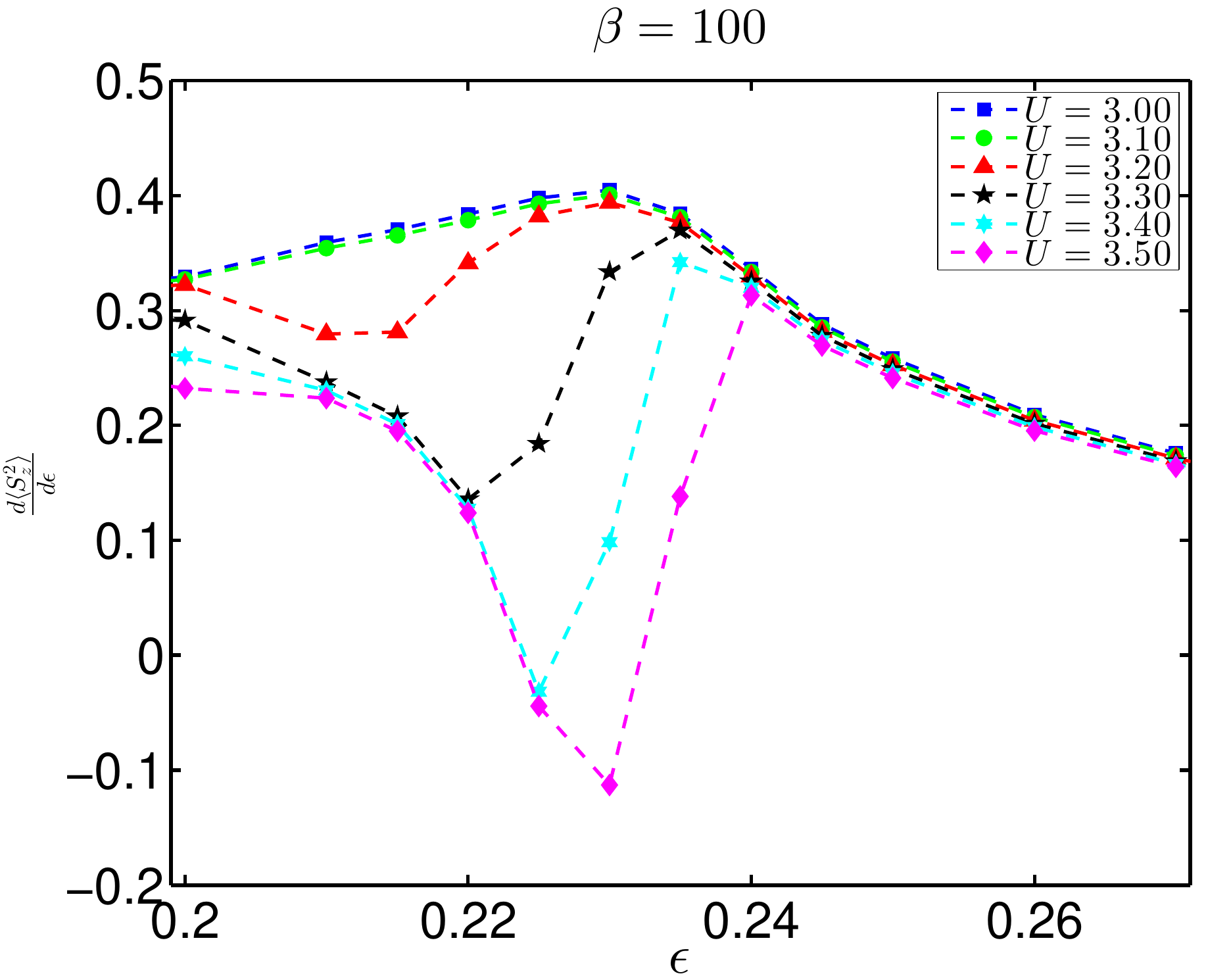}}
\caption{Plots of $\frac {d \langle S_{z}^{2} \rangle}{d\epsilon}$ as a function of strain $\epsilon$, for $\beta = 100$. The values of interaction $U$ are given in the legend.}
\label{fig_der_Z_vs_eps}
\end{figure}

This can be further substantiated by plotting the derivative of $\langle S_{z}^{2}\rangle$ with respect to strain $\epsilon$ ($\frac {d \langle S_{z}^{2} \rangle}{d\epsilon}$) , as a function of strain for various values of temperature as shown in Fig. \ref{fig_der_Z_vs_eps}. For low values of temperature, beyond $U > U_{c}(\epsilon)$, the singularity in the derivative points to the existence of a phase transition driven by the growing antiferromagnetic fluctuations. 
\section {Scenario beyond critical strain}
\begin{figure}[!ht]
\centering
\subfloat{\includegraphics[scale = 0.45]{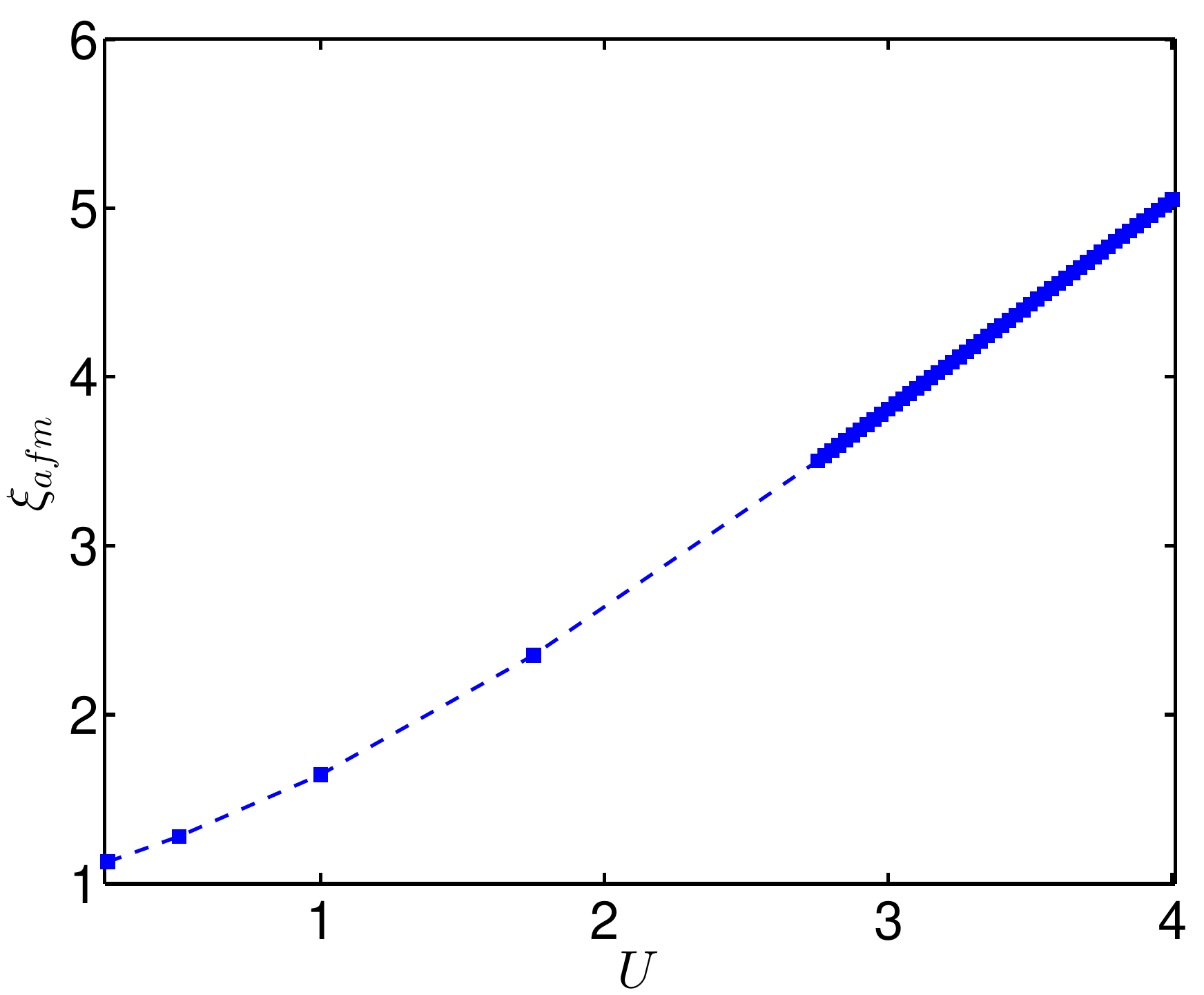}}
\caption{Plot of antiferromagnetic correlation length ($\xi_{afm}$) for $\epsilon = 0.3$, at $\beta = 100$. It can be clearly seen that the antiferromagnetic fluctuations do not grow large.}
\label{fig_prelim_results_beyond_eps_crit}
\end{figure}

\begin{figure}[!ht]
\centering
\subfloat{\includegraphics[scale = 0.3]{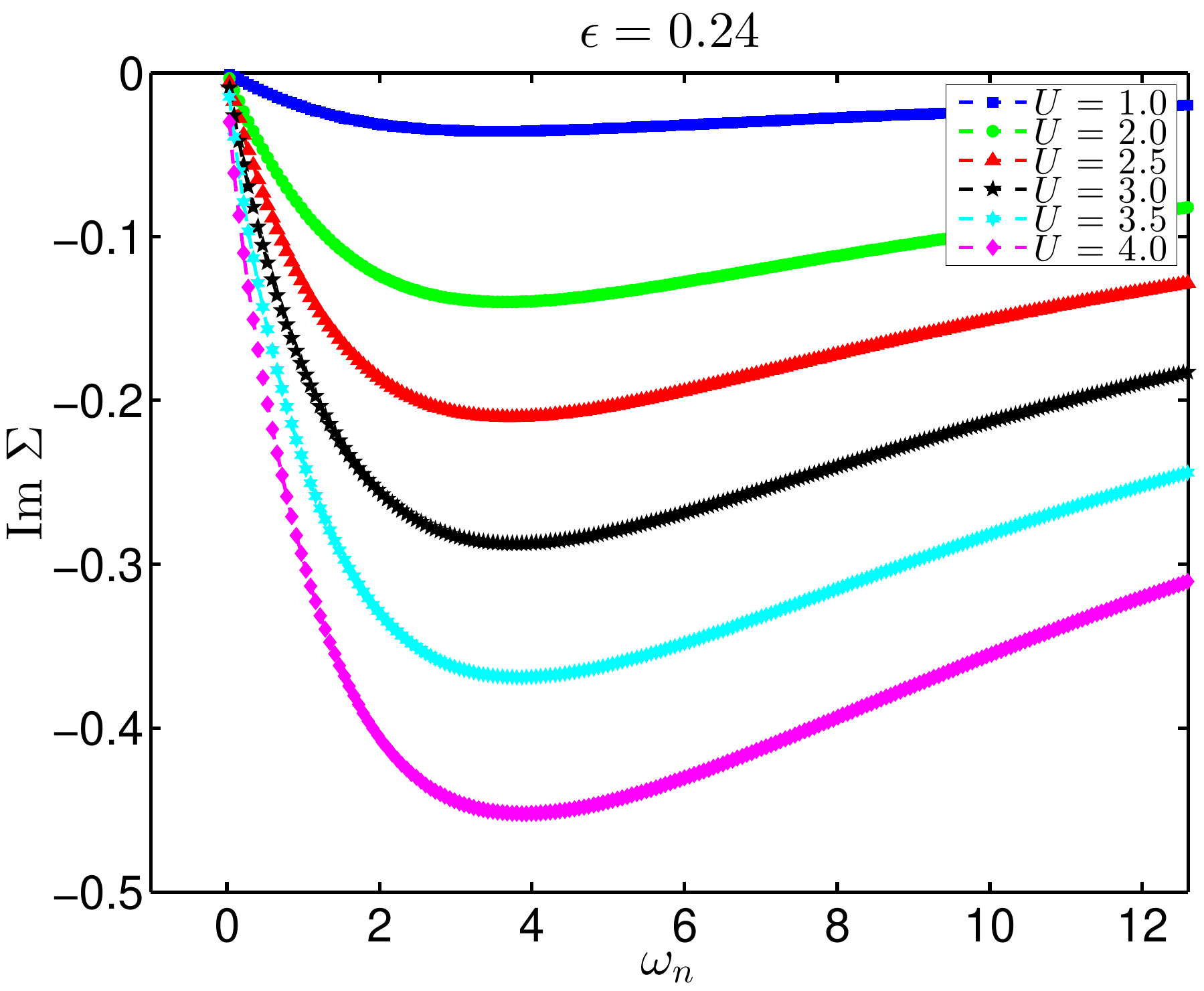}}
\subfloat{\includegraphics[scale = 0.3]{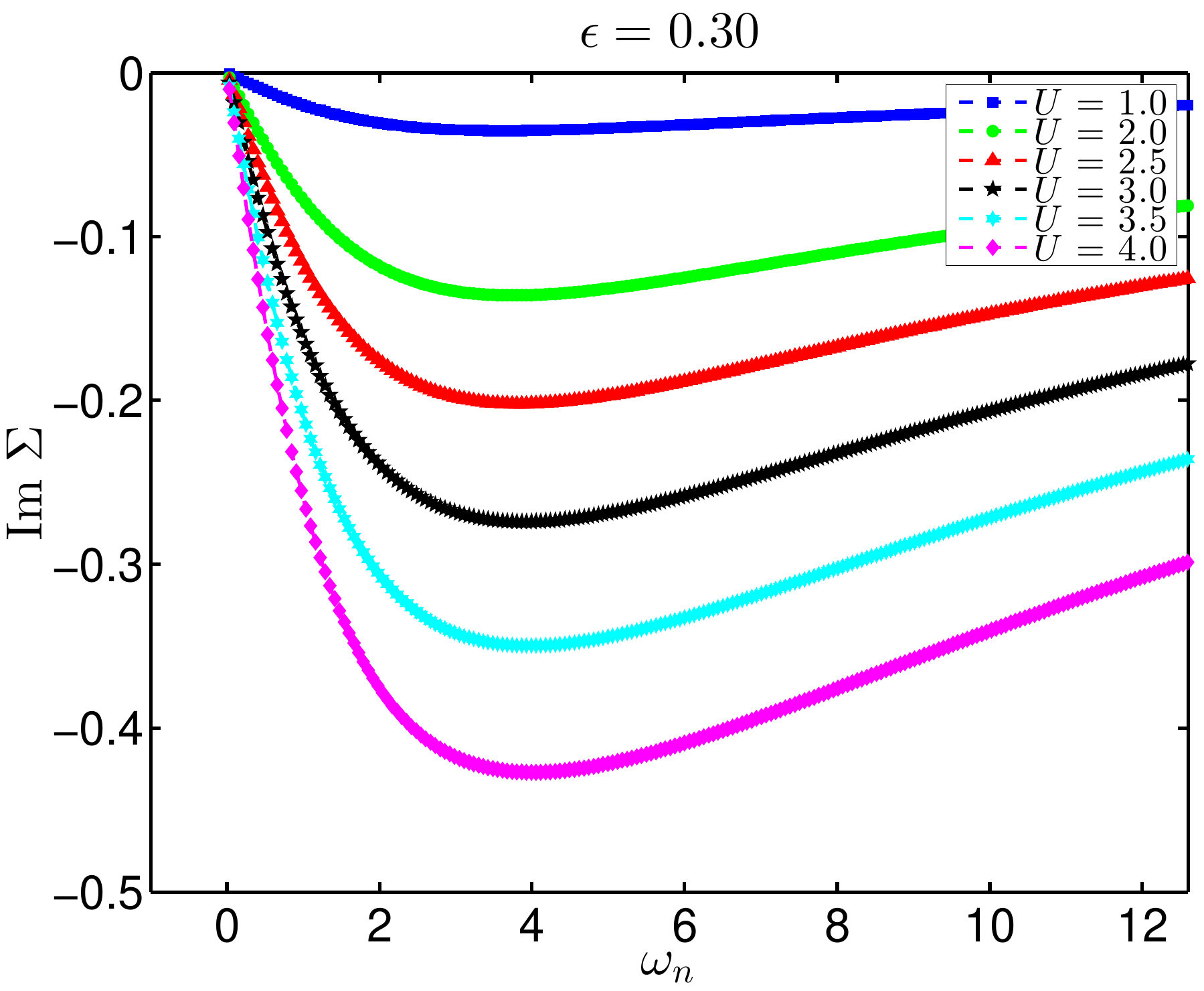}}
\subfloat{\includegraphics[scale = 0.3]{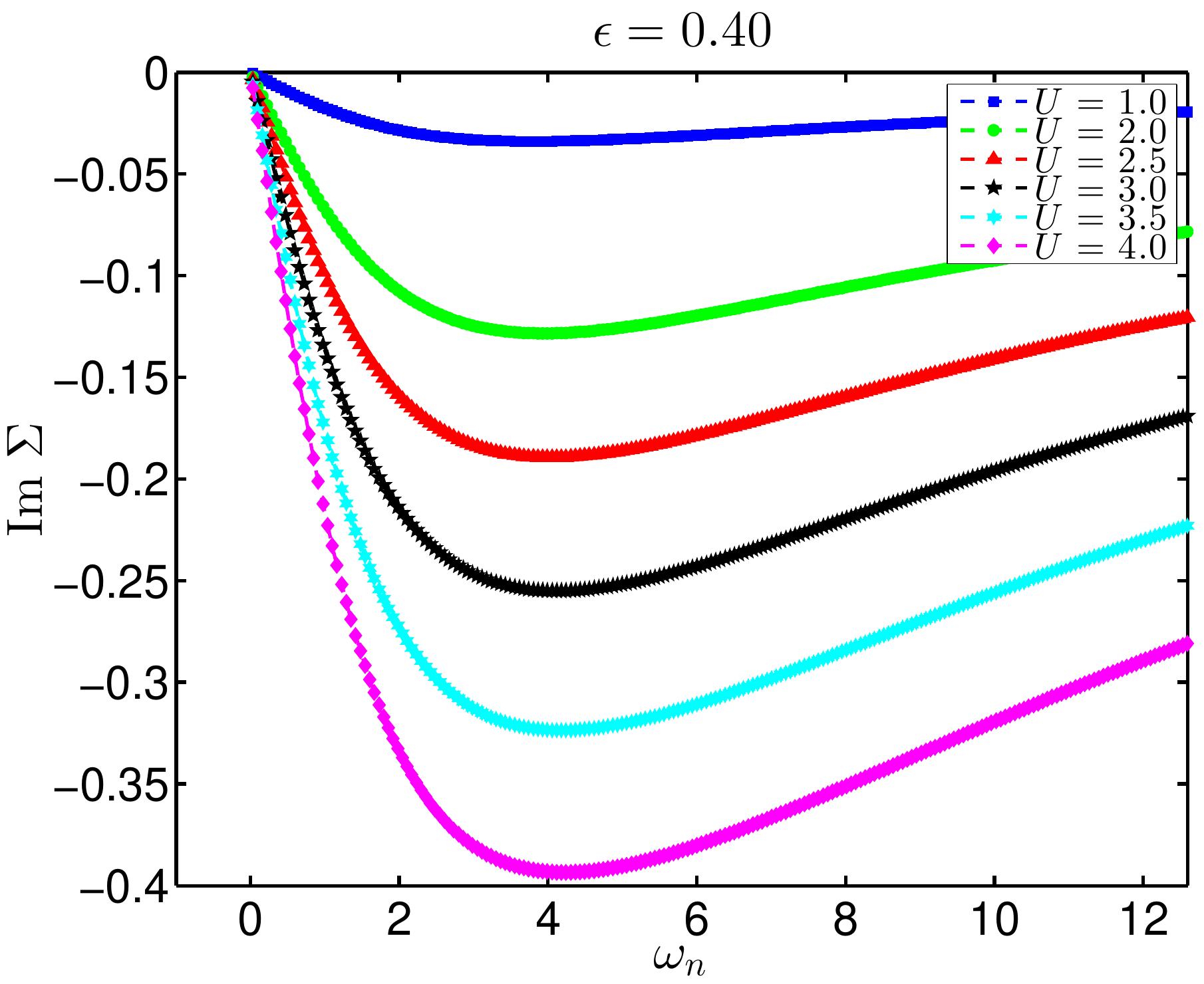}}
\caption{Imaginary part of the self energy $\Sigma$ plotted as a function of Matsubara frequency $\omega_{n}$, for various values of strain $\epsilon$ greater than critical strain for $\beta = 100$. The values of interaction $U$ are shown in the legend.}
\label{fig_se_beyond_eps_crit}
\end{figure}

\begin{figure}[!ht]
\centering
\subfloat{\includegraphics[scale = 0.3]{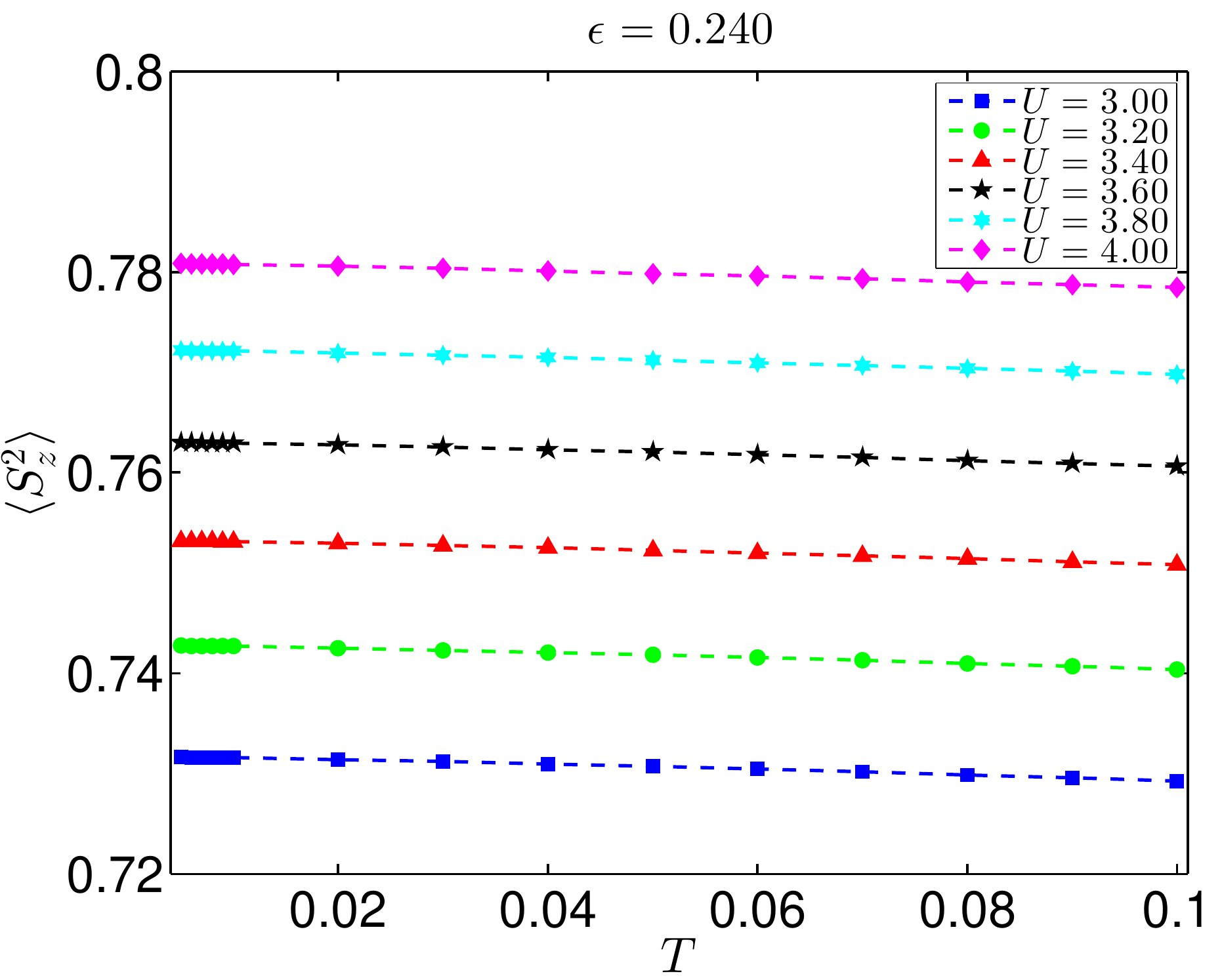}}
\subfloat{\includegraphics[scale = 0.3]{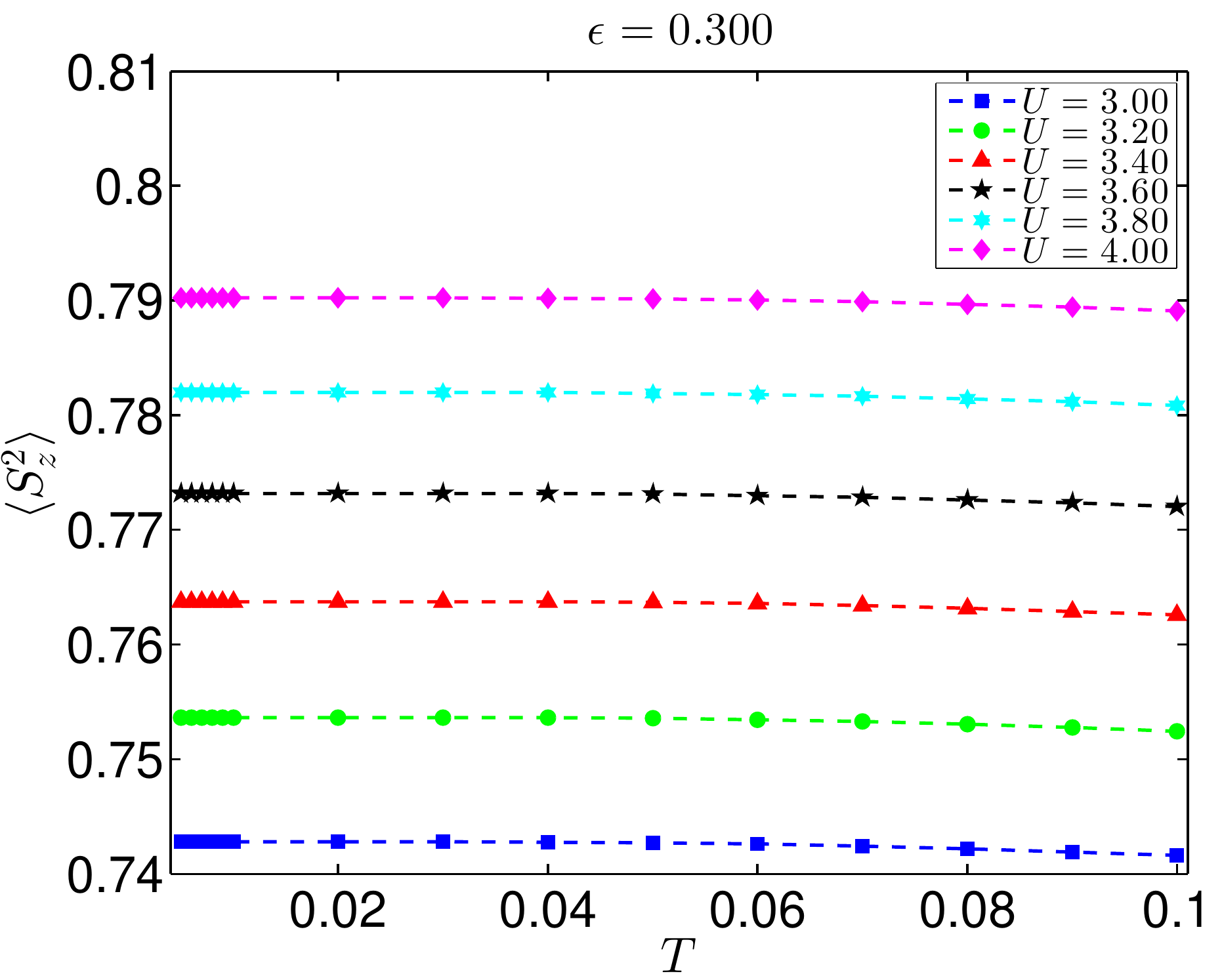}}
\subfloat{\includegraphics[scale = 0.3]{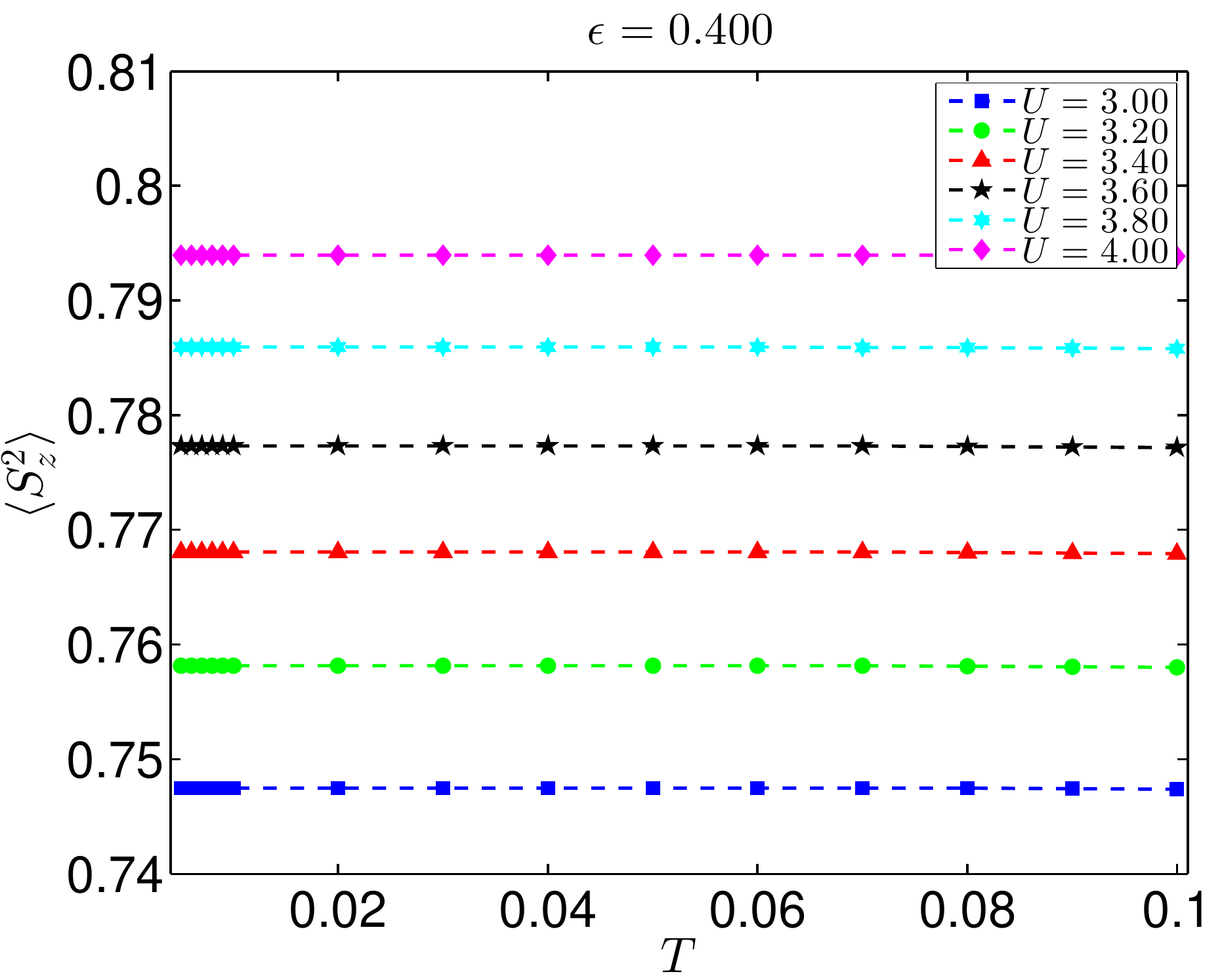}}
\caption{Plots of $\langle S_{z}^{2} \rangle = 1 - 2 \langle n_{\uparrow} n_{\downarrow} \rangle$ as a function of temperature $T$, for various values of strain $\epsilon$ greater than the critical strain. The legend indicates the values of interaction $U$.}
\label{fig_Z_beyond_eps_crit}
\end{figure}

\begin{figure}[!ht]
\centering
\subfloat{\includegraphics[scale = 0.3]{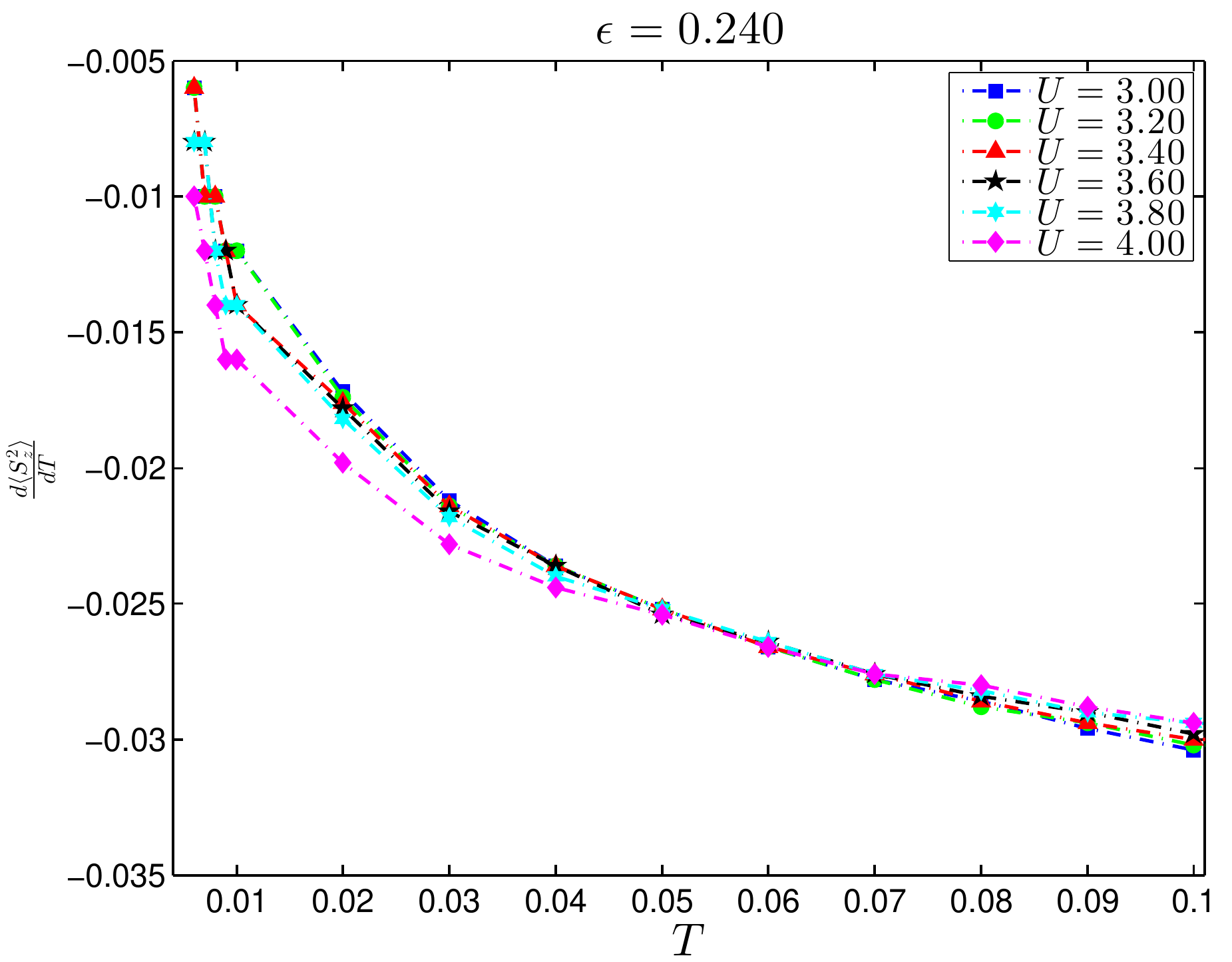}}
\subfloat{\includegraphics[scale = 0.3]{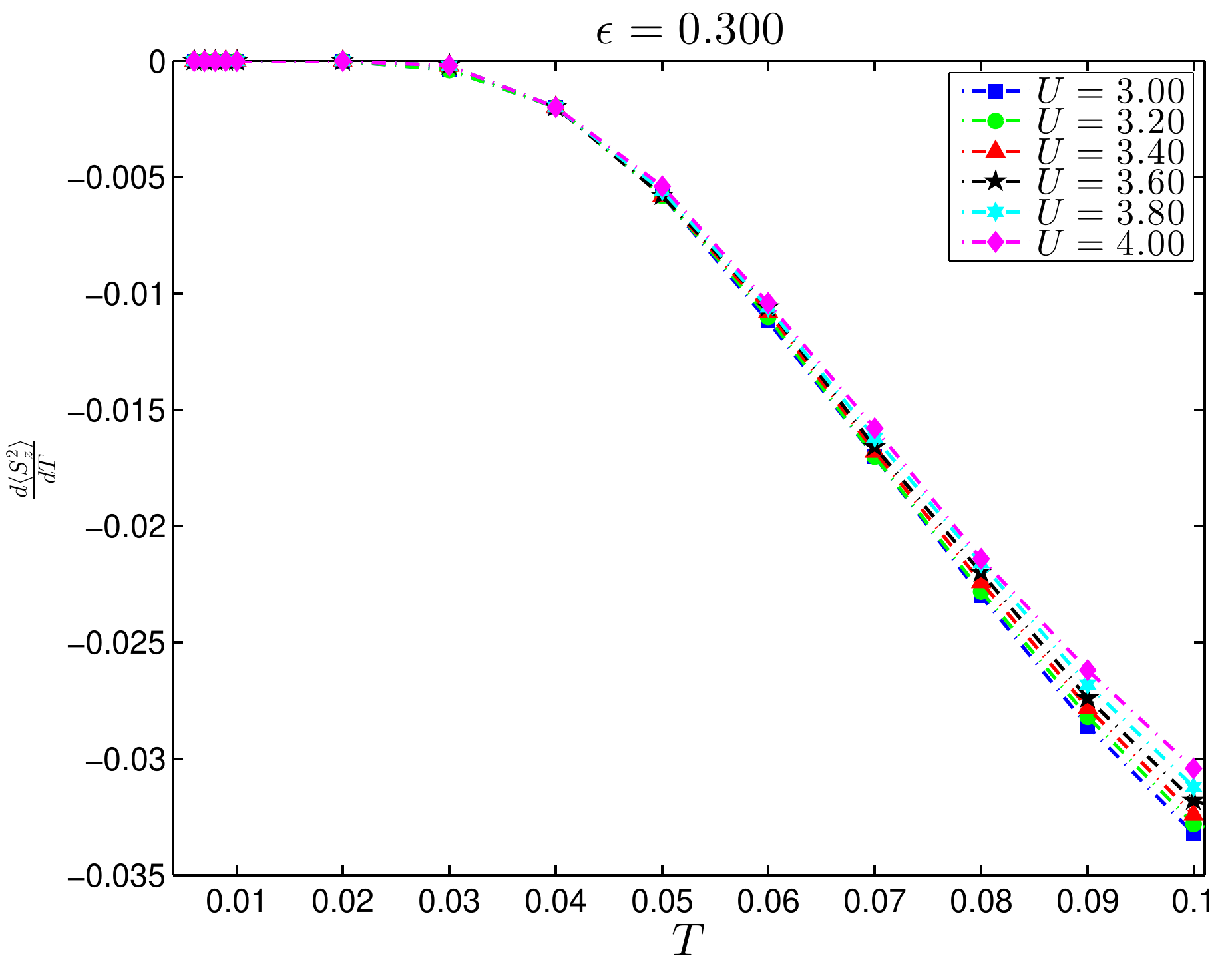}}
\subfloat{\includegraphics[scale = 0.3]{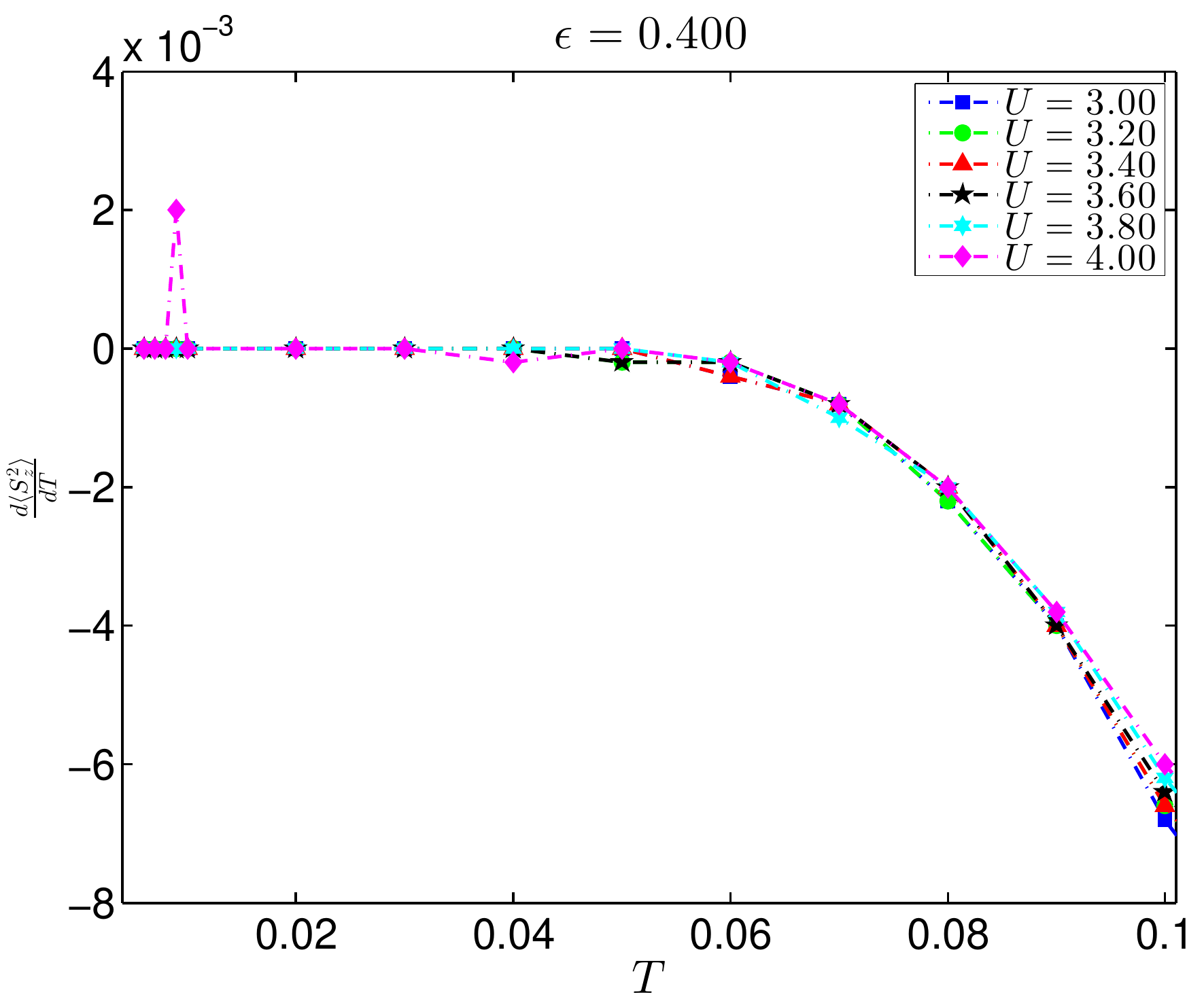}}
\caption{Plots of $\frac {d \langle S_{z}^{2} \rangle}{dT}$ as a function of temperature $T$, for various values of strain $\epsilon$ greater than the critical strain. The legend indicates the values of interaction $U$.}
\label{fig_der_Z_beyond_eps_crit}
\end{figure}

For values of strain beyond the critical strain $\epsilon \geq 0.24$, the system becomes a band insulator. The main result is that even at low temperatures, the antiferromagnetic fluctuations do not become huge as seen from the plots in Fig. \ref{fig_prelim_results_beyond_eps_crit}. These corroborate the statement in the main text that the features of quantum criticality are influenced by the growing antiferromagnetic fluctuations in the system.
Since the antiferromagnetic fluctuations do not become huge, we do not expect the self energy to show any anomalous feature as $\omega_{n} \rightarrow 0$. This can be seen clearly from Fig. \ref{fig_se_beyond_eps_crit} where we have plotted imaginary part of the self-energy as a function of the fermionic Matsubara frequency, for values of strain beyond the critical strain.
The plots of $\langle S_{z}^{2} \rangle$ in Fig. \ref{fig_Z_beyond_eps_crit} and the derivative of $\frac {d \langle S_{z}^{2} \rangle}{dT}$ in Fig. \ref{fig_der_Z_beyond_eps_crit} also give us the same information. From the plots of Fig. \ref{fig_Z_beyond_eps_crit}, we can see that $\langle S_{z}^{2} \rangle$  is almost temperature independent, for the same values of interaction at which we see large antiferromagnetic fluctuations in the system below critical strain. And, $\frac {d \langle S_{z}^{2} \rangle}{dT}$ does not diverge as $T \rightarrow 0$, and is almost independent of the value of interaction.
%


\end{document}